\newif\ifconf
\newif\iftr
\newif\ifEXT
\EXTfalse

\trtrue
\conffalse

\ifconf
\documentclass[conference]{IEEEtran}
\else
\documentclass[conference]{IEEEtran_custom}
\fi

\usepackage{cite}

\usepackage{amsmath,amssymb,amsfonts}
\usepackage{algorithmic}
\usepackage{graphicx}
\usepackage{textcomp}
\usepackage{xcolor}
\def\BibTeX{{\rm B\kern-.05em{\sc i\kern-.025em b}\kern-.08em
    T\kern-.1667em\lower.7ex\hbox{E}\kern-.125emX}}

\newif\ifsq     

\newif\ifsqCAP
\newif\ifsqVS
\newif\ifsqEN
\newif\ifsqTIT

\newcommand{\ignore}[1]{}

\sqfalse
\sqCAPfalse
\sqENfalse
\sqVSfalse
\sqTITfalse

\ifconf
\sqtrue
\sqCAPtrue
\sqENtrue
\sqVStrue
\sqTITtrue
\fi

\usepackage{balance}
\usepackage{epstopdf}
\usepackage{placeins}

\usepackage{graphicx}
\usepackage{float}
\usepackage{multirow}
\usepackage{rotating}
\usepackage{makecell}
\usepackage{tabulary}
\usepackage{parcolumns}
\usepackage{tikz}
\usetikzlibrary{tikzmark}

\usepackage{xpatch}
\expandafter\xpatchcmd
\csname pgfk@/tikz/every picture/.@cmd\endcsname
{\thepage}{\arabic{page}}{}{}

\tikzstyle{comment} = [draw, fill=blue!70, text=white, text width=3cm, minimum height=1cm, rounded corners, align=left, font=\scriptsize]
\tikzstyle{background_alg} = [draw, fill=blue!20, opacity=0.4, inner sep=4pt, rounded corners=2pt]

\usetikzlibrary{shapes}
\usetikzlibrary{plotmarks}
\usetikzlibrary{calc, fit}

\usepackage{enumitem}

\usepackage{amsthm}
\usepackage{amsmath,amsfonts}
\usepackage{mathtools}

\newtheorem{theorem}{Theorem}[section]

\newtheorem{definition}[theorem]{Definition}

\usepackage{soul}
\usepackage{fontawesome}
\usepackage{pifont}
\usepackage{textcomp}
\usepackage{booktabs}
\usepackage{url}
\usepackage{pbox}
\usepackage[normalem]{ulem}
\usepackage[10pt]{moresize}

\newcommand{\macb}[1]{\textbf{{#1}}}

\usepackage[font={normalfont, scriptsize}]{caption}
\usepackage[font={normalfont, scriptsize}]{subcaption}
\newcommand{\vspaceSQ}[1]{\ifsqVS\vspace{#1}\fi}
\newcommand{\enlargeSQ}[1]{\ifsqEN\enlargethispage{\baselineskip}\fi}

\ifsqTIT
\usepackage[compact]{titlesec}
\titlespacing*{\section}{0pt}{6pt}{3pt}
\titlespacing*{\subsection}{0pt}{4pt}{2pt}
\titlespacing*{\subsubsection}{0pt}{2pt}{1pt}
\fi

\usepackage{xcolor}
\definecolor{darkgrey}{RGB}{70,70,70}
\definecolor{lightgrey}{RGB}{200,200,200}
\definecolor{lyellow}{RGB}{255,255,100}
\definecolor{llyellow}{RGB}{250,250,180}
\definecolor{lgreen}{RGB}{144,238,144}
\definecolor{raphael_comments}{RGB}{13, 145, 24}

\usepackage[customcolors]{hf-tikz}
\hfsetbordercolor{white}
\hfsetfillcolor{vlgray}

\definecolor{vlgray}{rgb}{0.77 0.77 0.77}
\definecolor{ablack}{rgb}{0.2 0.2 0.2}
\definecolor{vllgray}{rgb}{0.9 0.9 0.9}
\definecolor{bblue}{rgb}{0.7 0.7 0.99}

\usepackage{colortbl}

\usepackage{listings}

\ifsq
\else
\fi

\definecolor{hlL}{rgb}{0.8 0.8 0.99}

\newcounter{highlight}

\newcounter{hlLR}

\newcounter{hlLIR}

\newcounter{hlLIIR}

\newcounter{Ahighlight}

\renewcommand{\epsilon}{\ensuremath\varepsilon}

\renewcommand{\phi}{\ensuremath{\varphi}}

\usepackage[hidelinks]{hyperref}

\renewcommand{\arraystretch}{1.0}

\DeclareMathOperator{\rotl}{rotl}

\begin{document}

\iftr
\title{EvalNet: A Practical Toolchain for Generation and Analysis of Extreme-Scale Interconnects \vspaceSQ{-1em}}
\fi

\ifconf
\author{\IEEEauthorblockN{Maciej Besta}
\IEEEauthorblockA{\textit{Department of Computer Science} \\
\textit{ETH Zurich}\\
Zurich, Switzerland\\
0000-0002-6550-7916}
\and
\IEEEauthorblockN{Patrick Iff}
\IEEEauthorblockA{\textit{Department of Computer Science} \\
\textit{ETH Zurich}\\
Zurich, Switzerland\\
0000-0001-5979-4915}
\and
\IEEEauthorblockN{Marcel Schneider}
\IEEEauthorblockA{\textit{Department of Computer Science} \\
\textit{ETH Zurich}\\
Zurich, Switzerland}
\and
\IEEEauthorblockN{Nils Blach}
\IEEEauthorblockA{\textit{Department of Computer Science} \\
\textit{ETH Zurich}\\
Zurich, Switzerland\\
0009-0003-0401-0388}
\and
\IEEEauthorblockN{Alessandro Maissen}
\IEEEauthorblockA{\textit{Department of Computer Science} \\
\textit{ETH Zurich}\\
Zurich, Switzerland\\
0009-0003-9239-8748}
\and
\IEEEauthorblockN{Salvatore Di Girolamo}
\IEEEauthorblockA{\textit{Department of Computer Science} \\
\textit{ETH Zurich}\\
Zurich, Switzerland\\
0000-0003-2197-8860}
\and
\IEEEauthorblockN{Jens Domke}
\IEEEauthorblockA{\textit{Supercomputing Performance Research} \\
\textit{RIKEN}\\
Wakō, Japan\\
0000-0002-5343-414X}
\and
\IEEEauthorblockN{Jascha Krattenmacher}
\IEEEauthorblockA{\textit{Department of Computer Science} \\
\textit{ETH Zurich}\\
Zurich, Switzerland\\
0009-0003-6516-9683}
\and
\IEEEauthorblockN{Kartik Lakhotia}
\IEEEauthorblockA{
\textit{Intel Labs}\\
Santa Clara, CA, USA\\
0000-0002-9414-8481}
\and
\IEEEauthorblockN{Laura Monroe}
\IEEEauthorblockA{
\textit{Los Alamos National Laboratory}\\
Los Alamos, NM, USA\\
0000-0002-7175-0103}
\and
\IEEEauthorblockN{Fabrizio Petrini}
\IEEEauthorblockA{
\textit{Intel Labs}\\
Santa Clara, CA, USA\\
0000-0002-4977-7107}
\and
\IEEEauthorblockN{Robert Gerstenberger}
\IEEEauthorblockA{\textit{Department of Computer Science} \\
\textit{ETH Zurich}\\
Zurich, Switzerland\\
0009-0003-3129-4689}
\and
\IEEEauthorblockN{Torsten Hoefler}
\IEEEauthorblockA{\textit{Department of Computer Science} \\
\textit{ETH Zurich}\\
Zurich, Switzerland\\
0000-0002-1333-9797}
}
\else
\author{Maciej Besta$^{1*}$,
Patrick Iff$^1$,
Marcel Schneider$^1$,
Nils Blach$^1$,
Alessandro Maissen$^1$,\\
Salvatore Di Girolamo$^1$,
Jens Domke$^2$,
Jascha Krattenmacher$^1$,
Kartik Lakhotia$^3$,
Laura Monroe$^4$,\\
Fabrizio Petrini$^3$,
Robert Gerstenberger$^1$,
Torsten Hoefler$^1$\\
{\small $^1$Department of Computer Science, ETH Zurich; $^2$RIKEN Center for Computational Science (R-CCS);}\\
{\small $^3$Intel Labs; $^4$Los Alamos National Laboratory}\\
{\small $^*$Corresponding author}
}
\fi

\maketitle

\begin{abstract}
The diversity of communication paths in a network—especially non-minimal paths—is a key enabler of performance at extreme scales. We present EvalNet, a toolchain for scalable generation and analysis of over 25 important network topologies, such as Slim Fly, PolarFly, and Orthogonal Fat Trees, with a strong focus on path diversity metrics. EvalNet provides an extensive and fine-grained analysis of shortest and non-shortest paths, including their multiplicities, lengths, and interference. It supports exact measurement and visualization of bandwidth and throughput between every router pair, enabling unprecedented insight into routing potential. EvalNet also includes detailed models for construction cost and power consumption, and interfaces seamlessly with established simulators, which we tune to support large-scale evaluations on low-cost hardware. Using EvalNet, we deliver the widest and most comprehensive path diversity study to date, demonstrating how path diversity underpins throughput and scalability, and facilitating progress towards new frontiers in extreme-scale network design.
%
%
\end{abstract}

\iftr
\begin{IEEEkeywords}
Network evaluation, scalable simulation, large-scale simulation, packet-level simulation, data center networks, HPC networks, htsim
\end{IEEEkeywords}
\fi

{
\noindent\macb{Code:} \url{https://github.com/spcl/EvalNet} \\
\ifconf
\noindent\macb{Extended technical report:} \url{https://arxiv.org/abs/2105.12663}
\else
\noindent\macb{IPDPS '26:} \url{https://ieeexplore.ieee.org/document/11575429}
\fi
}

\section{INTRODUCTION}
\label{sec:intro}

\iftr
The interconnection network plays an important role in today's large-scale
\fi
computing systems~\cite{besta2014slim, niranjan2009portland, handley2017re,
valadarsky2015, besta2020highrouting, di2019network, copik2021sebs, hoefler2020high, di2022building}. Large networks with tens of thousands of
nodes are deployed in warehouse-sized HPC and data centers~\cite{fu2016sunway,
liao2014milkyway}. Future supercomputers as well as mega data centers
will require even larger scales with hundreds of thousands of servers.

A modern network topology must satisfy many stringent requirements related to
different aspects of the interconnect design, such as high performance or low
construction cost. However, the evaluation of all these aspects is challenging.
First, there are many networks that must be considered for comparison when
designing a new topology, for example Fat Tree~\cite{leiserson1996cm5},
Hypercube~\cite{Dally:2003:PPI:995703},
Torus~\cite{Dally:2003:PPI:995703}, Dragonfly~\cite{dally08}, Slim
Fly~\cite{besta2014slim}, Xpander~\cite{ahn2009hyperx}, or
Jellyfish~\cite{singla2012jellyfish} -- to name only a few. On top of that,
many of these baseline networks, for example Xpander or Orthogonal Fat Trees, do not have
straightforward construction algorithms, making it challenging and time-consuming 
to compare to these baselines.

Similarly, there are many performance measures that should be considered for
overall evaluation, and these measures often require different tools. A core challenge in evaluating interconnects is the analysis of path diversity, which has emerged as a critical enabler of performance and scalability in modern networks~\cite{besta2020fatpaths}. As networks grow to extreme sizes, there is a growing need for fine-grained and scalable tools that can assess not only the existence of diverse paths but also their structure, interference, and utility for multipath routing. In parallel, several complementary performance metrics must also be considered, such as throughput and latency. Throughput is ideally evaluated using packet-level simulators, which offer high accuracy but often do not scale to large topologies. Flow-based simulations scale more easily, but they tend to produce less accurate predictions~\cite{casanova2008simgrid}. Furthermore, additional aspects like construction cost and power consumption must be analyzed. Unfortunately, the tools supporting these various metrics often rely on different formats for specifying networks, which significantly complicates integrated and efficient evaluation.

\iftr
To address all these issues, we propose EvalNet: a publicly available toolchain
that facilitates
the integrated and comprehensive evaluation of networks (\textbf{contribution~1}).
EvalNet consists of several modules, some of which were developed from scratch,
while others are interfaces to existing tools which facilitate seamless operation.
First, EvalNet enables generation of a very large number of network topologies.
We cover the state-of-the-art class of low-diameter
topologies developed by optimizing the network structure towards the Moore
Bound ({Slim Fly}~\cite{besta2014slim}, {PolarFly}~\cite{lakhotia2022polarfly},
{BundleFly}~\cite{bundlefly_2020}, {Xpander}~\cite{valadarsky2015}), {Dragonfly
topologies} ({fully-connected}~\cite{dally08}, {Cascade}~\cite{aries}), random networks
({Jellyfish}~\cite{singla2012jellyfish}), torus-related topologies ({Mesh},
{Torus}, {Hypercube}~\cite{Dally:2003:PPI:995703}, {Express
Mesh}~\cite{jain2017partitioning}, {HyperX}~\cite{ahn2009hyperx}, {Flattened
Butterfly}~\cite{kim2007flattened}, {Tofu}~\cite{ajima2018tofu,
ajima2012tofu}), {indirect topologies} ({Fat Tree}~\cite{leiserson1996cm5},
{Multi-Layer Full-Mesh}~\cite{kathareios2015cost}, {Orthogonal Fat Tree}~\cite{kathareios2015cost},
{eXtended Generalized Fat Tree}~\cite{ohring1995generalized}, {k-ary
n-tree}~\cite{petrini1997k}), and Kautz networks ({Undirected
Kautz}~\cite{li2004graph}, {Arrangement Network}~\cite{day1992arrangement}).
\fi

\iftr
\begin{figure*}[t]
\centering
\includegraphics[width=0.7\textwidth]{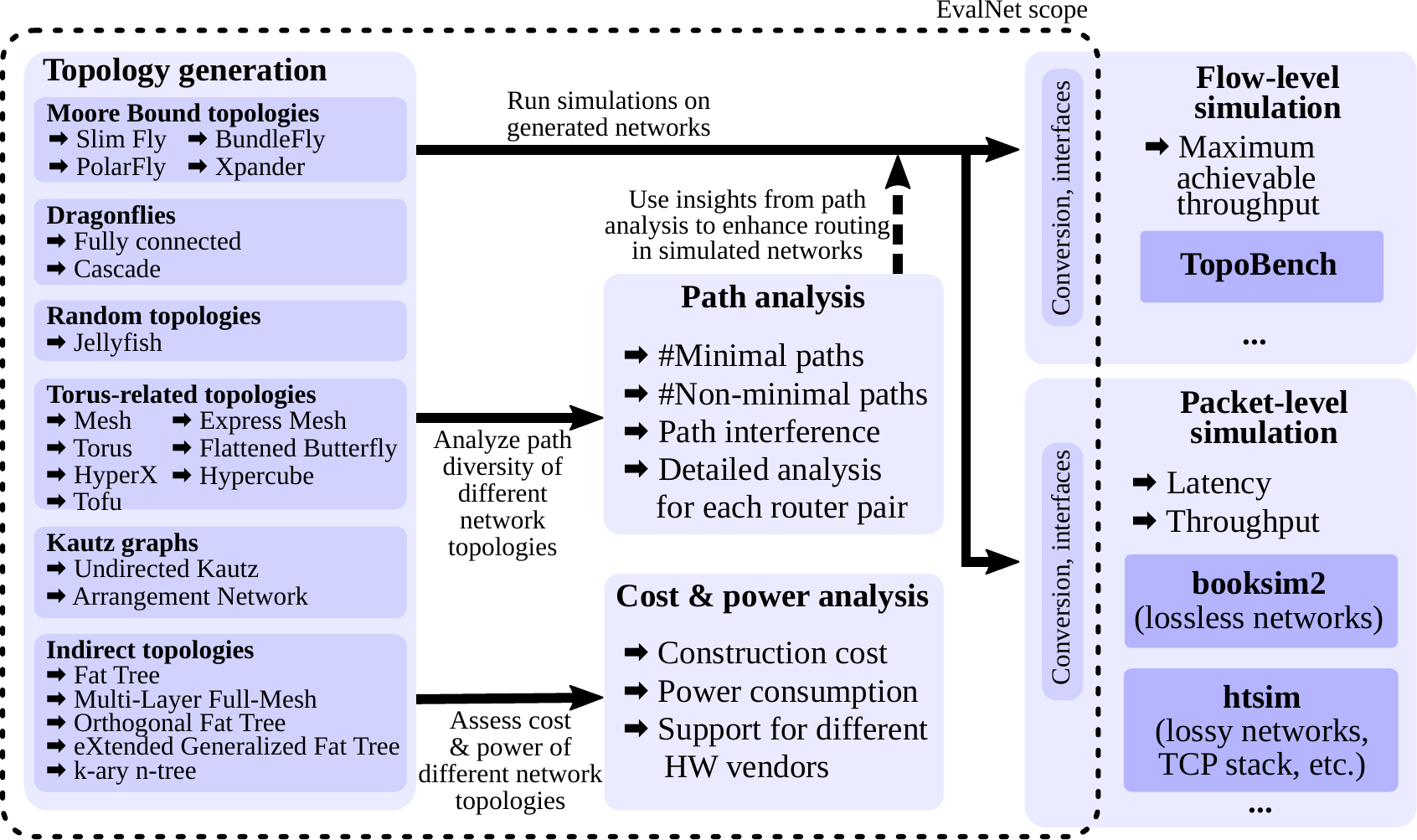}
\caption{An overview of EvalNet.}
\label{fig:overview}
\vspaceSQ{-1.0em}
%
%
\end{figure*}
\fi

Second, we observe that the high performance of network workloads is becoming increasingly dependent on \textit{effectively harnessing various forms of path diversity, particularly through non-minimal multipathing}~\cite{besta2020fatpaths, lakhotia2022polarfly}. This is especially critical in state-of-the-art low-diameter topologies, such as Slim Fly and PolarFly, which typically offer limited shortest-path diversity but compensate with a rich structure of non-minimal paths. In such networks, the ability to balance load and avoid congestion hinges on the capacity to exploit this non-shortest-path structure. To address this need, EvalNet provides a comprehensive suite of fine-grained tools for analyzing path diversity (\textbf{contribution~2}). It supports the computation of shortest and non-shortest path multiplicities, detailed distributions of path lengths, and interference-aware metrics that capture how routes interact with each other. EvalNet also includes visualization and measurement tools that expose
\iftr
the
\fi
exact bandwidth and multipathing potential between every pair of routers, enabling both coarse-grained insights and fine-grained optimization of routing and scheduling schemes~\cite{lakhotia2023network}. All path diversity analyses are designed to scale to very large topologies, supporting networks with more than 1 million servers, thus making them practical for evaluating future systems.

We design EvalNet to enable simulations of massive networks with different architectures even on low-cost commodity hardware (\textbf{contribution~3}).
For this, EvalNet harnesses two complementary packet-level simulators
(htsim~\cite{raiciu2010data} and Booksim2~\cite{jiang2016booksim}), which target, respectively, lossy
and lossless architectures (EvalNet does not compete with existing simulators as it does not introduce a new simulator. Instead, it tunes two of these simulators and harnesses them within its full comprehensive pipeline). We tune these simulators to support large-scale evaluations on low-cost hardware, \textit{achieving simulations of networks with more than 1 million endpoints on a commodity PC with just 16GB of RAM} without impacting the simulation accuracy. 
\iftr
On top of that, EvalNet also seamlessly interfaces with the flow-level evaluation tool
TopoBench~\cite{jyothi2016measuring} that enables approximating throughput for
arbitrary traffic patterns and comes with a specially designed worst-case scenario that
puts particularly large stress on the interconnect.
\fi
Finally, EvalNet also provides a construction cost and power consumption assessment module. The user can parametrize this module with arbitrary cost and power models for both routers and cables, considering arbitrary vendors and architectures such as Ethernet or InfiniBand.

Finally, we use EvalNet to conduct the widest-ever comparative analysis of interconnect topologies (\textbf{contribution~4}), focusing on the path diversity measures, but also combining them with classical metrics that include latency, global bandwidth, construction cost, and power consumption. We investigate path diversity in-depth because other measures (latency, bandwidth under flow/packet simulations, fault-tolerance) have already been extensively researched. However, path diversity has only been analyzed in a limited scope. We provide a total of 55 plots rich in path diversity information based on 4 different
\ifconf
metrics\footnote{\scriptsize Due to space limits, most plots are in the appendix of the extended technical report.},
\else
metrics,
\fi
combined with observations, analyses, and insights that can be used to develop better routing strategies.
%

\iftr
\begin{figure*}[t]
\centering
\includegraphics[width=1\textwidth]{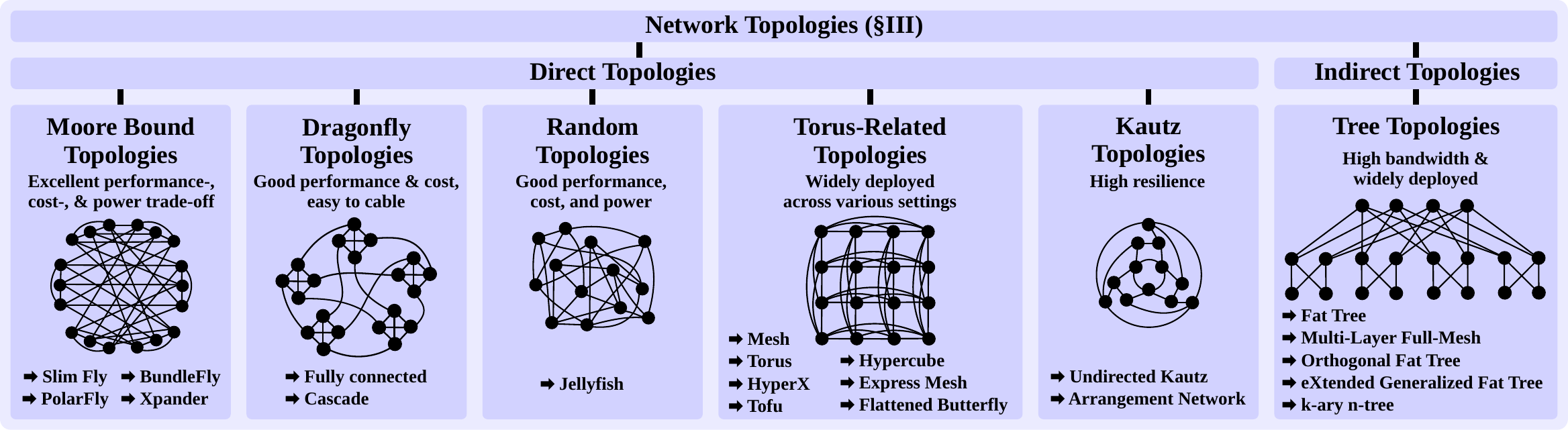}
\vspace{-1em}
\caption{An overview of network topologies offered by default in EvalNet. New topologies can be straightforwardly added to the pipeline.}
\label{fig:overview-topos}
\vspaceSQ{-1.5em}
%
\end{figure*}
\fi

\section{EVALNET TOOLCHAIN}

\ifconf
\begin{figure}[t]
\centering
\includegraphics[width=1\columnwidth]{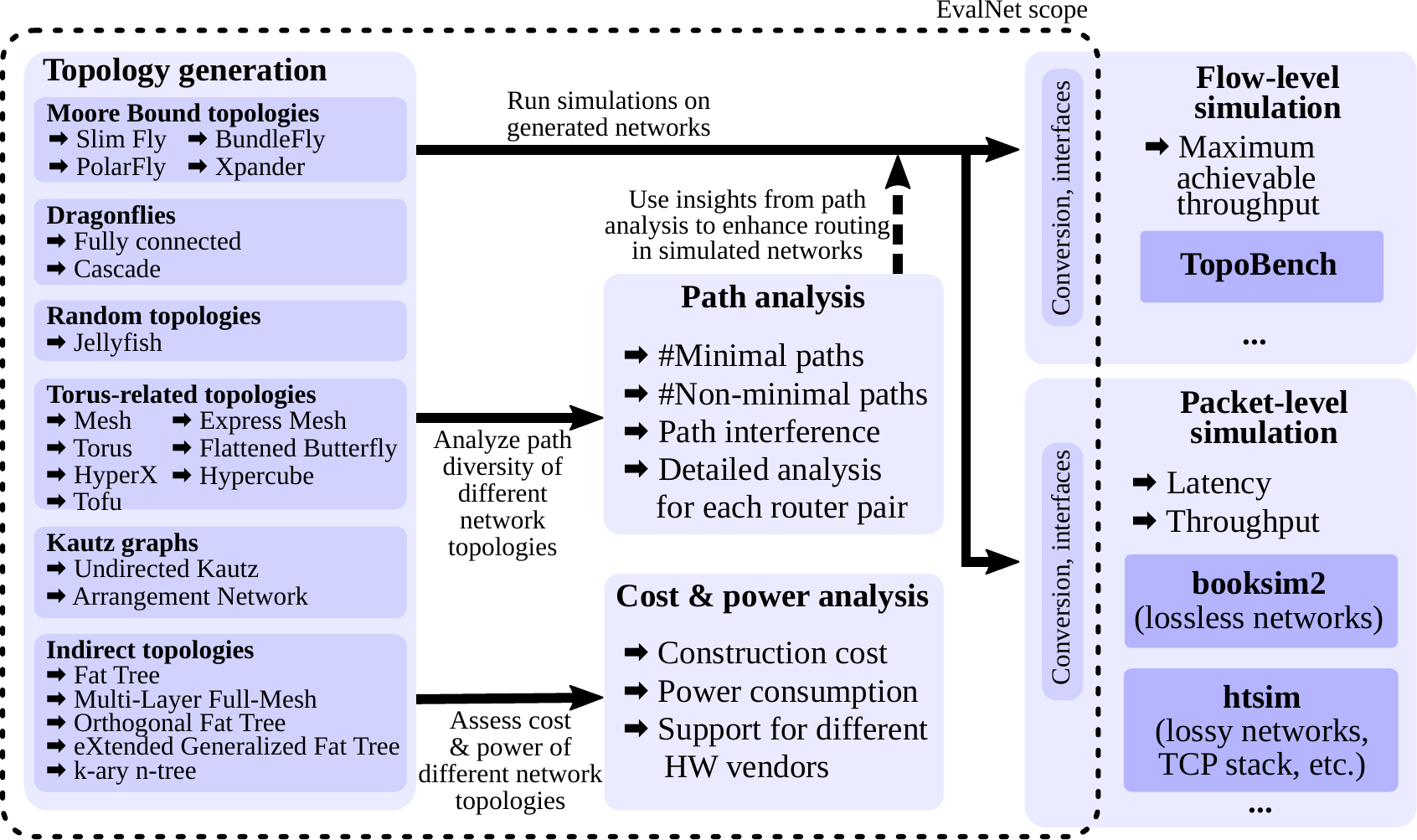}
\vspace{-1.0em}
\caption{An overview of EvalNet.}
\label{fig:overview}
%
%
\end{figure}
\fi

\ifconf
\begin{figure*}[t]
\centering
\includegraphics[width=1\textwidth]{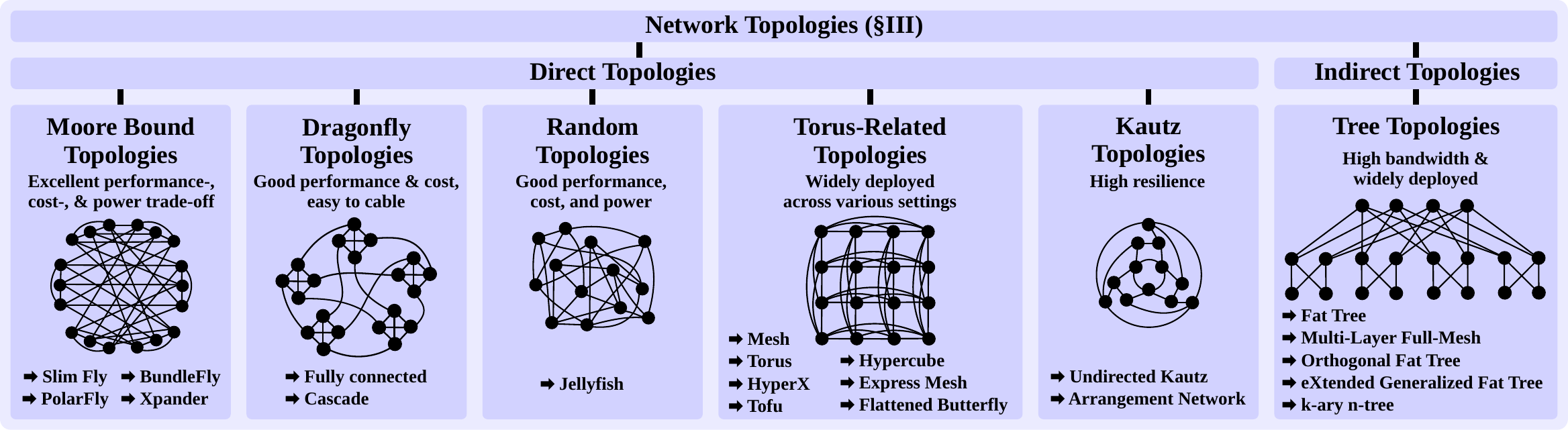}
\vspace{-1.5em}
\caption{An overview of network topologies offered by default in EvalNet. New topologies can be straightforwardly added to the pipeline.}
\label{fig:overview-topos}
\vspaceSQ{-1.0em}
%
\end{figure*}
\fi

\iftr
Figure~\ref{fig:overview} illustrates the overall structure of EvalNet.
%
%
%
First, the user selects a topology to be generated; it can be any of the
networks described in Section~\ref{sec:back}; the user can also provide 
a script generating their own interconnect. 
The user can then use a
module for analyzing the path diversity of a network, which can subsequently be used
to enhance routing schemes in simulations (an overview is in Section~\ref{sec:path-measures}, details and evaluation are in Section~\ref{sec:paths}).
The generated networks can also be evaluated using either flow or packet-based
simulations (overview is in Section~\ref{sec:sims}, details and evaluation are in Section~\ref{sec:flows-eval} and \ref{sec:sims-eval}, respectively). These simulations are conducted by external tools
(TopoBench~\cite{jyothi2016measuring} for flow-based and
htsim~\cite{raiciu2010data} as well as Booksim2~\cite{jiang2016booksim} for
packet-based simulations of lossy and lossless networks, respectively); we
provide input/output conversion routines to integrate these tools seamlessly.
Finally, the user can use our cost
and power models to assess the construction cost or static power consumption (Section~\ref{sec:cost-power}).

We now proceed to outline these factors (Sections~\ref{sec:back}--\ref{sec:evaluating-met}) and then detail each of them and provide extensive evaluations (Sections~\ref{sec:paths}--\ref{sec:sims-eval}).
We will conclude with a comprehensive summary comparing {all} considered interconnects and measures (Section~\ref{sec:takeaways}).
\fi

\section{RELEVANT NETWORK TOPOLOGIES}
\label{sec:back}

We first describe network topologies included in EvalNet -- see also Figure~\ref{fig:overview-topos}. The most important symbols are
listed in Table~\ref{tab:symbols}.

\begin{table}[h]
\centering
\scriptsize
\setlength{\tabcolsep}{2pt}
\ifconf
\renewcommand{\arraystretch}{0.7}
\fi
\caption{{The {most important symbols} used in this work.}}
\label{tab:symbols}
\vspaceSQ{-1.0em}
\begin{tabular}{@{}ll@{}}
\toprule
                    $V, E$ & Sets of vertices/edges (routers/links, $V=\{0,\dots,N_r-1\}$).\\
                    $N, N_r$& \#endpoints and \#routers in the network ($N_r = |V|$).\\
                    $p, k'$& \#endpoints attached to a router, \#channels to other routers.\\
                    $k$&\emph{Router radix} ($k = k' + p$).\\
                    $D, d$&Network diameter and the average path length.\\
                    $c_l(A,B)$ & \emph{Count of (at most $l$-hop) disjoint paths} between router sets $A$, $B$.\\
                    $c_\text{min}(s,t), l_\text{min}(s,t)$ & \emph{Diversity} and \emph{lengths of minimal paths} between routers $s$ and $t$.\\
                    $I_{ac,bd}$ & \emph{Path interference} between pairs of routers $a,b$ and $c,d$.\\
                    $\lambda, v$ & \textit{Flow arrival rate} [flows/s] and the \textit{flow volume} [bytes].\\
\bottomrule
\end{tabular}
\end{table}

\iftr
\subsection{{Network Model}}
\fi
We model an interconnection network as an undirected graph $G = (V,E)$; $V$ and
$E$ are sets of vertices and edges. A vertex models a {router\footnote{We
abstract away HW details and use ``router'' for both L2 switches and L3
routers.}} ($|V| = N_r$). An edge models a full-duplex inter-router physical
link. Servers (endpoints) are modeled implicitly. There are $N$ servers in total,
and $p$~endpoints are
attached to each router (\emph{concentration}). There are $k'$ channels between any two routers (\emph{network radix}). $k = p+k'$ is the total router \emph{radix}.
The diameter is $D$ while the average path length is $d$.
$\lambda$ and $v$ are the flow arrival rate [flows/s] and the flow volume
[bytes] of a used workload.

While EvalNet supports arbitrary configurations of $p$, $k$, and $k'$, in the
experiments, we usually set $p = \frac{k'}d$, as it maximizes throughput while
minimizing congestion and network cost (assuming a random uniform traffic
pattern)~\cite{besta2020fatpaths}. We also usually select network radix~$k'$
and router count~$N_r$ so that, for a fixed $N$, the compared topologies use
similar amounts of networking hardware and thus have similar construction
costs. We assume router-based topologies hereafter, i.e.,
no direct channels between any two endpoints.

\iftr
\begin{figure*}[t]
\centering
\includegraphics[width=1\textwidth]{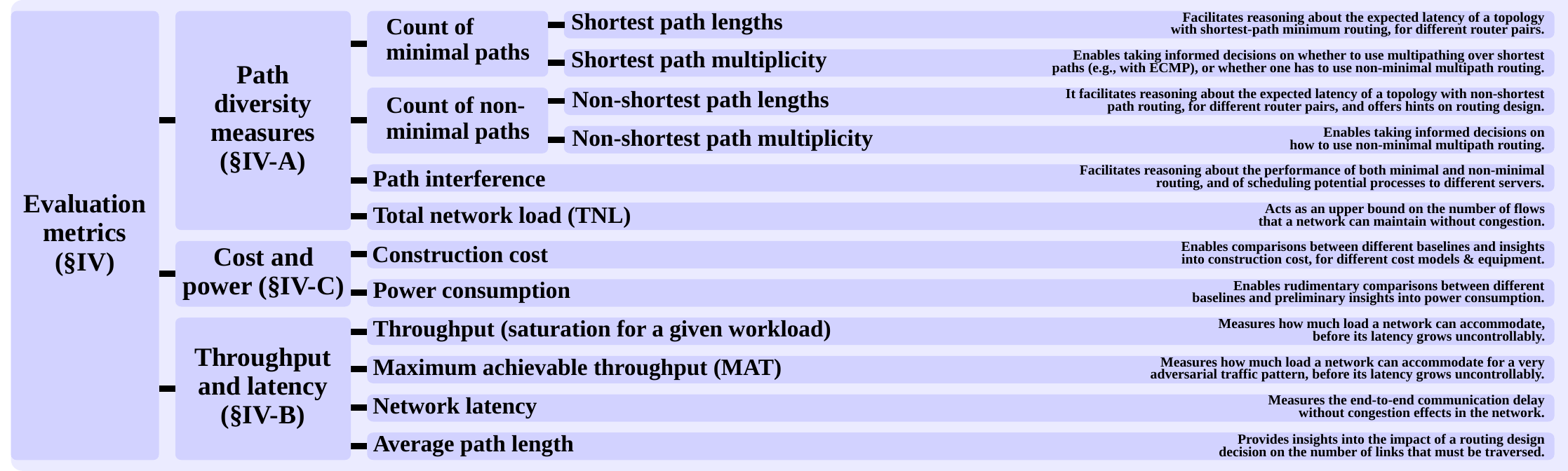}
\vspace{-1.5em}
\caption{An overview of evaluation measures of network topologies offered by default in EvalNet. New measures can be straightforwardly added to the pipeline.}
\label{fig:overview-metrics}
%
\vspaceSQ{-1.5em}
\end{figure*}
\fi

\iftr
\subsection{Moore Bound Topologies}
\fi
A recent class of low-diameter topologies explicitly optimizes the network
structure towards the Moore Bound~\cite{besta2014slim}. This maximizes the
number of routers and endpoints for a fixed diameter and radix, minimizing cost
and power consumption.
\iftr
These topologies are important as they have achieved
new frontiers in performance, cost, and power consumption.
\fi

\textbf{Slim Fly} (SF)~\cite{besta2014slim}, \textbf{PolarFly}
(PF)~\cite{lakhotia2022polarfly}, and \textbf{PolarStar}~\cite{lakhotia2024polarstar}
belong to this class, they are based on the
underlying McKay-Miller-Širáň and Erdős-Rényi polar graphs, respectively.
\textbf{Spectralfly} (SpF)~\cite{young2022spectralfly} is based on the
Ramanujan graph construction of Lubotzky, Phillips, and Sarnak (LPS).
Moreover, a $\tau$-regular \textbf{Xpander} (XP)~\cite{valadarsky2015}
construction starts with a complete graph of $\tau+1$ vertices, and is then
repeatedly ``{lifted}'' ($\frac{N_r}{\tau+1}$ times) to obtain the final graph
with good expansion properties. Finally, \textbf{BundleFly}
(BF)~\cite{bundlefly_2020} is using SF together with a Paley graph. The routers
in both graphs are linked via a multi-star product based on a bijection. 

\iftr
\subsection{Dragonfly Topologies}
\fi
The \textbf{Fully-Connected Dragonfly} (DF)~\cite{dally08} is
comprised of $g$ groups of nodes internally interconnected in an all-to-all
fashion. The topology among groups is an all-to-all, too.
A \textbf{Cascade Dragonfly} (C-DF)~\cite{aries} alters the original Dragonfly
by assuming a 16x6 2D HyperX for intra-group and an all-to-all graph for
inter-group connectivity.
\textbf{Megafly} (MF)~\cite{flajslik2018megafly} is another Dragonfly variant.
These networks are relevant as they come with good performance and cost, albeit achieving worse values for these metrics than low-diameter
networks, but have been found to be easier to cable.
They also offer relatively easy scheduling: their modular
structure facilitates achieving data locality for workloads~\cite{tate2014programming}.

\iftr
\subsection{Random Topologies}

Fully random networks are infeasible due to the unbounded radix.
The \textbf{Jellyfish}~\cite{singla2012jellyfish} is a random
graph with a pre-selected router radix~$k$.
We support these networks as they
come with advantageous performance, cost, and power consumption~\cite{singla2012jellyfish, besta2020fatpaths}.
\fi

\ifconf
\begin{figure*}[t]
\centering
\includegraphics[width=1\textwidth]{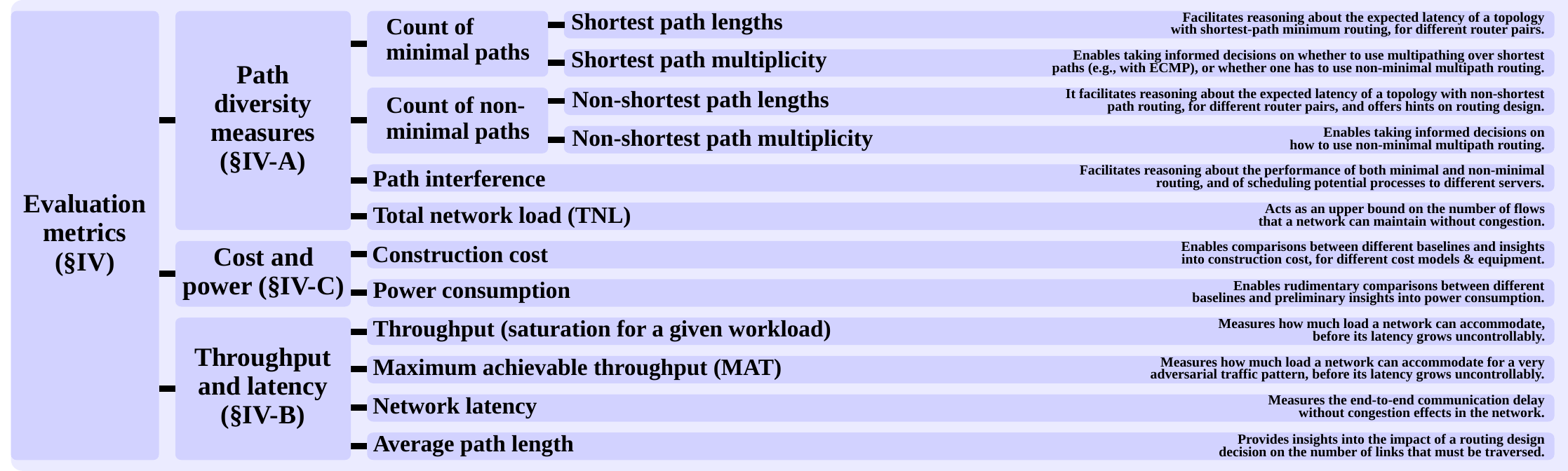}
\vspace{-1.5em}
\caption{An overview of evaluation measures of network topologies offered by default in EvalNet. New measures can be straightforwardly added to the pipeline.}
\label{fig:overview-metrics}
%
\vspaceSQ{-1.5em}
\end{figure*}
\fi

\iftr
\subsection{Torus-Related Topologies}
\fi
We consider traditional \textbf{Mesh} (MH), \textbf{Torus} (TR), and
\textbf{Hypercube} (HC) topologies. We also incorporate their more recent
enhancements: \textbf{HyperX} (HX)~\cite{ahn2009hyperx} (each router in \mbox{$n$-D}
grid connects to all other routers within the same dimension), \textbf{Express
Meshes} (EM)~\cite{jain2017partitioning} (cut-down versions of the HyperX),
\textbf{Flattened Butterfly} (FBF)~\cite{kim2007flattened} (a 2D variation of
HyperX), and \textbf{Tofu} (TF)~\cite{ajima2018tofu, ajima2012tofu} (a 6D torus
variant with inner 3D mesh/tori networks embedded as virtual routers).
These networks have been widely deployed across different settings and thus
are considered in EvalNet.

\iftr
\subsection{Indirect Topologies}
\fi
In indirect networks, 
some routers attach endpoints while other routers only forward packets. 
A \textbf{Fat Tree} (FT)~\cite{leiserson1996cm5} is a binary tree, with either
higher bandwidth links towards the root or multiple parallel links towards the
root to avoid communication bottlenecks. It is one of the most widely researched and used network topologies~\cite{katta2016hula, vanini2017let, alizadeh2014conga, katta2016clove, he2015presto, zhang2017resilient, huang2020tuning, alasmar2018polyraptor, montazeri2018homa, hwang2014deadline, lu2016sed, vamanan2012deadline, alizadeh2013pfabric, bai2014pias, raiciu2011improving, handley2017re, zhuo2016rack, he2016ac, alizadeh2010data, cardwell2017bbr, mittal2015timely, li2013openflow, perry2014fastpass, olteanu2016datacenter, park2019maxpass, benet2018mp, lu2018multi, widmer2001survey, callegari2014survey}.
The \textbf{k-ary n-tree} (AT)~\cite{petrini1997k} avoids infeasible high-radix
roots of FT by using the same number of routers per tree-level and
connecting them with increasing strides towards the top.
An \textbf{eXtended Generalized Fat Tree} (xGFT)~\cite{ohring1995generalized} is
a k-ary n-tree 
variant which allows reducing deployment cost.
\textbf{Orthogonal Fat-Trees} (OFT)~\cite{kathareios2015cost} are based on
k-ary 2-trees; the wiring pattern between the two levels is modified to
increase scalability.
Finally, a \textbf{Multi-Layer Full-Mesh} (MLFM)~\cite{kathareios2015cost}
consists of multiple 2D meshes (incl.~diagonal connections) stacked together,
but instead of direct links it uses global routers to link meshes.
EvalNet also supports indirect networks as they offer very high bandwidth;
FTs are also widely deployed.

\iftr
\subsection{Kautz Topologies}

Both \textbf{Undirected Kautz} (KA)~\cite{li2004graph}
and \textbf{Arrangement Network}~\cite{day1992arrangement}
have strings assigned to each router. Then, links between
routers are determined by similarities of the
corresponding router strings.
Kautz networks offer high resilience properties and are
an interesting alternative to other classes of interconnects.

\fi

\iftr
\subsection{Other Topologies}


EvalNet can be easily extended to model and evaluate interconnects from other domains, such as on-chip networks between different cores~\cite{ahn2016scalable, besta2018slim}, chiplets~\cite{iff2023hexamesh, iff2023rapidchiplet, iff2023sparse}, or FPGA interconnects~\cite{cai2015comparative}, due to its modular and flexible architecture. The toolchain's seamless integration with external simulators for packet-level and flow-level analysis ensures that new topologies, such as those designed for heterogeneous~\cite{hat-sbac-pad12} or spatial computing environments~\cite{gianinazzi2022spatial, baumann2024low}, can be evaluated with minimal effort, making EvalNet a versatile tool for broader interconnect research.
\fi

\section{EVALUATING NETWORK TOPOLOGIES}
\label{sec:evaluating-met}

We next describe the considered measures used to evaluate a network topology.
We analyzed a large body of related work to provide a comprehensive
overview~\cite{besta2020highrouting, dally08, kim2007flattened, besta2014slim,
prisacari2013fast, yuan2014lfti, yuan2013new, prisacari2014efficient,
prisacari2014randomizing, kathareios2015cost, prisacari2015performance,
chen2016evaluation, prisacari2013bandwidth, karacali2018assessing,
sehery2017flow, kassing2017beyond, Dally:2003:PPI:995703}. We
distinguish five classes of measures, used to analyze five aspects of a
network: path diversity, bandwidth, latency, construction cost, and
power consumption.
\iftr
All these metrics are also listed in Figure~\ref{fig:overview-metrics}.
\fi

\subsection{Path Diversity Measures}
\label{sec:path-measures}

Recent studies in high-performance interconnect design indicate that harnessing
different forms of path diversity, especially non-minimal multipathing, is a
key performance enabler, especially for state-of-the-art
low-diameter networks such as SF or PF. This is because these networks -- while
ensuring low latency, high throughput, as well as low construction cost and power
consumption -- have only \textit{one} shortest path between almost all router
pairs~\cite{besta2020fatpaths, besta2020highrouting}. Hence, one cannot use traditional
load-balancing mechanisms such as equal-cost multi-path routing
(ECMP)~\cite{hopps2000analysis} for more performance, and must resort to
multipath routing over non-minimal paths~\cite{besta2020fatpaths}.

To address this, EvalNet supports analyzing all recently proposed measures
for path diversity. First, it enables computing \textbf{counts of} both \textbf{minimal} and
\textbf{non-minimal} paths between any two routers. These two measures illustrate the
potential for multipathing over both traditional and recent networks.
Moreover, we use the \textbf{path interference (PI)} measure, which illustrates how
multipathing changes when considering interactions between different pairs
  of routers (i.e., some parts of a given network may be more or less advantageous
  for multipathing than other parts of the same network). This enables deriving 
  detailed insights into path diversity in the local network
  structure, facilitating the fine-grained design of routing protocols. 
Finally, EvalNet enables computing the \textbf{total network load (TNL)}, which is an
upper bound on the number of flows that a network can maintain without congestion;
TNL constitutes the maximum supply of path diversity offered by a given topology.
We define all these measures in more detail in Section~\ref{sec:paths} while
analyzing their values for different topologies.

Many of the above measures can be analyzed at \textbf{different levels of detail}. They
can be summarized with histograms, or they can
be computed and plotted separately for each individual router pair. EvalNet supports
both options.

\begin{figure*}[t]
\centering
\includegraphics[width=0.99\textwidth]{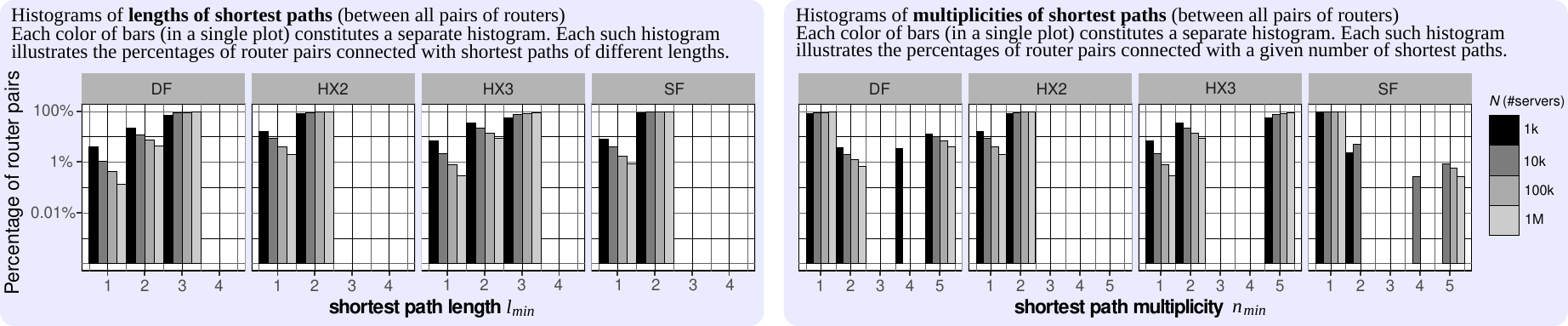}
\iftr
\vspaceSQ{-2em}
\fi
\iftr
\caption{\textbf{(left)} Histograms of shortest path lengths $l_\text{min}$.
\textbf{(right)}
Histograms of shortest path multiplicities $n_\text{min}$.
Considered networks: DF, SF, and two HX variants with $D \in \{2,3\}$.
}
\fi
\label{fig:len-mul-shortest-paths}
%
%
\vspaceSQ{-1.5em}
\end{figure*}

\subsection{Throughput \& Latency Measures}
\label{sec:sims}

In EvalNet, we offer seamless interfaces to established tools, facilitating
assessing \textbf{throughput} using packet-based simulations.
We define throughput to be the normalized rate of injecting input packets for which the network
saturates, i.e., the average latency of packets becomes higher than a
pre-specified threshold.
Moreover, we also consider \textbf{Maximum Achievable Throughput (MAT)}, i.e., an
approximation of throughput obtained using flow-based simulations conducted
with TopoBench. MAT is defined as the ratio of the flow that can be forwarded
between each pair of endpoints, to the demanded amount of flow for this pair.


\textbf{Network latency} is also obtained using packet-based simulations. The network latency is defined as the
latency achieved for a very low injection rate that is close to zero. Thus,
latency measures the end-to-end communication delay without congestion effects
in the network.
Moreover, the interfaced simulators enable delivering 
the \textbf{average path length}, i.e., the average number of inter-router links
traversed between any two communicated endpoints. This measure is useful
because it is related to the network topological structure, i.e., one can see
more clearly (than with latency) the impact of a routing design decision on the
number of links that must be traversed.

\subsection{Construction Cost \& Power Consumption}
\label{sec:cost-power}

EvalNet enables assessing the network \textbf{construction cost}.  The user
can plug in different cable and router cost models, coming from arbitrary
vendors such as Ethernet or InfiniBand, and incorporating different rack placement. 
EvalNet also offers a model assessing the base power
consumption required to power the network. This enables rudimentary comparisons
between different baselines and preliminary insights into power consumption;
detailed power models are notoriously hard to devise and vendors resort to
their in-house infrastructures.


\section{PATH DIVERSITY ANALYSIS}
\label{sec:paths}

We present path diversity measures and conduct analyses in EvalNet, obtaining interesting insights about various interconnects.
These measures are defined on pairs of routers~\cite{besta2020fatpaths}.
\iftr
Extended results are provided in the Appendix.
\fi
We start with basic definitions shared by all measures.

\begin{definition} The {neighborhood} $h(A)$ of a set of vertices $A \subseteq V$
contains all the vertices adjacent to vertices in $A$, i.e., $h(A) = \left\{ t \in V : \exists_{s \in A}(s,t) \in E \right\}.$
\end{definition}

\noindent
\iftr
Intuitively, $h(A)$ are all the routers connected to routers in set $A$.
\fi

\begin{definition}
The {$l$-step neighborhood} $h^l(A)$ (or a transitive hull) of $A
\subseteq V$ contains all the vertices reachable by a path of length $l$ from
some vertex in $A$. Hence, 
$h^l(A) = \underbrace{h(\cdots h(}_{l \text{\ times}} A)\cdots).$
\end{definition}

\noindent
Intuitively, $h^l(A)$ are all the routers reachable with $l$ hops from routers in set $A$.

\begin{definition}
For $s, t \in V$, the number of length-$l$ paths $n_l(s,t)$ from $s$ to
$t$ is recursively defined as
$n_l(s,t) = \sum_{u \in V} n_{l-1}(s,u) \cdot n_1(u,t),$ 
$n_1(s,t) = 1 \text{ iff } (s,t) \in E, \text{ else } 0.$
\end{definition}
%

%
%

\begin{figure*}[hbtp]
\centering
\includegraphics[width=0.92\textwidth]{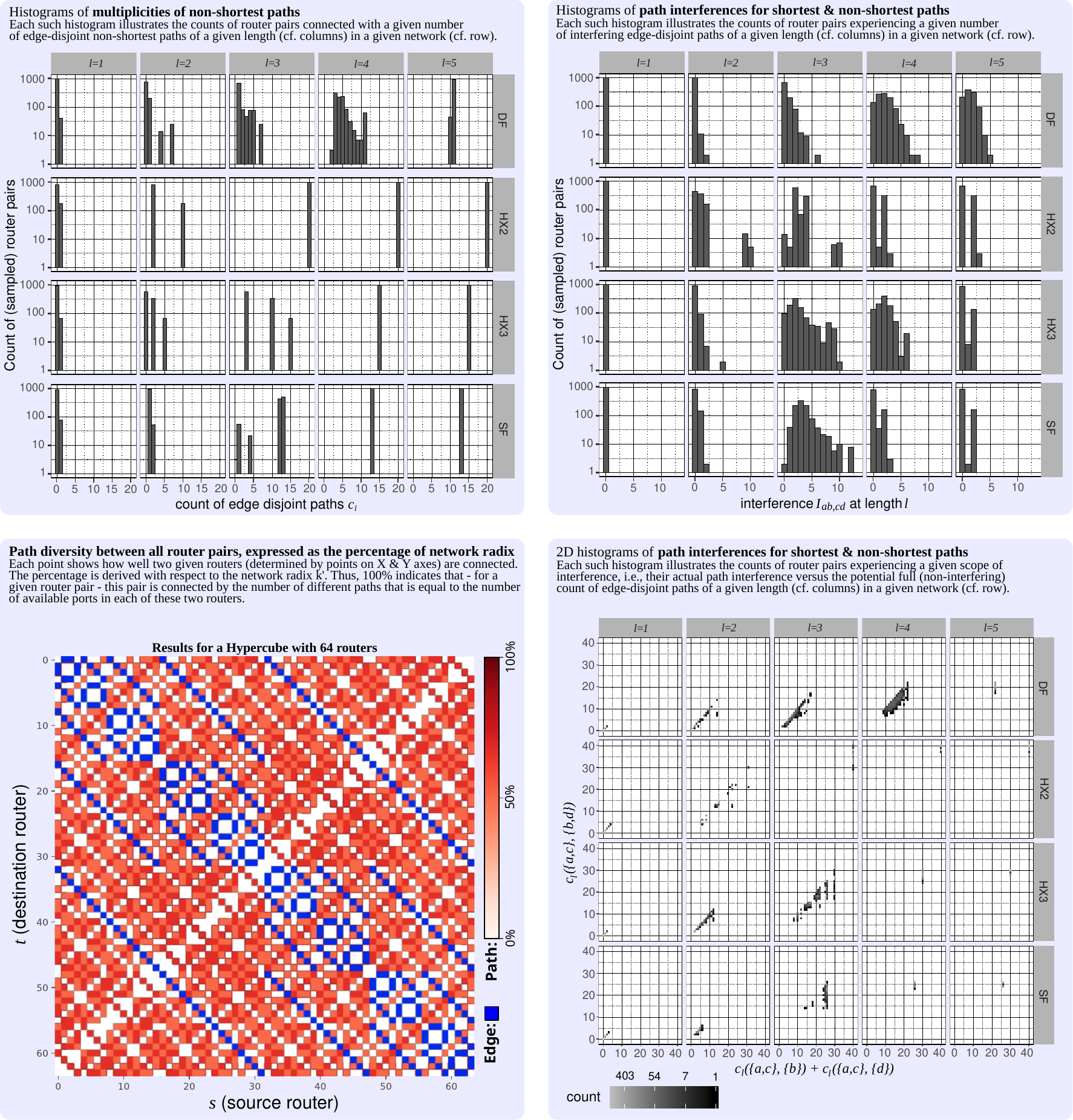}
\iftr
\vspaceSQ{-1em}
\fi
\caption{The analysis of non-shortest path diversity and path interference.}
\label{fig:non-shortest-analysis}
\vspaceSQ{-1em}
\end{figure*}

\subsection{Diversity of Shortest Paths}

First, we consider the lengths and multiplicities of shortest paths.
%
%
Understanding the former facilitates reasoning about the expected latency of a topology with
shortest-path minimum routing, for different router pairs.
\iftr
Knowing the latter, on the other hand, enables taking informed decisions on whether to use multipathing over
\fi
shortest paths (e.g., with ECMP), or whether one has to use non-minimal multipath routing~\cite{besta2020fatpaths, besta2020highrouting}.

\iftr
\begin{definition}
\label{def:shortest_path_length}
For $s,t \in V$, the length of the shortest path $l_\text{min}(s,t)$ between them is
$$ l_\text{min}(s,t) = \min  \left\{ i \in \mathbb{N} : t \in h^i(\{s\}) \right\}.$$
\end{definition}
\fi

\begin{definition}
\label{def:shortest_path_count}
The shortest path multiplicity between $s,t \in V$ is the
number of shortest paths between $s$ and $t$ and is defined as
\ifconf
{
\setlength{\abovedisplayskip}{0pt}
\setlength{\belowdisplayskip}{0pt}
\fi
$$ n_\text{min}(s,t) = n_{l}(s,t) \text{ with } l=l_\text{min}(s,t).$$
\ifconf
}
\fi
\end{definition}

%

We provide example results in Figure~\ref{fig:len-mul-shortest-paths}, for
both lengths of shortest paths (the left panel) and for their multiplicities
(the right panel), for
different sizes of four selected networks (indicated with different colors).
\iftr
The left panel illustrates the rich information about lengths of shortest
paths.
For example, most pairs are connected with a shortest path
having two hops.
\fi
Specifically, more router pairs in SF are connected with a 2-hop shortest path than
in HX2 for all topology sizes. This motivates a focus on making best-possible routing
decisions over 2-hop paths in SF; 1-hop paths are negligible (especially for large networks).
The right panel shows that, for most router pairs, the multiplicity of DF, HX2,
and SF is low ($n_\text{min} \leq 2$) compared to HX3.
We also observe that nearly all router pairs in SF have only 1 shortest path between them,
making non-minimal multipathing crucial. Still, note that DF and SF contain
some outlier router pairs connected with large counts of shortest paths. Finally, most router pairs in HX have many shortest
paths between them, and this ratio is growing with $N$, making ECMP effective.



\subsection{Diversity of Non-Shortest Paths}

Next, we are interested in the number of edge disjoint paths between two
vertices $s,t \in V$, that are not necessarily shortest. 
It facilitates reasoning about the expected latency of a topology with
non-shortest path routing, for different router pairs, and offers hints on routing design.

\iftr
For convenience,
we define this measure for two \emph{groups} of vertices $A,B$:
\fi

\begin{definition}
\label{Def:edge_disjoint_paths}
The count of edge disjoint paths $c_l(A,B)$ at length $l$ between 
sets of routers $A \subset V$, $B \subset V$ is the smallest number of edges that can be removed such that
$A$ and $B$ are not connected by any path of length at most
$l$.
\end{definition}

To obtain this count for two specific routers, we set $A = \{s\}, B = \{t\}$.
The above definition lets us easily derive the number of shortest \emph{edge disjoint}
paths between two vertices:

\begin{definition} The count of shortest edge disjoint paths $c_\text{min}(s,t)$ between $s,t \in V$ is defined as 
\ifconf
{
\setlength{\abovedisplayskip}{0pt}
\setlength{\belowdisplayskip}{0pt}
\fi
$$ c_\text{min}(s,t) = c_{l}(\{s\},\{t\}) \text{ with } l=l_\text{min}(s,t)$$
\ifconf
}
\fi
\end{definition}

$c_\text{min}(s,t)$, in contrast to $l_\text{min}(s,t)$, does not
count shortest paths that share any edges. Hence, $c_\text{min}(s,t) \le
l_\text{min}(s,t)$.

\iftr
We show example results in Figure~\ref{fig:non-shortest-analysis}.
The top-left panel contains the histograms of $c_{l}$ at various lengths for
DF, SF, and two HX instances. The values of $c_{l}$ of DF are in general lower
than in the low dimensional HX instances (HX2 and HX3) and SF. Accordingly,
$c_{l}$ of the HX and SF instances attains the network radix at lower lengths.
HX2 has very high path diversity that is also ``stable'' (20 for all router
pairs) for all higher path lengths ($l=3,4,5$). This facilitates the design for complex adaptive
routing that routes around hotspots and uses complex paths. Next, DF has relatively large
diversity of paths for $l=5$, suggesting similar insights as for HX2, but the count is 11, so less
than 20 by nearly 50\%. Then, for $l=4$ or $l=3$, there is quite some diversity but the counts for
different pairs may differ significantly, requiring attention when developing routing.
\fi

We further investigate $c_{l}$ with a detailed view of the count of edge
disjoint paths for \emph{every router pair} in a small SF, showing pairs with
$\frac{c_{l}}{k'} \leq 75\%$ (the bottom-left panel).  This plot reveals all
pairs of routers that do \textit{not} come with a large path diversity between them. This
enables a fine-grained performance debugging of routing and job scheduling
schemes.
The results show that the majority of router pairs are connected by a number of different
edge-disjoint paths that is larger than 1/3 of all their available network ports (indicated with
dark shade). Such pairs do not require a lot of attention when selecting a path to
route. However, certain pairs do not come with large path diversity (cf. light shade). These
specific pairs are not good candidates for scheduling communication-intense jobs.

\subsection{Path Interference}

\iftr
So far, we considered measures that only take into account paths between two vertices.
\fi
To capture the interactions of paths (of different lengths) between various
pairs of vertices, we now define path interference.
Here, paths between two router pairs $a,b$ and $c,d$ \textit{interfere} if their total count $c_{l}(\{a,c\},\{b,d\})$
of edge disjoint paths at length $l$ is lower than the
sum of the single destination count of edge disjoint paths
$c_{l}(\{a,c\},\{b\}) + c_{l}(\{a,c\},\{d\})$.


\begin{definition}
\label{Def:interference}
Path interference $I^{l}_{ac,bd}$ between two router pairs $a,b$ and $c,d$ is defined as 
\ifconf
{
\setlength{\abovedisplayskip}{0pt}
\setlength{\belowdisplayskip}{0pt}
\fi
$$ I^{l}_{ac,bd}  = c_{l}(\{a,c\},\{b\}) + c_{l}(\{a,c\},\{d\}) - c_{l}(\{a,c\},\{b,d\})$$
\ifconf
}
\fi
\end{definition}

Understanding path interference facilitates reasoning about the performance of both minimal and 
non-minimal routing, and the scheduling of potential processes to different servers.

Note that $I^{l}_{ac,bd} \geq 0 $, where equality is achieved when the disjoint
paths from $a,c$ to $b$ do not share any edges with the disjoint paths from
$a,c$ to $d$. Thus, if $I^{l}_{ac,bd} = 0$, then connections
can use fully the available link bandwidth. On the other hand, the higher the value
of $ I^{l}_{ac,bd}$, the more edges overlap so that the connections have to
share this bandwidth.

Figure~\ref{fig:non-shortest-analysis} (the top-right panel)
shows histograms of interference $I^{l}_{ac,bd}$ up to length $l=5$ for SF, DF, and
two instances of HX. At length $l =4$ and $l=5$ DF
shows higher interference than HX2, HX3 and SF in contrast to length $l=3$
where interference of DF is lower. This insight could be used when
deciding on non-minimal adaptive routing decisions: taking paths of length 3
is more beneficial than 4 or 5.
%
Also, DF, HX3, and SF have almost no interference at two hops ($l=2$).
Thus, one does not have to pay much attention to how to schedule pairs of communicating
jobs, assuming they are executed on servers attached to routers separated by the distance
$l=2$ (or $l=1$). Moreover, DF has also little interference between router pairs separated by 3 hops.
Finally, at $l=3$ (i.e., for 3 inter-router hops), HX2, HX3, and SF have significant amounts
of path interference, calling for care if devising routing/scheduling with such path lengths.

\iftr
We further show a more detailed view of interference up to length $l=5$ on the
bottom-right panel. Each single histogram illustrates the counts of router
pairs experiencing a given scope of interference, i.e., their actual path
interference versus the potential full (non-interfering) count of edge-disjoint
paths, for a given length (as determined by the column) and for a given network
(as determined by the row).
Here, our pipeline also enables delivering various performance related insights.
In general, we observe that the further a data point is from the ``diagonal'' of each histogram,
the more path interference this data point indicates. Such far data points denote a large
difference between the ideal amount of path diversity (on the X axis) versus the
actual amount of diversity (on the Y axis).
No topology is much imbalanced in
this respect, as all data points far from the diagonal are dark (indicating few router pairs
characterized a given difference between ideal and actual path diversity). Such pairs still
require care when scheduling communication-intense jobs.
\fi

\subsection{Computing Path Diversity Efficiently}

In EvalNet, we primarily use algebraic algorithms to compute path diversity metrics, as they offer better scalability and parallelizability compared to purely combinatorial approaches, particularly for large, low-diameter networks~\cite{besta2017slimsell, solomonik2017scaling, solomonik2011communication, solomonik2014tradeoffs}. For example, for counting the number of paths of fixed length between all node pairs, we use adjacency matrix exponentiation, where the $\ell$-th power of the adjacency matrix $A$ gives the number of length-$\ell$ paths via $(A^\ell){i,j}$. This computation is efficient for small diameters and can be adapted for next-hop tracking using semiring-based variants. For length-restricted connectivity, we extend the algebraic path propagation algorithm of Cheung et al.~\cite{cheung2011graph}. 
\iftr
For other metrics, we also explore max-flow-based approaches such as Push-Relabel and Ford-Fulkerson with BFS path constraints (used for sampled length-limited disjoint path analysis), as well as classic Gomory-Hu trees for all-pairs edge connectivity. While these combinatorial algorithms serve as correctness baselines or targeted tools for small-scale samples, EvalNet's core design centers on the algebraic methods due to their efficiency, extensibility, and compatibility with parallel computation.
\fi

\begin{figure}[t]
\centering
\includegraphics[width=\linewidth]{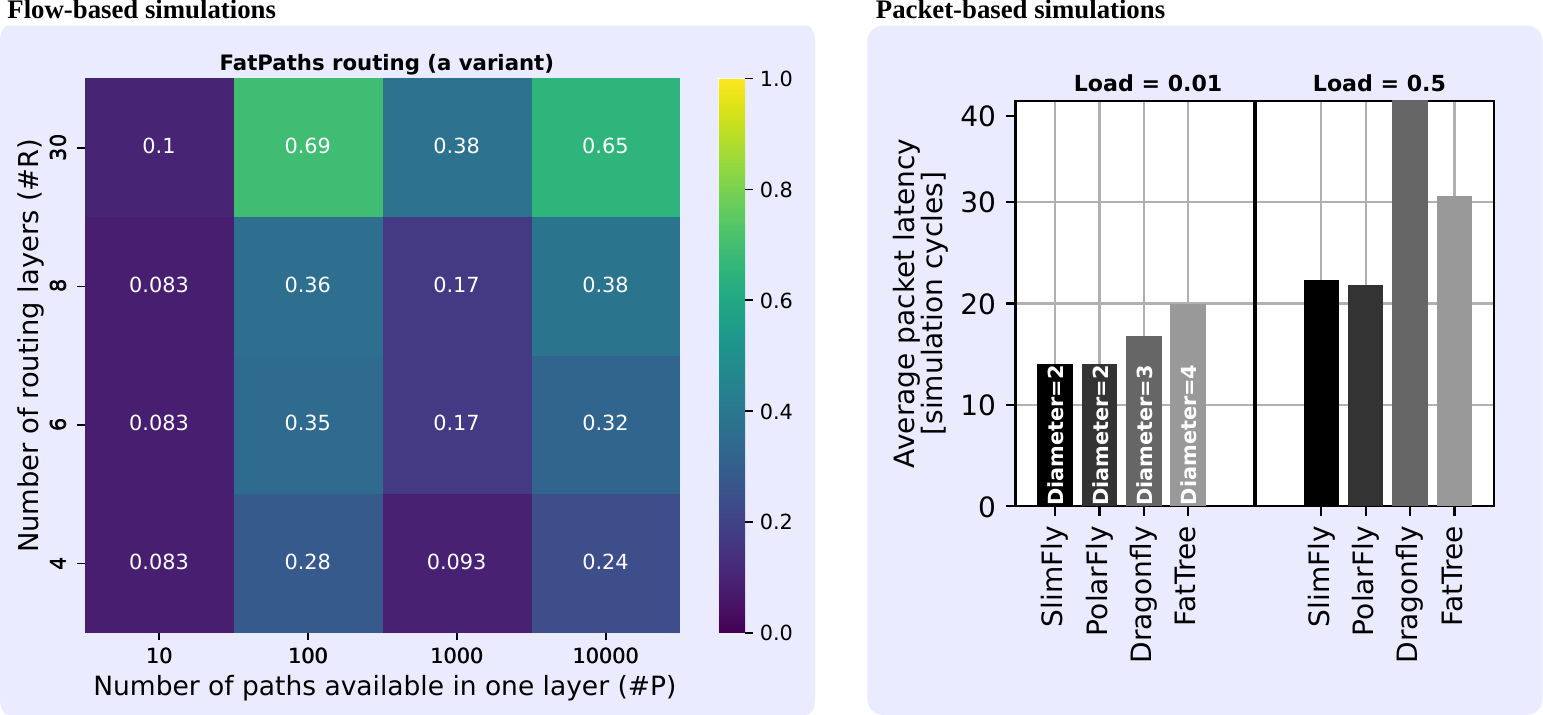}
\caption{\textbf{(left)} Heatmap showing the Maximum Achievable Throughput (MAT), obtained with TopoBench,
for a SF with 10k endpoints and a variant of FatPaths routing with a worst-case traffic pattern.
%
%
\textbf{(right)}
Packet-based simulation runtimes [cycles] with Booksim2, for networks of 50k endpoints,
for two injection rates, 0.01 and 0.5.
As expected, networks with lower diameter usually provide lower latency (load=0.01).
For load=0.5, the network global bandwidth starts to play an important role; here,
DF has higher latency than FT3 because it has much lower global bandwidth.
}
\label{fig:theory}
%
%
\vspaceSQ{-2em}
\end{figure}

\section{FLOW-LEVEL SIMULATIONS}
\label{sec:flows-eval}

Flow-level simulators enable evaluation of large-scale systems, at a cost of a
coarse design that may hinder realistic insights into performance.
Flow-level simulations assume a continuous, real-valued flow model.  Each flow
has a rate $\mu \in \mathbb{R^+}$. The sum of all flows on each link must not
exceed the link capacity. Without any additional constraints, this leads to
constrained max-flow solutions in the topology graph.  Other constraints (e.g.,
on the routes that can be taken by specific flows) lead to more realistic
upper-bounds for performance of various routing schemes on specific topologies.
However, since these models do not represent time accurately, they are unable
to estimate latency or provide realistic insights into the impact of routing
choices on flow control. As such, they are a useful tool for initial insights
when designing a network.

\iftr
For flow simulations, we provide an interface to TopoBench, an evaluation
tool~\cite{jyothi2016measuring} that uses linear programming (LP) to derive
MAT.
\fi
Our TopoBench version supports state-of-the-art
{layered routing} schemes such as FatPaths~\cite{besta2020fatpaths} or extended layered routing~\cite{blach2023high}, built specifically for enabling non-minimal multi-path routing.
%
%
\iftr
Example results are in the left part of Figure~\ref{fig:theory}, where we analyze a variant of FatPaths that offers non-minimal
multipath routing.
\fi
\iftr
That part contains a heatmap showing MAT
\fi
for a SF with 10k endpoints and a variant of FatPaths routing with a worst-case traffic pattern.
``\#R'' (Y axis) as well as ``\#P'' (X axis) are parameters of the FatPaths
routing scheme;
\iftr
they can be tuned for a specific topology.
Usually, the higher \#R is, the better
the throughput becomes
(because more layers entail more potential for multipathing~\cite{besta2020fatpaths}).
Simultaneously, higher \#R indicates more HW costs because one needs larger forwarding tables.
\fi
Then, higher \#P does not always enhance MAT (as confirmed by the results), because certain
value ranges may cause larger path interference across routing layers, ultimately diminishing
performance~\cite{besta2020fatpaths}.
Overall, for certain configurations, FatPaths on SF achieves very high MAT.

\section{PACKET-LEVEL SIMULATIONS}
\label{sec:sims-eval}

\iftr
%
Packet-level simulators offer a more detailed packet-based model of the network
and routing.
All actions in the simulation are modeled as \emph{events}, which are scheduled
to happen at a given point in time. For each event, some code is executed,
which can update the state of the simulation, and produce new events that are
scheduled to happen in the future. The main simulation loop keeps the scheduled
events in a priority queue and advances the simulation by dequeuing and
executing successive events.
\fi

\subsection{Which Simulator to Use?}

A large number of simulators exist, and it is challenging to select the one best suited for
one's purposes.
%
%
%
%
%
%
\iftr
We analyze a broad selection
of available simulators, and show that none scale to the desired sizes of
hundreds of thousands of servers. The overview is in
Table~\ref{tab:simulators}. We analyzed their codebases, investigating
which (1) facilitate modifications that could enhance scalability, (2)
are easy to use and extend without requiring distributed-memory infrastructure support, and (3) cover a large number of protocols and settings.
Based on that analysis, we selected two simulators:
htsim~\cite{raiciu2010data} for lossy TCP/IP stacks and
Booksim2~\cite{jiang2016booksim} for lossless HPC architectures.
\fi
\iftr
Other simulators do not meet these criteria. For example, SST and ROSS natively target HPC clusters and we concluded that they are relatively complex to set up, use, and extend on low-cost PCs.
\fi

\textbf{htsim }
The \emph{htsim} simulator~\cite{raiciu2010data} provides a lightweight
infrastructure for lossy networks. It only models the transport layer: there is no model for
links or routers. Instead, the route for each packet is pre-computed as a
sequence of queues that the packet will pass through. Such a route
specification is attached to the packet. Each queue has a finite service rate,
which models the link capacity.
Due to this design, htsim is highly flexible, and new topologies can be added
straightforwardly: the topology only affects the route computation, which is
explicitly called for every flow during its setup.
\iftr
The disadvantage is that
many adaptive routing schemes cannot be modeled, since there is no per-router state
that could affect the routing.
\fi
Furthermore, all flows are initialized before
the first event is processed, and all routes are kept in memory. 
{EvalNet offers a seamless interface to htsim.}
\iftr
We select htsim as it targets TCP stacks; it supports by default modern TCP-based
schemes such as NDP~\cite{handley2017re}, and it can be easily extended with new routing schemes.
\fi

\textbf{Booksim2 }
\iftr
Booksim2
\fi
targets lossless networks. It explicitly models
input-queued routers, thus one can easily extend the simulator with new routing
protocols. Moreover, there is support for flow control unit (flit) packets, and
thus for schemes such as wormhole routing or virtual cut-through flow control.
Virtual channels are also supported. One can specify the total buffering/port,
router delay for credit processing, delays for channel latency, router
and VC allocations, and processing in a crossbar.
\iftr
As with htsim,
EvalNet offers modules that enable seamless evaluation of any
generated topologies within the Booksim2 setting.
\fi

\begin{table}
\centering
\iftr
\vspaceSQ{-0.5em}
\fi
\scriptsize
\setlength{\tabcolsep}{12pt}
\renewcommand{\arraystretch}{0.6}
\caption{The comparison of available packet-level network simulators. ``Scalability'' is
the largest size of a topology (simulated with a specific simulator) that we
were able to find in the literature. ``Design'' includes the details of
the compute platform used for simulations in the ``Scalability'' column: ``SM'' (shared-memory design), ``DM'' (distributed-memory design).
}
\label{tab:simulators}
\vspaceSQ{-0.5em}
\begin{tabular}{llc}
\toprule
\textbf{Simulator} & \textbf{Scalability} [\#servers] & \textbf{Design} \\
\midrule
%
%
LogGOPSim~\cite{hoefler2017spin} & 
1024~\cite{hoefler2017spin} & 
SM \\

htsim~\cite{raiciu2010data} & 
128~\cite{raiciu2010data} & 
SM \\ 

NS2~\cite{issariyakul2012introduction} &
$<$1,000 & 
SM \\

booksim~\cite{jiang2016booksim} & 
10,000~\cite{besta2014slim} & 
SM \\

FOGSim~\cite{garcia2014fogsim} & 
16,512~\cite{benito2015ethernet} & 
SM \\

OMNet++\cite{varga2008overview} & 
100,000~\cite{gamer2009large} &
SM \\

\midrule

NS3~\cite{ns3web} & 
5,000~\cite{pelkey2011ns3} & 
DM \\

NS4~\cite{Fan:2017:NPN:3123878.3132002} & 
N.A. & 
DM \\

BigSim~\cite{zheng2004bigsim} & 
46,656~\cite{jain2016evaluating} & 
DM \\

SST~\cite{rodrigues2016structural} & 
110,592~\cite{groves2016sai} & 
DM \\

xSim~\cite{bohm2011xsim} & 
2,097,152~\cite{engelmann2014scaling}& 
DM \\

ROSS/CODES~\cite{cope2011codes} & 
\makecell[l]{1,000,000~\cite{wolfe2016slimfly} (Slim Fly), \\50,000,000~\cite{mubarak2012dragonfly} (Dragonfly)} & 
DM \\

\midrule
\textbf{htsim~[this work]} & 
\textbf{1,000,000} &
\textbf{SM} \\

\textbf{Booksim2~[this work]} & 
\textbf{50,000} &
\textbf{SM} \\
%
\bottomrule
\end{tabular}
\vspaceSQ{-1.5em}
\end{table}

\begin{table*}[t]
\centering
\renewcommand{\arraystretch}{0.8}
\scriptsize
\caption{Estimation of the number of elements of a data center Slim Fly simulation. The numbers and thus required resources vary widely depending on a chosen configuration. ``Range'' lists parameter bounds considered in this work, in the following columns we show numbers for example configurations. Memory estimates are based on htsim. The chosen workload parameters keep the simulations feasible on a laptop (configuration details are in Section~\ref{sec:packet-eval}).
$^1$Negligible for relevant configurations.
$^2$Data depends on the routing scheme.
$^3$Objects are transient and accounted for in the flow counts.
$^4$Inter-router cables + server links.
$^5$The average over the flow size distribution, excluding re-transmissions.
}
\label{tab:estimates}
\vspaceSQ{-0.5em}
\begin{tabular}{ll|rrr|rrr|rrr|rrr}
\toprule
\multicolumn{2}{l|}{\textbf{Element}} &%
\multicolumn{3}{c|}{\textbf{Range (lower--upper bounds)}} &%
\multicolumn{3}{c|}{\textbf{10,000 servers}} &%
\multicolumn{3}{c|}{\textbf{100,000 servers}} &%
\multicolumn{3}{c}{\textbf{1,000,000 servers}} \\
\midrule
& &%
\makecell[c]{\textbf{Count}\\\textbf{(relative)}} & %
\makecell[c]{\textbf{Count}\\\textbf{(total)}} & %
\makecell[c]{\textbf{Memory}\\\textbf{(per item)}} & %
\makecell[c]{\textbf{Count}\\\textbf{(relative)}} & %
\makecell[c]{\textbf{Count}\\\textbf{(total)}} & %
\makecell[c]{\textbf{Memory}\\\textbf{(total)}} & %
\makecell[c]{\textbf{Count}\\\textbf{(relative)}} & %
\makecell[c]{\textbf{Count}\\\textbf{(total)}} & %
\makecell[c]{\textbf{Memory}\\\textbf{(total)}} & %
\makecell[c]{\textbf{Count}\\\textbf{(relative)}} & %
\makecell[c]{\textbf{Count}\\\textbf{(total)}} & %
\makecell[c]{\textbf{Memory}\\\textbf{(total)}} \\
\midrule
\multirow{4}{*}{\begin{turn}{90}Network\end{turn}}%
& router &        1 & $10^2$--$10^4$  & ($^1$)    
         &        1  & 242             &  ($^1$)          
         &        1  & 1,058            &  ($^1$)          
         &        1  & 5,618            &   ($^1$)         
\\
& cable (per router) & 10--1000 & $10^4$--$10^6$  & ($^1$)    
         & ($^4$) 8.5+40   & 11,737           &    ($^1$)        
         & ($^4$) 17.5+90  & 113,735          &     ($^1$)       
         & ($^4$) 39.5+200 & 1,345,511         &    ($^1$)        
\\
& server (per router) & 10--1000 & $10^4$--$10^6$  & 5kB       
         & 40       & 9,680            & 47MB      
         & 90       & 95,220           & 460MB     
         & 200      & 1,123,600         & 5.4GB     
\\
& routing entries
         & $O\left(N_r^2\right)$  & $10^4$--$10^8$  & ($^2$) 100B      
         & $O\left(N_r^2\right)$         & 58,564           & 6MB       
         & $O\left(N_r^2\right)$         & 1,119,364         & 110MB     
         & $O\left(N_r^2\right)$         & 31,561,924        & 3GB       
\\
\midrule
\multirow{4}{*}{\begin{turn}{90}Workload\end{turn}} %
& task (per server)  & 1-10     & $10^2$--$10^5$   & ($^1$)    
         & 1        & 968,0            &      ($^1$)      
         & 1        & 95,220           &        ($^1$)    
         & 1        & 1,123,600         &        ($^1$)    
\\
& flow (per task)  & 1-100    & $10^2$--$10^7$  & 2kB       
         & 100      & 968,000            & 1.8GB      
         & 10       & 952,200           & 1.8GB      
         & 1        & 1,123,600         & 2.1GB     
\\
& path (per flow)  & 1-10    & $10^2$--$10^8$   & 600B      
         & 5        & 4,840,000         & 2.7GB       
         & 5        & 4,761,000         & 2.7MB      
         & 5        & 5,618,000         & 3.2GB     
\\
& packet (per flow) & 10-1000  & $10^3$--$10^{10}$& ($^3$)  
         & ($^5$) 110 & 106,480,000           &  ($^3$)        
         & ($^5$) 110 & 104,742,000          &  ($^3$)        
         & ($^5$) 110 & 123,596,000        &   ($^3$)       
\\
\bottomrule
\end{tabular}
\vspaceSQ{-0.5em}
\end{table*}

\subsection{Can We Simulate Massive Networks on Low-Cost HW?}
\label{sec:packet-feas}

\iftr
Next, we
conduct a feasibility analysis in which we argue that million-server
packet-level simulators should {in theory be achievable on a low-cost PC}.
For this, we analyze the approximate memory and time cost of targeted large-scale
simulations.
\fi


\iftr
\textbf{Number of Simulation Elements }
We estimate the number of elements of a simulated topology (e.g., servers) and
\fi
of a simulated workload (e.g., flows which require flow control state). The results
of the analysis are in Table~\ref{tab:estimates}.
The number of network elements does not pose the most serious scalability
problems. 
%
%
Instead, \emph{the number and corresponding size of elements related to
workloads dominates the memory usage and the running time}.

\iftr
\textbf{Memory and Time Requirements }
The total memory usage and simulation time depend on the simulation software
\fi
and on \emph{granularity}, i.e., what elements of the simulated workload are
explicitly stored and simulated. We estimate these numbers in
Table~\ref{tab:estimates}, which also shows what may \emph{not} be feasible in the
largest-scale simulations. Specifically, we cannot simulate large workloads on
large networks, and must not store any per-packet state. 
%


%
\emph{The lowest required amount of memory is determined by the number of
simulated flows}, because we want to simulate end-to-end flow control
for each flow. Thus, the flow control algorithm state for each flow needs to be
stored. We also account for the packets in flight on a per-flow basis, since
  the memory occupied by packets is tightly coupled to the number of concurrent
  flows. Memory consumed is about 2kB/flow plus 600B/path,
  which could probably be further reduced.  To make large simulations
  feasible, it is crucial to keep the per-flow overhead low. Special attention
  needs to be paid to monitoring and debugging code, which can easily dominate
  the per-flow state.  For very large networks, the quadratically increasing
  size of the routing tables starts to dominate, but as
  Table~\ref{tab:estimates} shows, this was not a problem for our simulations.


\emph{The lowest amount of time is determined by the amount of packet
forwarding.} Each packet forwarding has to be processed as at least one event,
and these events constitute the majority of simulation time. In the modern
low-diameter networks, each packet has to be forwarded $\approx$2--4 times to
reach its destination, and we observed around 60 events processed for each
transmitted packet
\iftr
(this includes entering and leaving queues on the path
through the network, for the data packet as well as the ACK packet and
potential retransmissions).
\fi
We observe event rates of around $10^6$ per second
on one CPU core. Since each event likely causes at least one cache miss, we
cannot expect much higher rates without distributed simulation or middle- to high-cost PC stations, which are
outside the scope of this work.
%

%
%




\subsection{How to Simulate Massive Networks on Low-Cost HW?}
\label{sec:packet-eval}

Finally, we show how to use htsim and
Booksim2 to enable more scalability.
%
%
%
In Booksim2, the key limiting factor that we identified was processing related to single
packets. This processing is usually related to gathering statistics, which
is not relevant for the actual simulation outcomes. In Booksim2, eliminating this processing was
the main enabler for being able to simulate \textbf{networks with 10k to 1M endpoints
and more}.
%
%
%
In htsim, on the contrary, the main bottleneck was the storage related to
flows. Specifically, we had to add a more efficient, routing table based scheme
for arbitrary topologies. We do not use the provided logging solution, which
  tends to produce too much output at the simulation scales that we consider.
  Instead, the statistics of each flow are printed in text format to standard
  output when the flow finishes. Another scalability obstacle was the
  \texttt{net\_paths} structure present in the sample programs, a preallocated
  $N^2$ size matrix of routes, which would only be sparsely populated for
  permutation workloads and would dominate memory use for large networks.

\subsection{Representative Simulation Results}

\iftr
We simulate networks with $\approx$10k, $\approx$100k, and $\approx$1M servers using a laptop with 16GB RAM and an Intel Core i7-8550U CPU; each run takes 1–3 hours. Networks are 5$\times$ oversubscribed, as in configurations studied in~\cite{kassing2017beyond}, and include topologies larger than those used in prior large-scale SF simulations. Due to the cost of simulating large networks, we capture only a few milliseconds of activity—sufficient for reliable relative performance comparisons. We focus on flow completion time (FCT) distributions as a function of flow size, using the pFabric web search distribution (average $\approx$1MB) and injection rates of 40–60 flows/server/s. Results are shown as mean, 10th, and 99th percentiles. To mitigate artifacts from the fixed injection window, we carefully interpret short and long flows. 
\fi

\begin{figure*}
\ifconf
\centering
\includegraphics[width=0.28\textwidth]{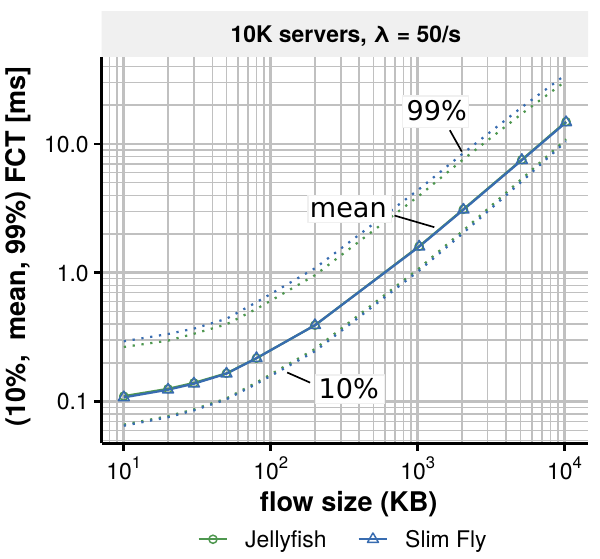}%
\includegraphics[width=0.28\textwidth]{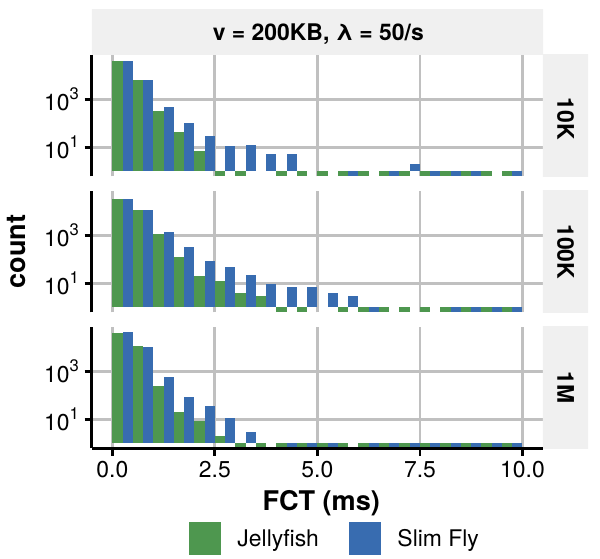}%
\includegraphics[width=0.28\textwidth]{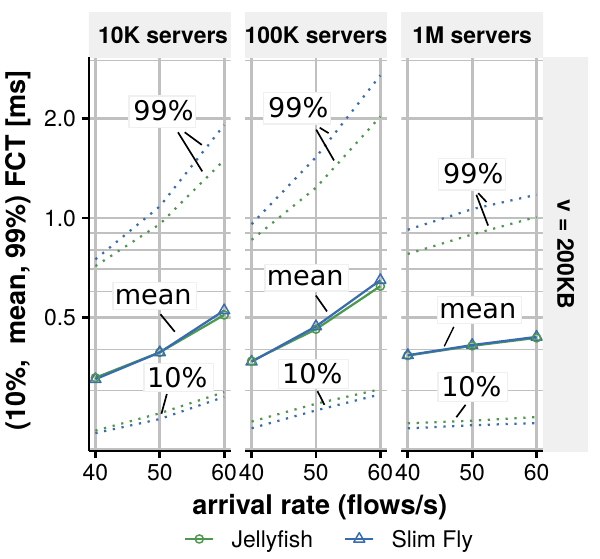}%
\else
\includegraphics[width=0.33\textwidth]{bysize___e.pdf}%
\includegraphics[width=0.33\textwidth]{histogram___e.pdf}%
\includegraphics[width=0.33\textwidth]{bylambda___e.pdf}%
\fi
\caption{Results of large-scale simulations.}
\label{fig:results2}
%
%
\vspaceSQ{-1.5em}
\end{figure*}

As an example, we analyze 10k, 100k, and 1M server simulations, see
Figure~\ref{fig:results2}. The left plot shows FCT as a function of size. For
long flows ($v>200$kB), FCT is limited by bandwidth, while for shorter flows,
latency dominates. The middle plot illustrates the distribution of FCTs for
middle-size flows: here, the better tail behavior of the randomized Jellyfish
topology is clearly visible. In the right plot, we show the impact of the flow
arrival rate $\lambda$. Even a small increase in load causes a large
degradation in FCT in the considered networks, due to the used
oversubscription. The 1M server simulation is less affected due to the limited
warmup time.

\textbf{Our analysis illustrates that million-server packet-level simulations on a
simple low-cost laptop are feasible. }
\iftr
However, they require some compromises
on the quality of simulation, most importantly on the number of allowed
flows and on the methodology (i.e., simulating a fixed amount of flows instead
of a fixed time window after reaching a steady-state).
\fi

\iftr
\section{\mbox{\hspace{-0.65em}COMPREHENSIVE ANALYSIS \& TAKEAWAYS}}
\fi
\label{sec:takeaways}

\begin{table}[ht]
\centering
\footnotesize
\scriptsize
\iftr
\setlength{\tabcolsep}{5.5pt}
\fi
\renewcommand{\arraystretch}{0.6}
\caption{{A general comparison of different network topologies.}
\textbf{Lat.}: latency.
\textbf{GlB.}: global bandwidth (i.e., the saturation point for the random uniform traffic pattern).
\textbf{Ext.}: extensibility (i.e., how far can we extend a given concrete network with new servers?). Note that
fixed diameter networks have lowest extensibility, because they have strict upper bounds on how many new
servers can be attached; the lower the diameter is, the lower this bound is.
Contrarily, networks such as torus can attach arbitrarily many new servers, because their
diameter grows to infinity.
\textbf{C\&P}: construction cost and static power consumption. They are assessed as proportional to the total number of ports in a network, for a fixed network size $N$
(i.e., for fixed $N$, networks with higher radix tend to have higher cost and power consumption).
\textbf{MS}: diversity of shortest paths. It is assessed as a weighted average of shortest path diversities at different lengths.
\textbf{MnS}: diversity of non-shortest paths. It is assessed as a weighted average of non-shortest path diversities at different lengths.
%
The battery symbols serve as a rudimentary measure of relative
comparison between networks.
\faBattery[0]: worst,
\faBattery[1]: bad,
\faBattery[2]: medium,
\faBattery[3]: good,
\faBattery[4]: excellent.
They always have a positive meaning, e.g., ``\faBattery[4]'' used for
cost means that the cost of a given network
is \emph{very low} compared to other networks.
}
\vspaceSQ{-0.5em}
\label{tab:networks-summary}

\begin{tabular}{@{}lllllll@{}}
\toprule
\textbf{Topology} & \textbf{Lat.} & \textbf{GlB.} & \textbf{Ext.} & \textbf{C}\&\textbf{P}&\textbf{MS} & \textbf{MnS} \\
\midrule
{Slim Fly}~\cite{besta2014slim} & \faBattery[4] & \faBattery[4] & \faBattery[0] & \faBattery[3] &  \faBattery[1] & \faBattery[4] \\
PolarFly~\cite{lakhotia2022polarfly} & \faBattery[4]  & \faBattery[4] & \faBattery[0] & \faBattery[4] & \faBattery[0] & \faBattery[4] \\
Xpander~\cite{valadarsky2015} & \faBattery[3]  & \faBattery[4] & \faBattery[1] & \faBattery[2] &  \faBattery[1] & \faBattery[3] \\
BundleFly~\cite{bundlefly_2020} & \faBattery[3]  & \faBattery[3] & \faBattery[1] & \faBattery[2] & \faBattery[1] & \faBattery[3] \\
\midrule
Dragonfly~\cite{dally08} & \faBattery[3]  & \faBattery[3] & \faBattery[1] & \faBattery[2] & \faBattery[0] & \faBattery[2] \\
Cascade Dragonfly~\cite{aries} & \faBattery[2]  & \faBattery[3] & \faBattery[1] & \faBattery[2] & \faBattery[1] & \faBattery[1] \\
\midrule
{Jellyfish}~\cite{singla2012jellyfish} & \faBattery[2]   & \faBattery[2] & \faBattery[1] & \faBattery[2] & \faBattery[1] & \faBattery[3] \\
\midrule
{Mesh} & \faBattery[0]  & \faBattery[0] & \faBattery[3] & \faBattery[0] &  \faBattery[0] & \faBattery[1]   \\
Torus 2D & \faBattery[0]  & \faBattery[0] & \faBattery[3] & \faBattery[0] &  \faBattery[0] & \faBattery[1]  \\
Torus 3D & \faBattery[1]  & \faBattery[1] & \faBattery[3] & \faBattery[1] &  \faBattery[1] & \faBattery[2]  \\
Torus 4D & \faBattery[2]  & \faBattery[2] & \faBattery[3] & \faBattery[1] & \faBattery[2] & \faBattery[3] \\
Torus 5D & \faBattery[3]  & \faBattery[3] & \faBattery[4] & \faBattery[2] & \faBattery[3]  &  \faBattery[3] \\
Torus 6D & \faBattery[3]  & \faBattery[3] & \faBattery[4] & \faBattery[2] &  \faBattery[3] & \faBattery[4] \\
Hypercube & \faBattery[3]  & \faBattery[3] & \faBattery[4] & \faBattery[2] & \faBattery[3] & \faBattery[4] \\
HyperX (2-dimensional)~\cite{ahn2009hyperx} & \faBattery[4]  & \faBattery[4] & \faBattery[1] & \faBattery[2] &\faBattery[3] & \faBattery[4] \\
HyperX (3-dimensional)~\cite{ahn2009hyperx} & \faBattery[3]  & \faBattery[4] & \faBattery[2] & \faBattery[2] & \faBattery[3] & \faBattery[4]  \\
Express Mesh~\cite{jain2017partitioning} & \faBattery[2] & \faBattery[3] & \faBattery[3] & \faBattery[3] & \faBattery[2] & \faBattery[3]  \\
Flattened Butterfly~\cite{kim2007flattened}  & \faBattery[4] & \faBattery[4] & \faBattery[1] & \faBattery[2] & \faBattery[3] & \faBattery[4]  \\
\midrule
Fat Tree (2-level)~\cite{leiserson1996cm5} & \faBattery[4]   & \faBattery[4] & \faBattery[1] & \faBattery[3] & \faBattery[4] & \faBattery[0] \\
Fat Tree (3-level)~\cite{leiserson1996cm5} & \faBattery[3]   & \faBattery[2] & \faBattery[2] & \faBattery[2] & \faBattery[4] & \faBattery[0] \\
{k-ary n-tree}~\cite{petrini1997k}  & \faBattery[3] & \faBattery[4] & \faBattery[4] & \faBattery[2] & \faBattery[4] & \faBattery[0]  \\
eXtended Generalized Fat Tree~\cite{ohring1995generalized}  & \faBattery[3] & \faBattery[3] & \faBattery[4] & \faBattery[2] & \faBattery[4] & \faBattery[0] \\
Orthogonal Fat-Trees~\cite{kathareios2015cost} & \faBattery[4]   & \faBattery[4] & \faBattery[1] & \faBattery[4] & \faBattery[4] & \faBattery[0] \\
Multi-Layer Full-Mesh~\cite{kathareios2015cost}  & \faBattery[4] & \faBattery[3] & \faBattery[1] & \faBattery[2] & \faBattery[3] &  \faBattery[2] \\
\midrule
Undirected Kautz~\cite{li2004graph}  & \faBattery[2]  & \faBattery[2] & \faBattery[1] & \faBattery[2] & \faBattery[1] & \faBattery[2]  \\
Arrangement Network~\cite{day1992arrangement}  & \faBattery[1]  & \faBattery[2] & \faBattery[1] & \faBattery[2] & \faBattery[0] & \faBattery[1] \\
\bottomrule
\end{tabular}
\vspaceSQ{-2.5em}
\end{table}

EvalNet enables multifaceted analysis of numerous network topologies. We list the considered networks in Table~\ref{tab:networks-summary}, and provide an assessment of their important properties (performance, cost \& power consumption, path diversity), obtained using EvalNet.
\iftr
The table compiles the results described in the previous sections of this paper, as well as further results illustrated in the Appendix in Figures~\ref{fig:appendix_001}--\ref{fig:appendix_052}.
\fi

\subsection{General Observations \& Patterns}

A central observation is the critical role of path diversity (both shortest-path (MS) and non-shortest-path (MnS)) in determining a network's performance and scalability. Topologies such as Xpander and OFT exhibit high MS, meaning they provide many equal-cost paths between node pairs. This allows routing schemes like ECMP to distribute traffic efficiently, reducing congestion and delivering both low latency and high global bandwidth under minimal routing. These networks are thus well-suited for environments that rely on deterministic routing, where fault tolerance and load balancing are achieved through structural redundancy alone.

In contrast, Moore-bound diameter-2 networks like SF and PF attain excellent performance through topological minimalism (i.e., by minimizing average path length and diameter). However, their MS is low: most router pairs have only a single shortest path, rendering traditional ECMP ineffective. These networks compensate by offering very high MnS, which enables advanced non-minimal multipathing strategies such as Valiant load balancing and FatPaths, allowing traffic to be dispersed over longer, disjoint paths to avoid congestion. This tradeoff is critical: although SF, PF and other such networks lack diversity in minimal paths, their rich non-minimal structure, if properly exploited, enables them to outperform MS-rich networks under high loads—provided the routing stack supports sufficient adaptivity and complexity (e.g., congestion-aware routing, larger tables, multipath tracking).

This divergence between MS and MnS has a broader design tradeoff. Networks with high MS (e.g., Xpander) typically have slightly larger diameters but benefit from simplicity and robustness of minimal routing. Networks with low MS but high MnS (e.g., SF, PF) require intelligent routing to achieve high performance but can then perform remarkably well. On the other hand, low-dimensional regular topologies such as torus and mesh suffer from low values of both MS and MnS. While these networks are easily extensible in terms of size, their limited path diversity leads to high latency. Without structural augmentation (e.g., express links or randomized shortcuts), their performance does not scale with node count.

Tree-based networks, including Fat Trees, k-ary n-trees, and eXtended Generalized Fat Trees, offer a moderate level of both MS and MnS due to the presence of multiple upper-layer routes. These designs are non-blocking when fully provisioned and can deliver high global bandwidth by supporting ECMP and basic multipathing. However, they suffer from moderate-to-high latency due to their hierarchical depth, and may lack resilience of the more modern high-MS or high-MnS topologies that have good expansion properties.

\iftr
These insights tie into performance, cost, and scalability tradeoffs. While SF and PF achieve top-tier latency and bandwidth, their fixed diameter fundamentally limits extensibility. Namely, they cannot grow beyond a certain size without losing their optimal structural properties. In contrast, torus and mesh networks offer near-unlimited extensibility due to their non-constant diameter, but this comes at the cost of poor latency and reduced throughput. Topologies with very high radix, like HyperX and Multi-Layer Full Mesh, provide rich connectivity and sometimes high MS/MnS, but incur substantially higher construction and power costs, as they require more expensive switching hardware due to very high radix. Dragonfly and Cascade Dragonfly offer a middle ground, balancing latency, cost, and path diversity, which explains their popularity in practical deployments.
\fi

\iftr
Ultimately, while our findings highlight that each topology occupies a different point in the design tradeoff space, low-diameter networks such as Slim Fly and PolarFly emerge as particularly desirable candidates. Their minimal path lengths enable very low latency and bandwidth, which are critical for modern high-performance workloads. Although they offer limited shortest-path diversity, their rich structure of non-minimal paths allows them to sustain high throughput under load—provided that routing schemes can exploit this flexibility. EvalNet enables these tradeoffs to be quantified and visualized precisely, supporting informed design decisions based on latency, throughput, scalability, cost-efficiency, and path diversity.
\fi

\subsection{Example Detailed Insights}

We also provide a few selected more detailed insights that could be used, e.g., as a basis for enhancements of routing strategies.
\iftr
Full results and observations are in the Appendix.
\fi

\iftr
\textbf{Topologies based on algebraic expanders minimize contention for multipathing.} For example, Spectralfly's path interference is nearly zero, even matching random Jellyfish variants, implying that one can exploit non‐minimal detours almost for free, boosting available parallel paths without contention.
\fi
    
\textbf{Express Meshes have bimodal latency.} Express Meshes have a ``gap'' in their shortest-path multiplicity: most node-pairs have very few ($\le$2) or a large number ($\ge$5) of disjoint shortest paths, with almost no pairs in between. Hence, under uniform ECMP, about half the traffic has rich bandwidth, while the other half is starved, creating a bimodal latency distribution (unless one employs adaptivity).

\iftr
\textbf{Megafly has hop-bound tail-latency.} Megafly variants have little path interference for up to 3 hops, yet develop a small set ($\approx$10) of high-interference outliers at more than 3 hops. Hence, one can exploit non-minimal multipathing up to 3 hops with near-ideal bandwidth, but more hops risk large tail-latency spikes.
\fi

\iftr
\textbf{PolarStar variants have performance trade-offs.} PolarStars with Paley-subgraphs have marginally higher shortest-path multiplicity at 5 hops and only moderate path interference, whereas PolarStars with BDF-subgraphs yield lower overall interference (smoother latency) but slightly fewer total paths. Thus, one can tune PolarStar for either peak throughput (Paley) or tighter latency tails (BDF) based on application's burstiness.
\fi

\section{CONCLUSION}

We introduce EvalNet, the first toolchain that enables comprehensive generation and in-depth analysis of extreme-scale interconnection networks, with a strong focus on path diversity as a core enabler of performance and scalability. EvalNet integrates topology generation, fine-grained path diversity analysis, performance modeling, and cost evaluation into a unified and extensible framework. Using EvalNet, we conduct the widest comparative study to date across 25 network topologies, revealing how shortest- and non-shortest-path diversity relate to key performance and deployment metrics such as latency, throughput, cost, and power. Our analysis shows that the ability to exploit non-minimal multipathing is essential for maintaining high throughput in low-diameter, scalable networks. EvalNet interfaces seamlessly with established simulators and serves as a practical and versatile tool for guiding future innovations in large-scale network design in data centers and supercomputers.

\section*{Acknowledgment}
We thank Ankit Singla for help during the early stages and Haoran Zhao for his code contributions.
We thank Colin McMurtrie, Mark Klein, Angelo Mangili, and the whole CSCS team granting access to the Ault and Daint machines. 
We thank Timo Schneider for help with infrastructure at SPCL.
This paper has been assigned the LANL identification number LA-UR-26-20454.
This project received funding from the European Research Council (Project PSAP, No.~101002047), and the European High-Performance Computing Joint Undertaking (JU) under grant agreement No.~955513 (MAELSTROM). 
This project received funding from the European Union's HE research and innovation programme under the grant agreement No. 101070141 (Project GLACIATION).
This project was supported by JSPS KAKENHI Grant Number JP19H04119.
We gratefully acknowledge Polish high-performance computing infrastructure PLGrid (HPC Center: ACK Cyfronet AGH) for providing computer facilities and support within computational grants no.~PLG/2024/017103 and PLG/2025/018259.

\bibliographystyle{IEEEtranS}
\bibliography{references.complete}

\iftr
\appendix
\section{Appendix}
\label{sec:appendix}

\begin{figure*}
\centering
\includegraphics[width=0.95\textwidth]{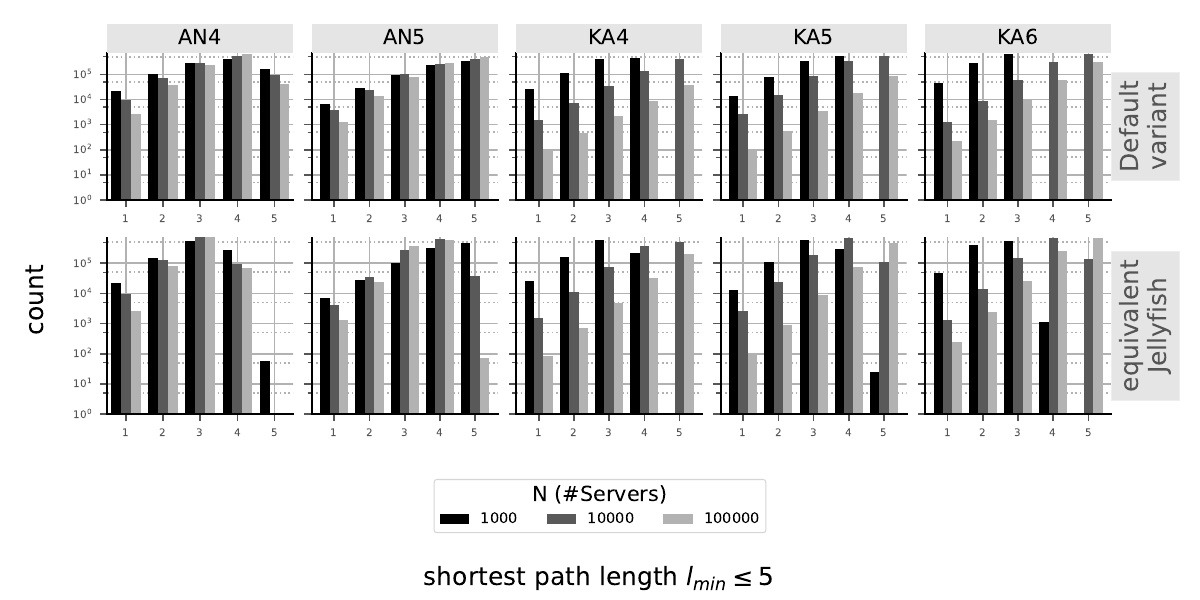}
\caption{Shortest path plot for Kautz and Arrangement graphs.
Kautz graphs have lower shortest paths compared to their Jellyfish equivalent for small networks.
They also have lower shortest paths when compared to a small Arrangement network. A small Kautz
graph has only a diameter of 3, while the Arrangement graph has already a diameter $\geq$ 5. The
analysis of the Arrangement graphs compared to their Jellyfish equilibrium, which is a Jellyfish topology built with the same hardware, looks fairly similar.
The Jellyfish equilibrium has less shortest paths of higher lengths $l$ = 5. The amount of shortest
paths of low length (between 1 and 3) gets smaller, the bigger the network gets for all topologies.
For Kautz graphs, these drop-offs are much more pronounced. There are about 100 times fewer
shortest paths of lengths 1 when we compare a small Kautz graph with a large Kautz graph. In
comparison, Arrangement networks have a drop-off between 5 and 10 times.}
\label{fig:appendix_001}
\end{figure*}

\begin{figure*}
\centering
\includegraphics[width=0.95\textwidth]{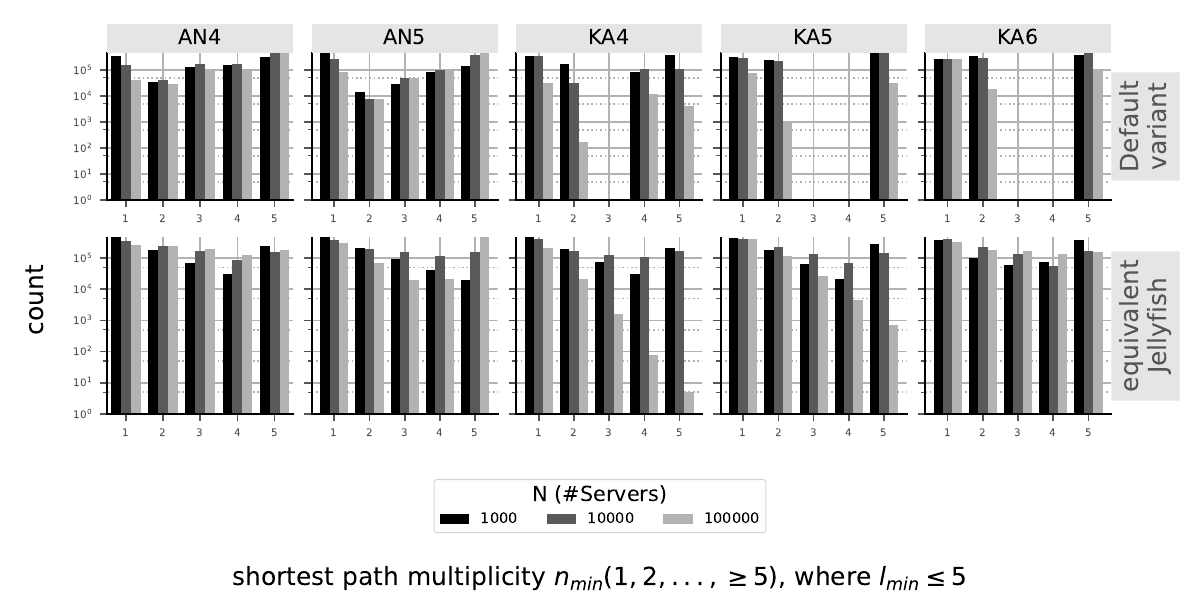}
\caption{Multiplicity plot for Kautz and Arrangement graphs. For Arrangement networks, the
shortest path density first decreases, then increases. In contrast, Kautz graphs have about the
same amount of path multiplicity, no matter the value of $n_{min}$. In the large Kautz network, we
can see a significant drop off in the amount of multiplicity. This drop off is less significant for
$K_{6,n}$. For Arrangement networks, these numbers of multiplicities stay about the same. There is
only a small shift, for $n_{min}$ = 1 the number of shortest path multiplicities sinks for higher N, but
for $n_{min}$ = 4 or 5 they only sink slightly. Kautz graphs have a gap in the multiplicity counts. They
either have a low amount (1 or 2) of shortest paths between any two nodes or a lot of overlap
($\geq$ 4) in the multiplicity. The analysis of the Jellyfish equivalent looks similar to the Arrangement
networks. The Jellyfish equivalent of the Kautz graphs on the other hand do not have this gap in
the multiplicity. For $K_{4,n}$ and $K_{5,n}$ the drop off in the multiplicity count can also be seen in the
Jellyfish equivalent for N $\approx$ 100,000. This cannot be seen for the Jellyfish equilibrium for $K_{6,n}$.}
\label{fig:appendix_002}
\end{figure*}

\begin{figure*}
\centering
\includegraphics[width=1.0\textwidth]{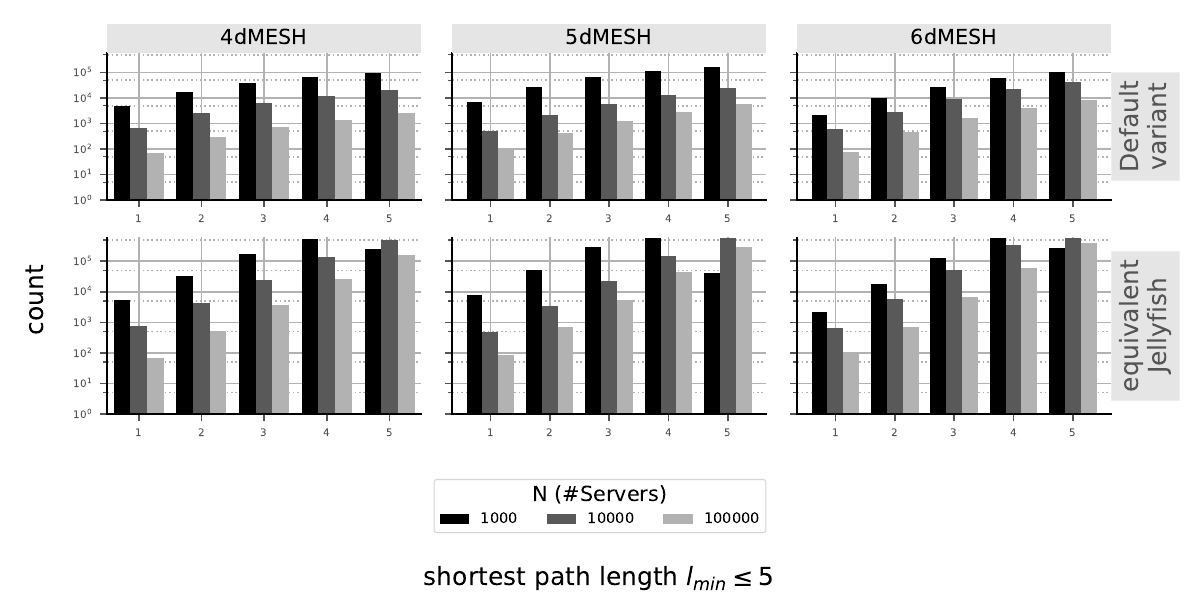}
\caption{Shortest path plot for mesh topology of different dimensions. The number of shortest
paths increases while $l_{min}$ increases. For bigger networks, there are fewer shortest paths with a
small value. The Jellyfish equivalent analysis looks fairly similar compared to the meshes. Only for
$l_{min}$ = 5, the Jellyfish network with higher N does not strictly have fewer amounts of shortest
paths.}
\label{fig:appendix_003}
\end{figure*}

\begin{figure*}
\centering
\includegraphics[width=1.0\textwidth]{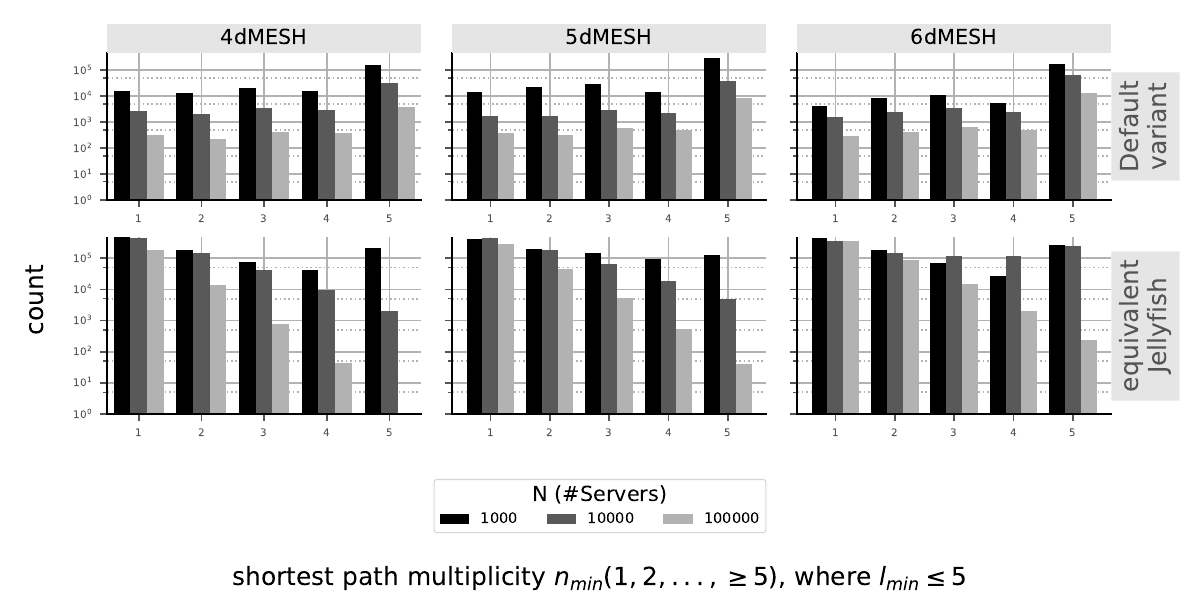}
\caption{Multiplicity plot for meshes of different dimensions. As expected, the
analysis looks fairly similar across all dimensions. For all sizes and all dimensions, the multiplicity
count is the highest for $n_{min} \geq$ 5. For bigger network sizes, the amount of shortest path
multiplicity decreases with roughly the same factor, independent of $n_{min}$. In contrast, the Jellyfish
equivalent networks have almost no difference in the multiplicity count for $n_{min}$ = 1. For bigger
values of $n_{min}$, the shortest path multiplicity drops even more compared to the original
mesh, when we increase the network size.}
\label{fig:appendix_004}
\end{figure*}

\begin{figure*}
\centering
\includegraphics[width=1.0\textwidth]{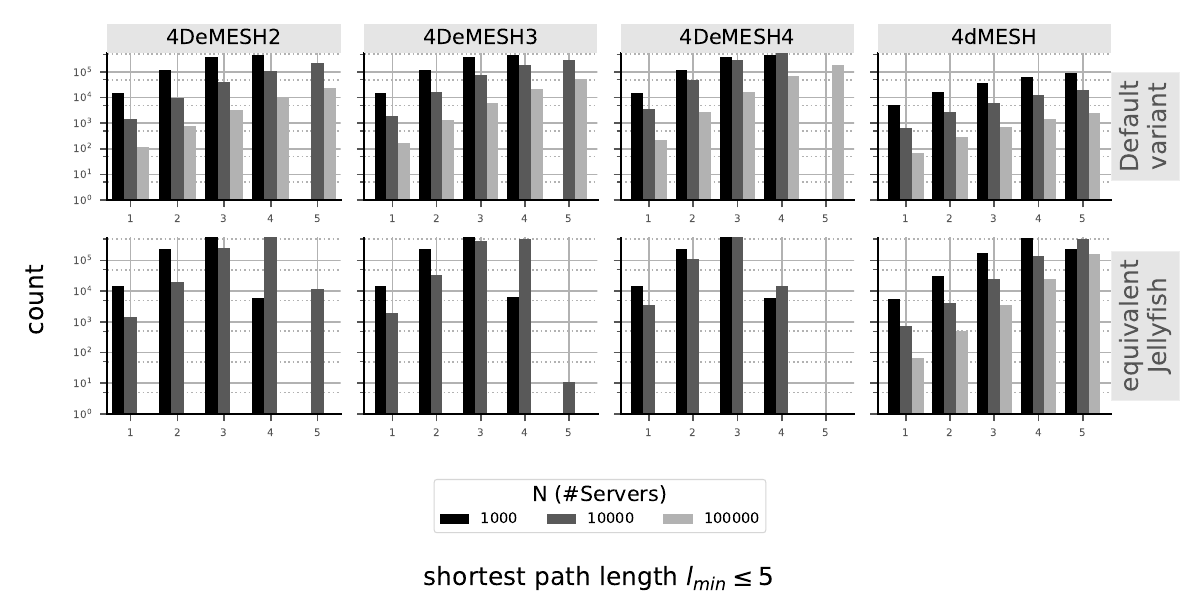}
\caption{Shortest path plot for the 4D Express Meshes with different gap sizes. We choose
the lowest g, such that for each Express Mesh 4DeMESHx we have at least x additional express
connections per dimension. As one can expect, the Express Meshes have lower shortest paths than the
original mesh due to their express connections. For some network sizes, the Express Meshes have only
a diameter of 4, while the standard 4D mesh has a diameter of $\geq$ 5. While the number of shortest
paths increase for bigger $l$ in the Express Meshes, their equivalent Jellyfishes don't show a similar
pattern. They increase first before they decrease for the biggest value of $l$ where there still exists a
shortest path.}
\label{fig:appendix_005}
\end{figure*}

\begin{figure*}
\centering
\includegraphics[width=1.0\textwidth]{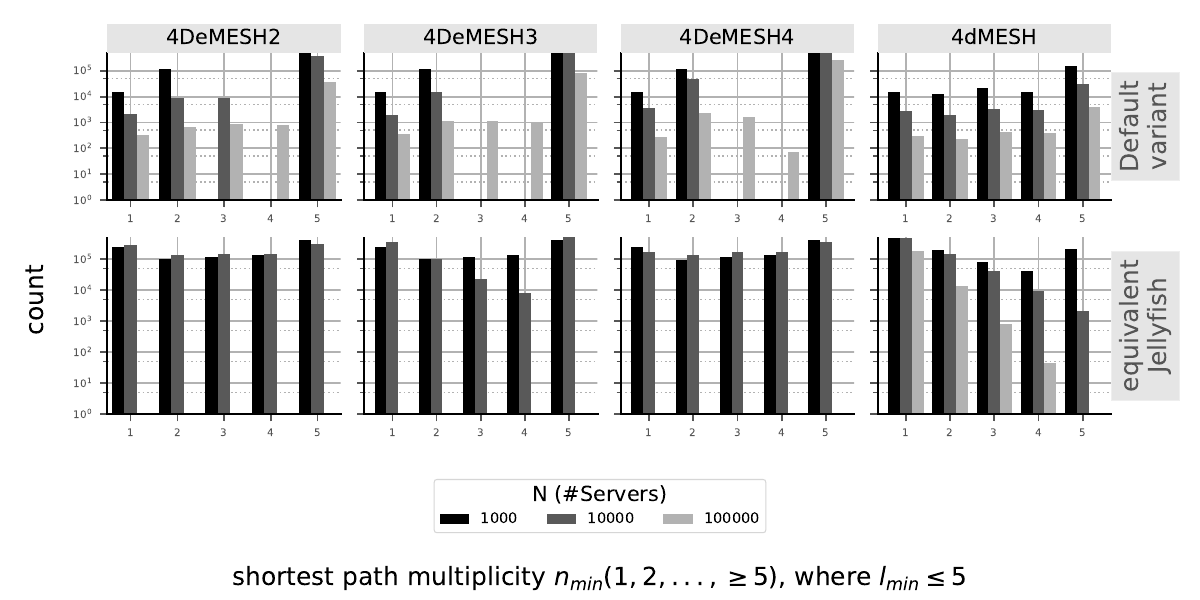}
\caption{Multiplicity plot for Express Meshes of dimension 4. For small and medium-sized
networks, there is a gap in the multiplicity counts. The length of the shortest path for any two
nodes exists either once or twice, or $\geq$ 5 times. For large Express Meshes, this gap is not visible
anymore. Express Meshes keep the high multiplicity of shortest paths, even for higher N,
due to their express connections, especially for lower gap sizes. For all network sizes, the Express
Meshes have the highest shortest paths multiplicity count for $n_{min} \geq$ 5. There is a big difference
in the multiplicity of the meshes compared to the equivalent Jellyfish. In contrast to the base
Express Meshes, the Jellyfish equilibriums are well distributed along all values for $n_{min}$ for small
and medium networks.}
\label{fig:appendix_006}
\end{figure*}

\begin{figure*}
\centering
\includegraphics[width=1.0\textwidth]{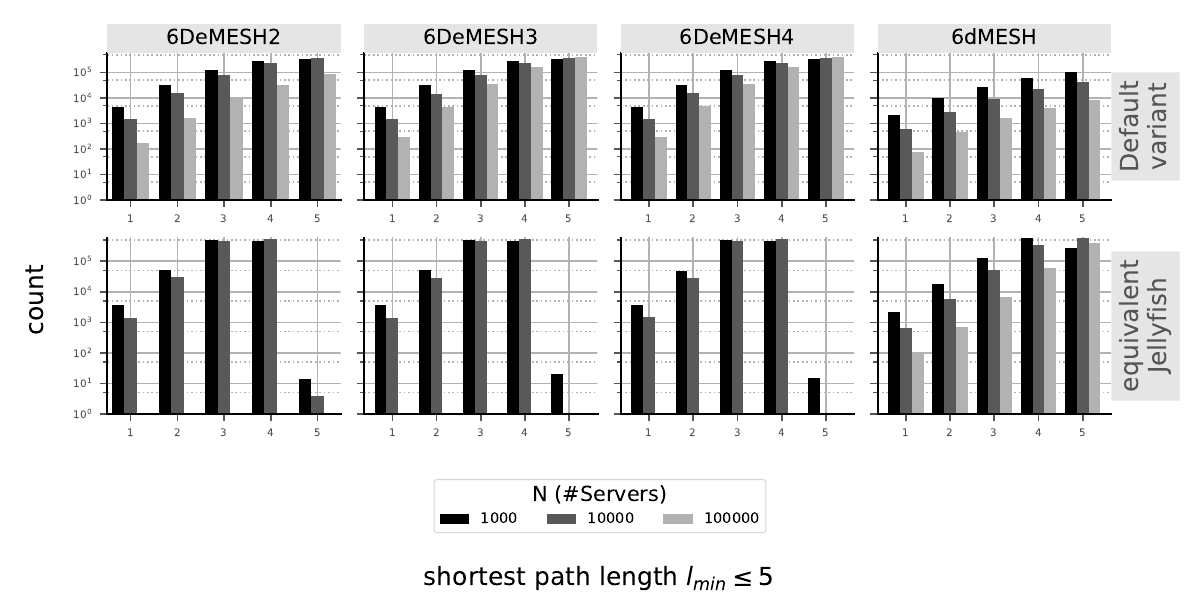}
\caption{Shortest path plot for 6D Express Meshes with different gap sizes. We chose the
lowest g, such that for each Express Mesh 6DeMESHx we have at least x additional express
connections per dimension. The shortest path analysis of these Express Meshes look similar to the
analysis of their 4D counterpart. Express connections do not allow the 6D graph to become a
diameter $\leq$ 5 network, since any path only traverses over one dimension. The Jellyfish equilibrium
is also fairly similar when comparing the 6D to the 4D analysis. The amount of shortest paths
grows as $l$ grows, before dropping down for some outliers that have a shortest path length of
5.}
\label{fig:appendix_007}
\end{figure*}

\begin{figure*}
\centering
\includegraphics[width=1.0\textwidth]{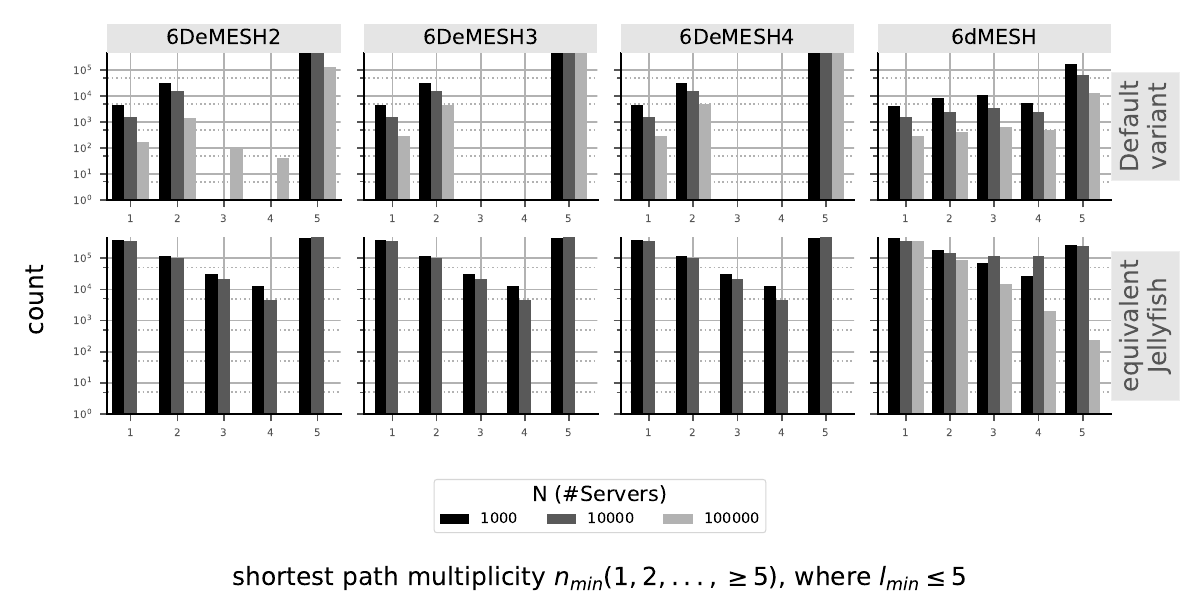}
\caption{Multiplicity plot for Express Meshes of dimension 6. We can see again the gap in the
amount of shortest path multiplicity. There either exist very few paths of length equal to the
shortest path between two nodes, or there exists a lot of them (more than 4). Only the large
Express Mesh with two additional express connections per dimension does not show this gap. Same
as in the 4D case, most shortest paths have a multiplicity $\geq$ 5. In case of the Jellyfish equilibrium,
the shortest path multiplicity count drops at higher $n_{min}$. Just for $n_{min}$ = 5, the multiplicity
count increases, as they have as well the most shortest paths with multiplicity $\geq$ 5.}
\label{fig:appendix_008}
\end{figure*}

\begin{figure*}
\centering
\includegraphics[width=1.0\textwidth]{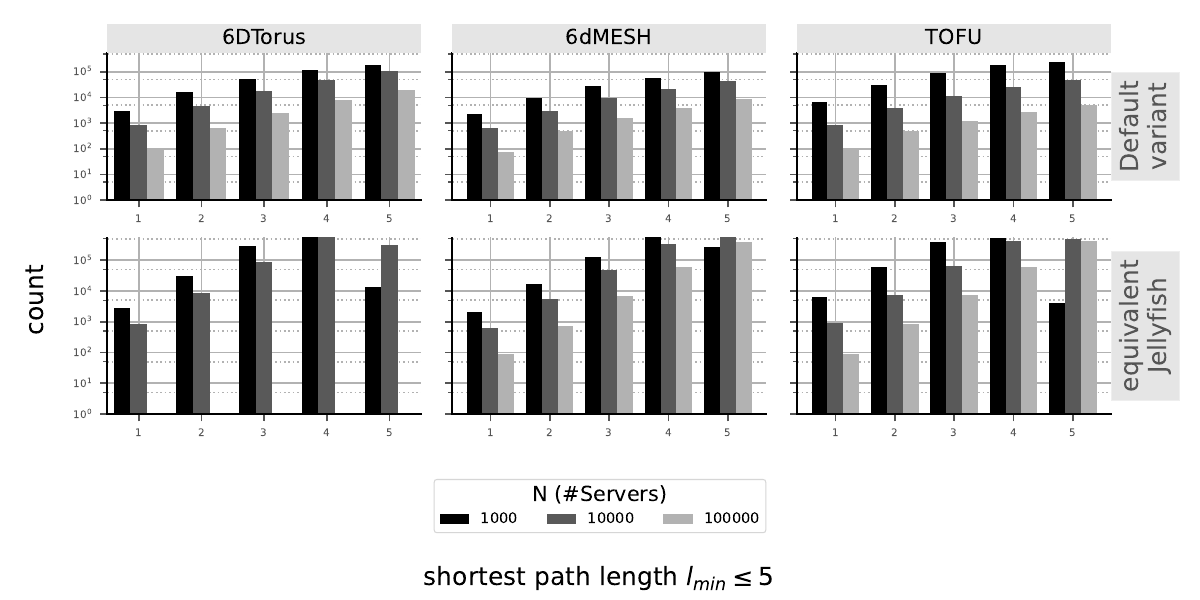}
\caption{Shortest path plot for the Tofu topology, 6D torus and 6D mesh. As we have already
seen in the analysis of the meshes, the amount of shortest paths sinks as the network size gets
bigger. The bigger $l$ gets, the more shortest paths exist of length $l$. All networks show similar patterns.
Tofu has more shortest paths $l_{min} \leq$ 5 compared to all the other meshes for
small networks, but is worse for bigger networks. The same is true when we compare Tofu to the
equivalent Jellyfish.}
\label{fig:appendix_009}
\end{figure*}

\begin{figure*}
\centering
\includegraphics[width=1.0\textwidth]{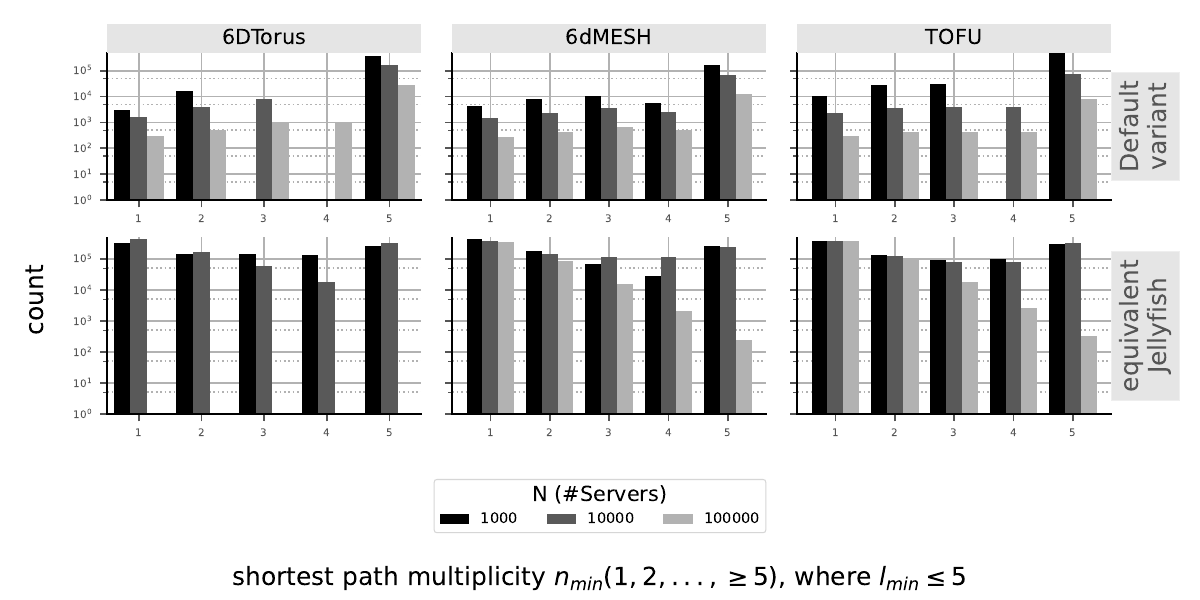}
\caption{Multiplicity plot for the Tofu network, 6D torus and 6D mesh. Similar as for the
meshes, all the topologies have the most multiplicity count where $n_{min} \geq$ 5. The Tofu network
has even the highest multiplicity count for $n_{min} \geq$ 5 for all small networks. In medium and
large networks, Tofu does not have the most multiplicity count anymore. The bigger the network
sizes are, the smaller the multiplicity counts get. In the analysis of the torus, we can find the
already mentioned gap in the shortest path multiplicities for small and medium-size networks.
Tofu has this gap only for small networks. The Jellyfish equilibrium of the Tofu network is well
distributed along all values for $n_{min}$ for small and medium networks. For the large network, there
is a significant drop-off and the multiplicity is about 30 times smaller for $n_{min} \geq$ 5 compared to
the small or medium network.}
\label{fig:appendix_010}
\end{figure*}

\begin{figure*}
\centering
\includegraphics[width=1.0\textwidth]{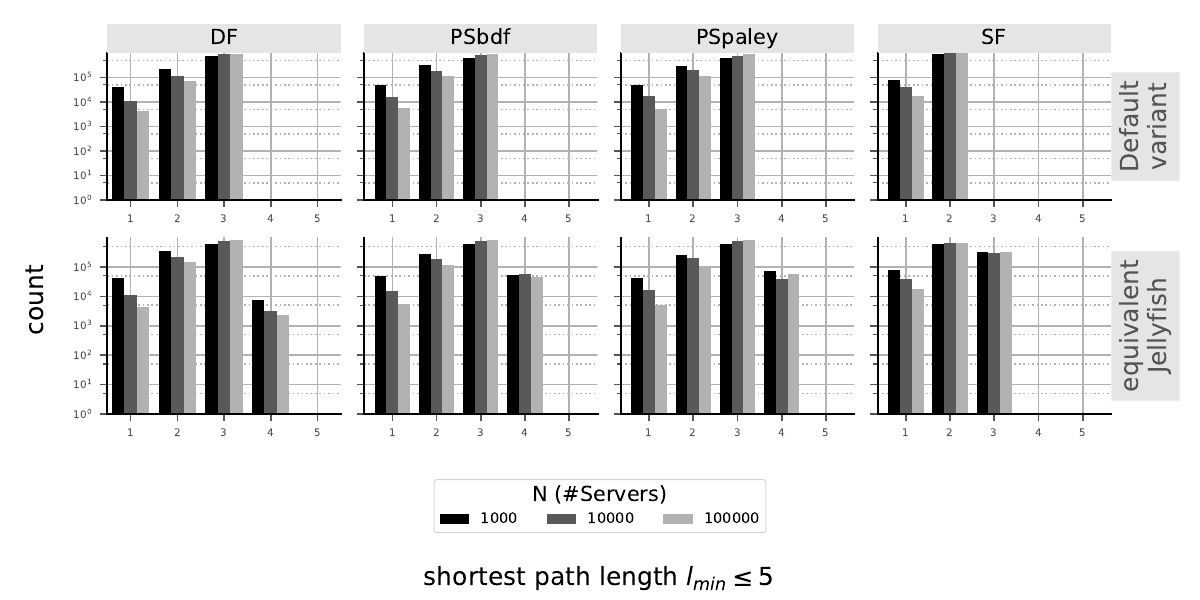}
\caption{Shortest path plot for PolarStar compared to Dragonfly and Slim Fly.
We can clearly see that all are diameter $\leq$ 3 topologies. All of the topologies have in general
lower smallest paths compared to their equivalent Jellyfish graph. Slim Fly has only shortest paths
of length 2. All the others look fairly similar, having fewer shortest paths with $l_{min}$ = 1 compared
to $l_{min}$ = 2 or 3. The amount of shortest paths of length 1 and 2 decreases as the network size
increases. The shortest paths of length 3 slightly increase though. There is almost no difference
for the analysis between PolarStar with BDF or Paley subgraphs.}
\label{fig:appendix_011}
\end{figure*}

\begin{figure*}
\centering
\includegraphics[width=0.9\textwidth]{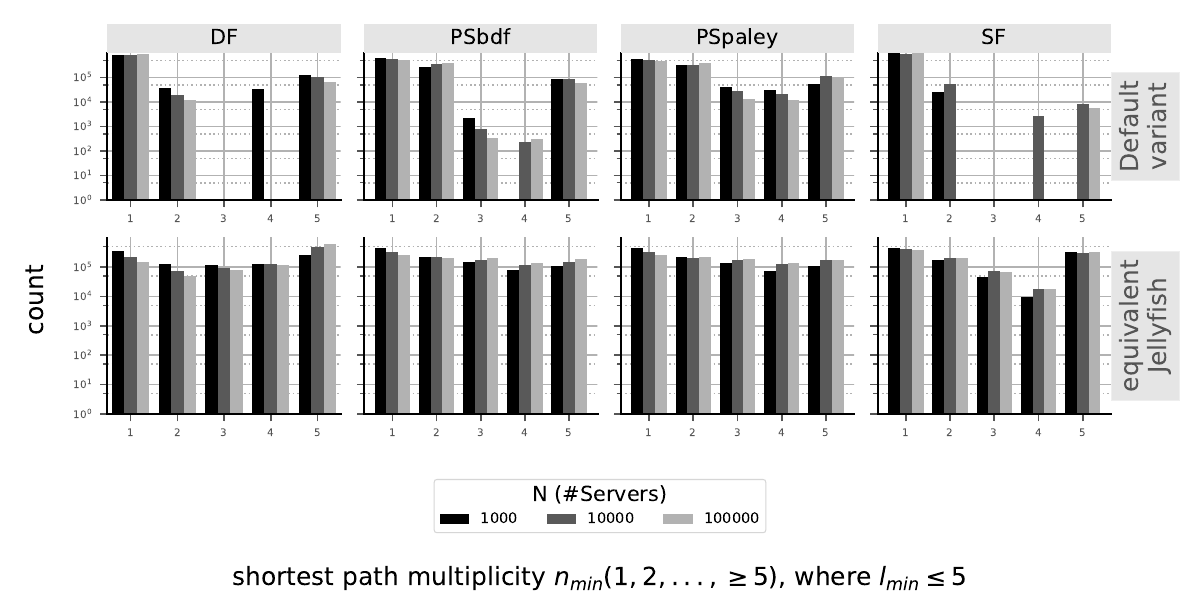}
\caption{Multiplicity plot for PolarStar, Dragonfly and Slim Fly. Slim Fly has the
least multiple shortest paths of all topologies. Only for medium and large networks, there
are pairs of nodes with shortest path multiplicity bigger than 3. All topologies have the most
shortest paths only once between any two nodes. PolarStar with BDF subgraphs
has only a few shortest path multiplicity count for $n_{min}$ = 3 or 4, while PolarStar with the
Paley subgraphs has a well-distributed count for all values of $n_{min}$.}
\label{fig:appendix_012}
\end{figure*}

\begin{figure*}
\centering
\includegraphics[width=0.9\textwidth]{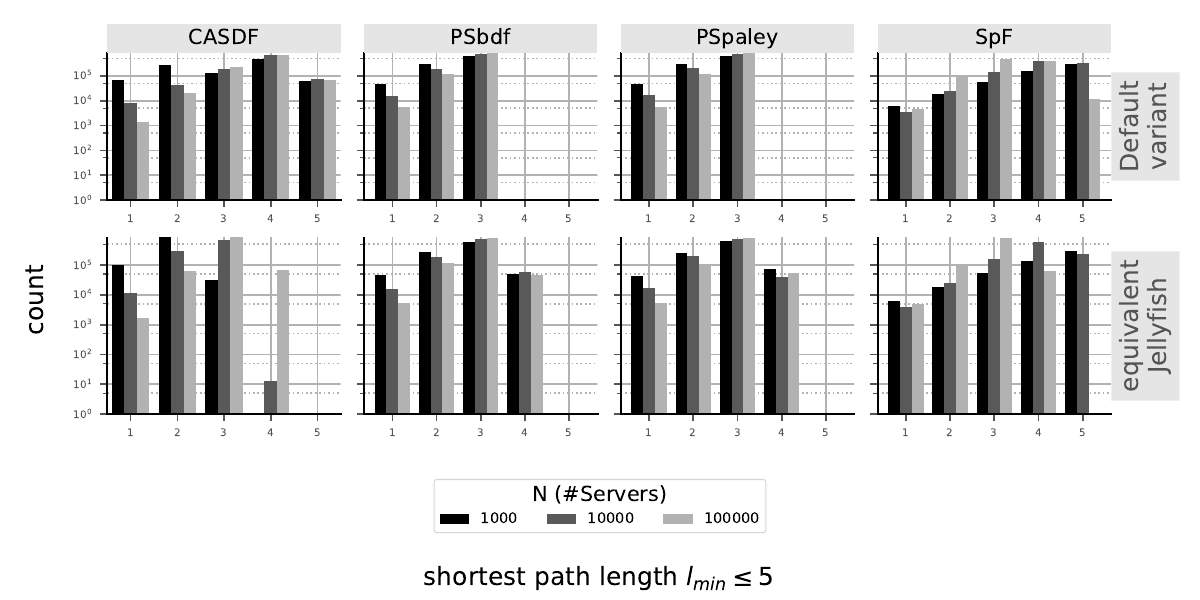}
\caption{Shortest path plot for Cascade Dragonfly, PolarStar and Spectralfly.
The amount of shortest paths of length 1 and 2 decreases, if the network gets bigger, because
a single node in a group can only reach 21 nodes from another group with at most 2
jumps. Meanwhile, all 95 nodes in the same group can be reached with 2 gaps. The amount
of node pairs that have a shortest path $\geq$ 3 stays almost the same. Spectralfly has the least
amount of shortest paths of length 1 for all network sizes. The amount of shortest paths of low
length stays the same or gets even better for higher-size networks. The Jellyfish equivalent of
the Cascade Dragonfly has only a diameter of 3 or 4 (depending on the network size), while Cascade
Dragonfly is a diameter 5 topology. Spectralfly's Jellyfish equilibrium on the other hand looks
similar to its base counterpart. Only for large network sizes, the Jellyfish graph shows no shortest
paths of length 5, while the Spectralfly graph still has some.}
\label{fig:appendix_013}
\end{figure*}

\begin{figure*}
\centering
\includegraphics[width=1.0\textwidth]{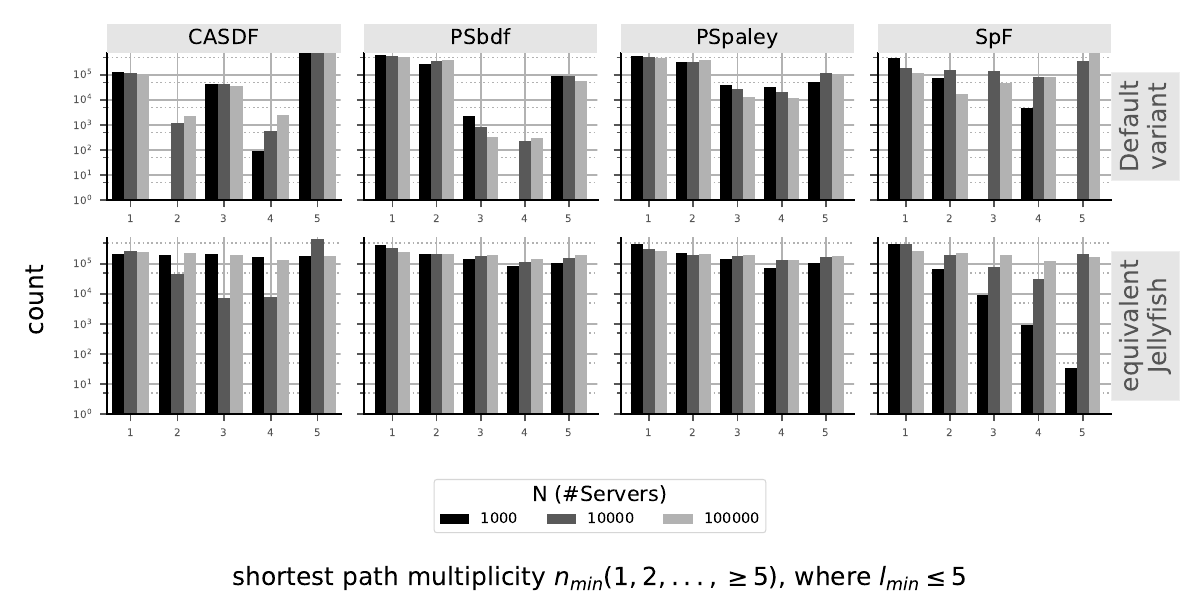}
\caption{Multiplicity plot for the Cascade Dragonfly, PolarStar and Spectralfly networks. Most
node pairs in the Cascade Dragonfly topology have multiple ($\geq$ 5) shortest paths. The multiplicity
count either stays about the same or gets bigger, as the Cascade Dragonfly graph increases its size.
The small Spectralfly network has only multiplicity counts $\leq$ 4. For large and medium networks,
the Spectralfly topology has slightly more multiplicity count for $n_{min}$, however only by a small margin.
For small and medium networks, the equivalent Jellyfish graph of the Cascade Dragonfly has as
well the most shortest path multiplicity for $n_{min} \geq$ 5. The large Jellyfish has about the same
multiplicity count over all values of $n_{min}$. While the equivalent Jellyfish of Spectralfly drops in
multiplicity counts for higher values of $n_{min}$, for medium and large networks it increases.}
\label{fig:appendix_014}
\end{figure*}

\begin{figure*}
\centering
\includegraphics[width=1.0\textwidth]{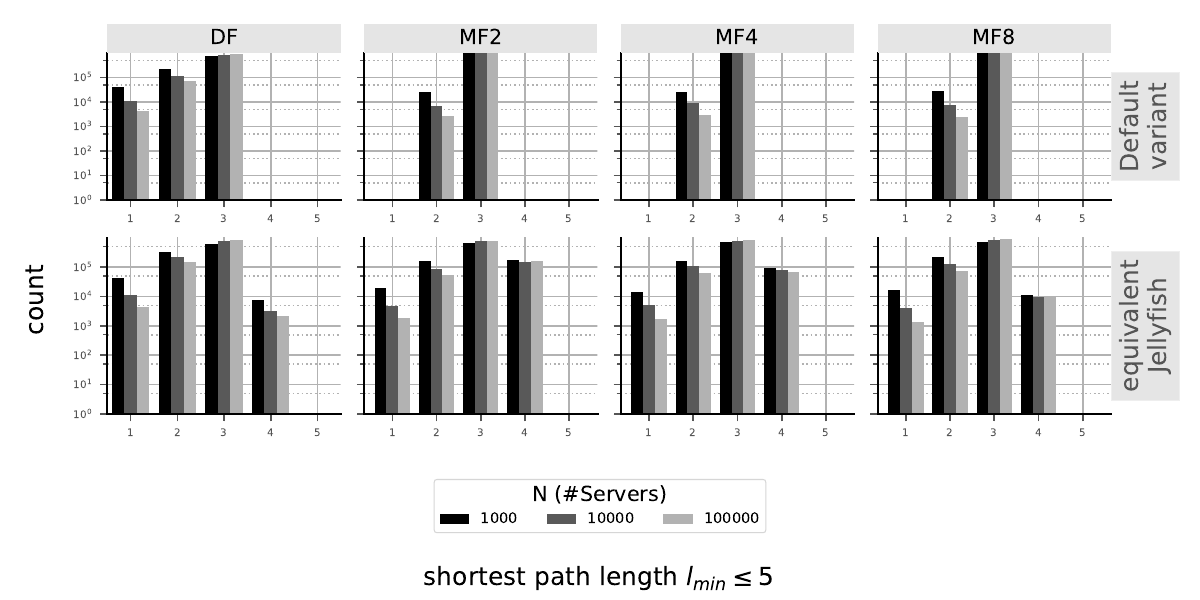}
\caption{Shortest path plot for Megafly with different values for g and Dragonfly.
We can see, that Megafly is a diameter 3 network as well. Megafly has no shortest path of
length 1, since every leaf router is only connected to spine routers, and not to another leaf router.
Because two nodes have only shortest path of length $l$ = 2 if they are in the same group, most
shortest paths are $l_{min}$ = 3 for Megafly. All topologies have the same amount of shortest paths
of length 3 for all network sizes. The analysis of the equivalent Jellyfishes looks pretty similar
compared to the base Megafly network, where Megafly has shortest paths of length $l$. The
Jellyfish graphs have additional shortest paths of length 1 and sometimes length 4.}
\label{fig:appendix_015}
\end{figure*}

\begin{figure*}
\centering
\includegraphics[width=1.0\textwidth]{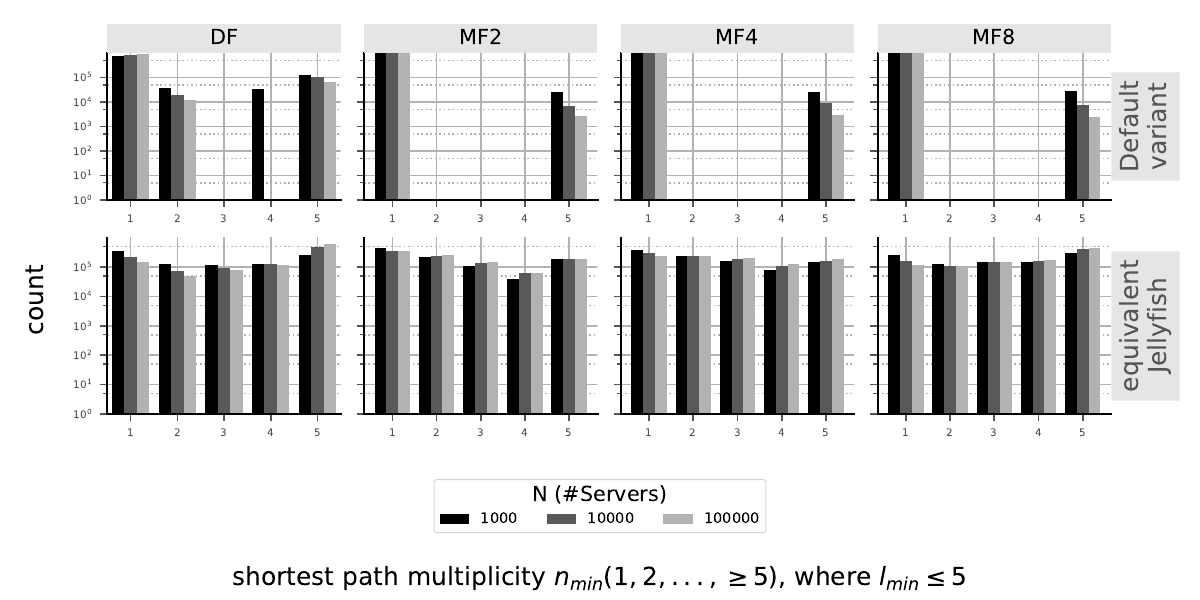}
\caption{Multiplicity plot for the Dragonfly and Megafly topologies with different values for
g. All topologies have mostly only one shortest path between any two nodes. Between any two
leaf nodes in Megafly, there is mostly only one shortest path. There is either one shortest path
between the nodes or a bunch ($\geq$ 5) of them. Dragonfly has some shortest paths of different
multiplicity, but most lie in the extremes $n_{min}$ = 1 or $n_{min} \geq$ 5. Their equivalent Jellyfish on the
other hand have well-distributed shortest path multiplicities.}
\label{fig:appendix_01}
\end{figure*}

\begin{figure*}
\centering
\includegraphics[width=1.0\textwidth]{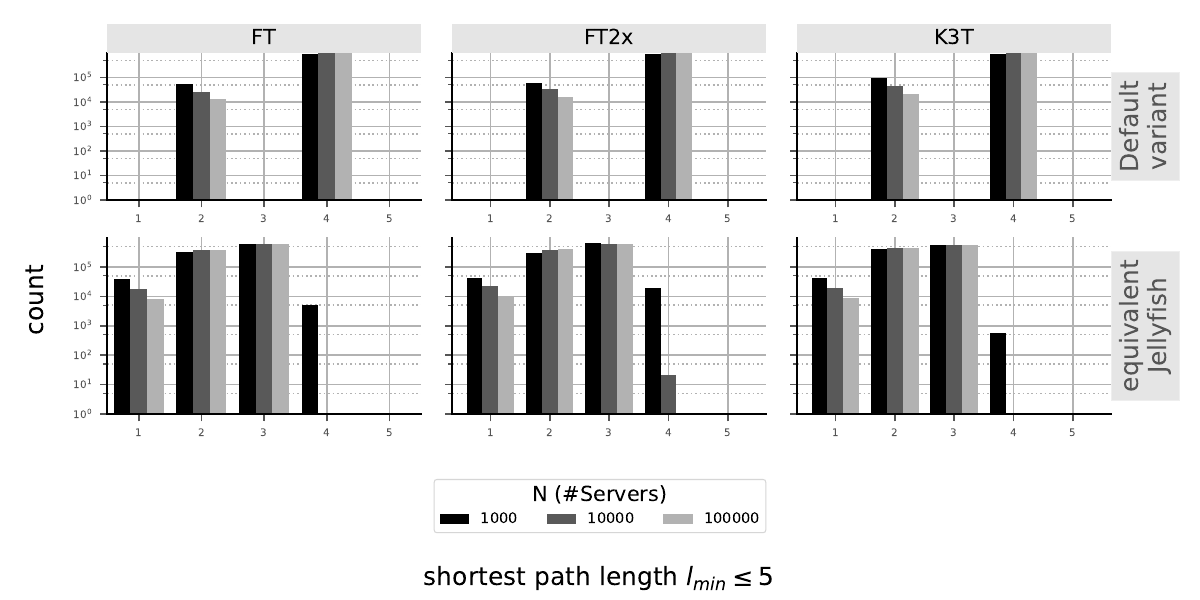}
\caption{Shortest path plot for Fat Trees and the k-ary 3-tree. All topologies exhibit the same patterns for the
shortest paths of a given length. While the shortest paths of length 4 stay the same for any network size,
the number of paths of length 2 decreases as the network gets bigger.}
\label{fig:appendix_017}
\end{figure*}

\begin{figure*}
\centering
\includegraphics[width=1.0\textwidth]{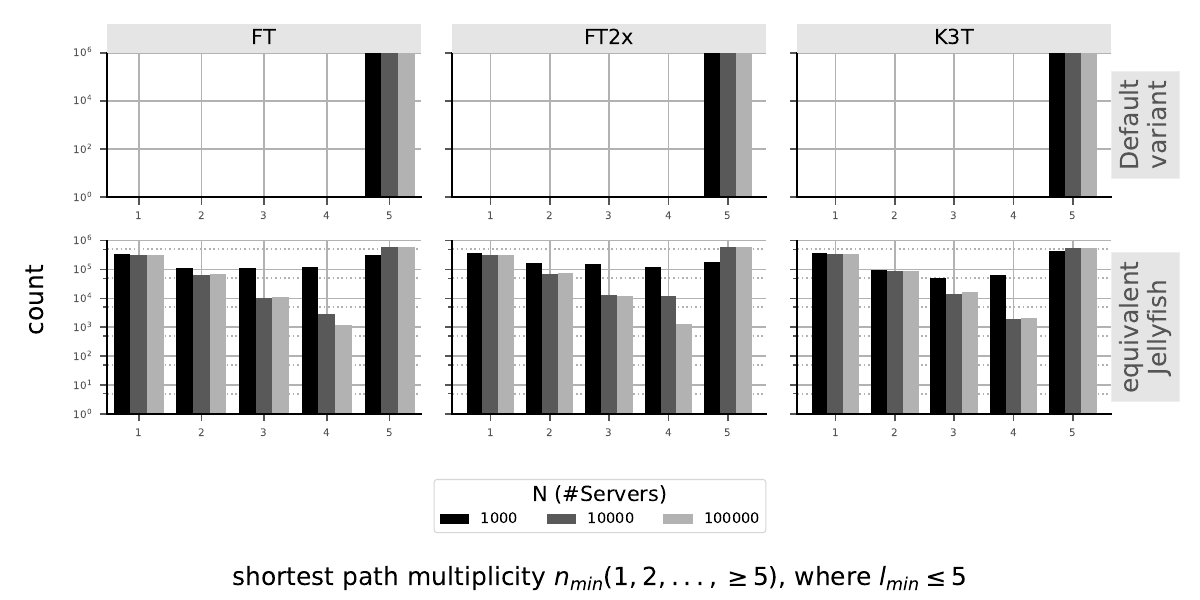}
\caption{Multiplicity plot for Fat Trees and the k-ary 3-tree. All networks and all sizes have a
path multiplicity count of 5 or higher.}
\label{fig:appendix_018}
\end{figure*}

\begin{figure*}
\centering
\includegraphics[width=1.0\textwidth]{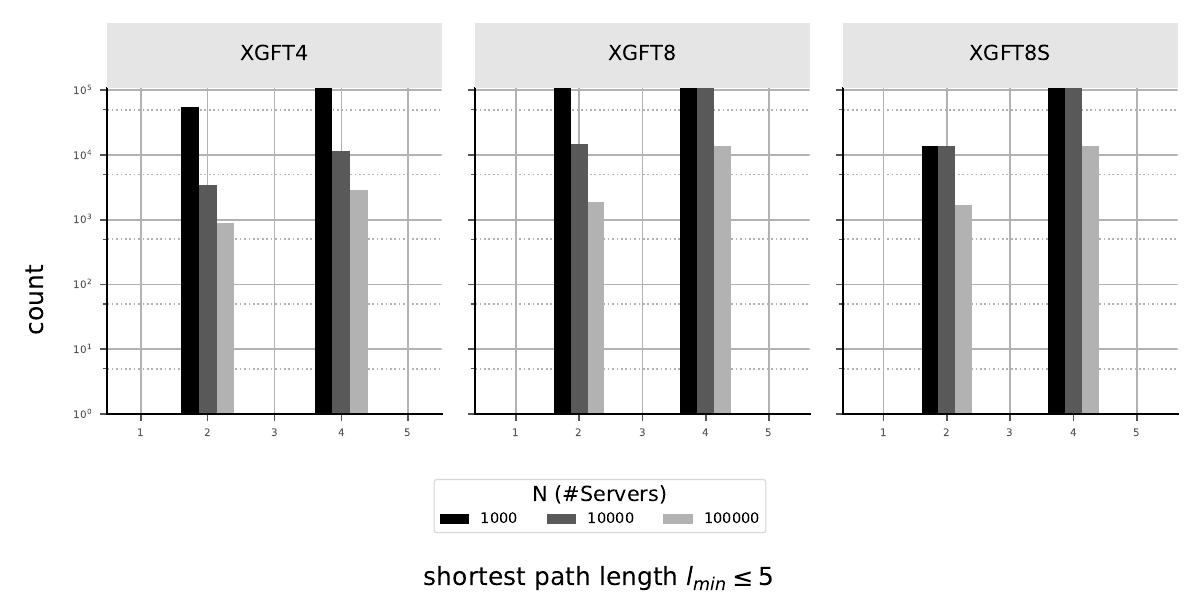}
\caption{Shortest path plot for length $l_{min}$ up to $l$ = 5 for multiple variants of xGFTs.
There are only shortest paths of even length. There exist more shortest paths of length
4 than length 2 in general. As the network size increases, the number of shortest paths
at each length decreases.}
\label{fig:appendix_019}
\end{figure*}

\begin{figure*}
\centering
\includegraphics[width=1.0\textwidth]{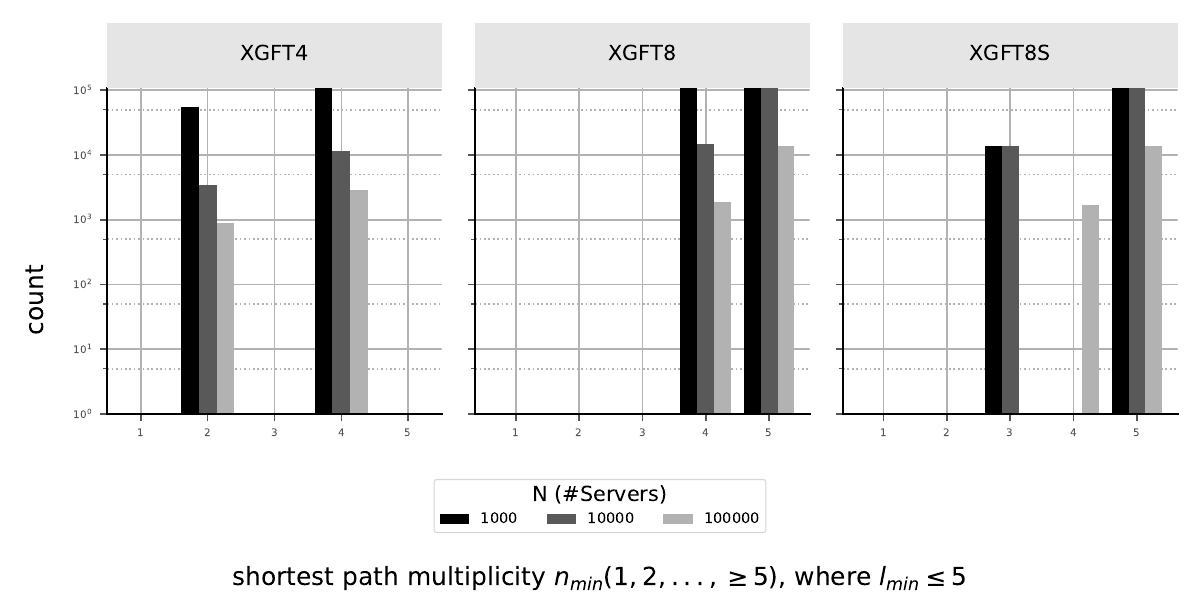}
\caption{Multiplicity plot for multiple variants of xGFTs. Variant 4 has the lowest amount of
multiplicity count. It only has a multiplicity count of 2 and 4. For large network sizes, the 8S
variant has even more multiplicity than the smaller networks of that variant.}
\label{fig:appendix_020}
\end{figure*}

\begin{figure*}
\centering
\includegraphics[width=0.6\textwidth]{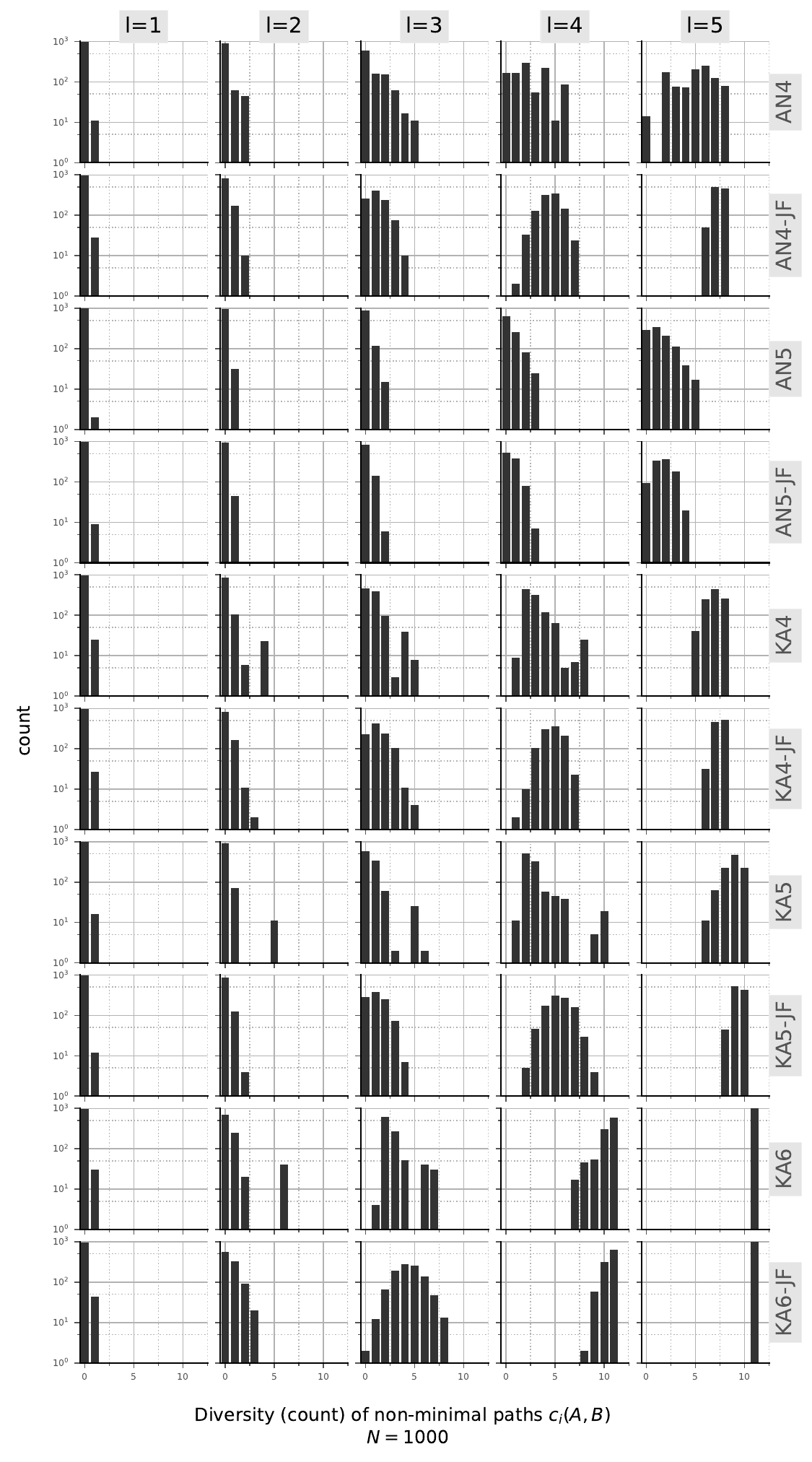}
\caption{Edge disjoint path plots for Kautz and Arrangement graphs of small sizes. The
connectivity for $A_{n,4}$ is better than the connectivity for $A_{n,5}$. Both have in general fewer edge
disjoint paths than the Kautz topologies. $K_{6,n}$ has the best connectivity for these topologies.
Furthermore, the Kautz networks analysis looks similar to its equivalent Jellyfish.}
\label{fig:appendix_021}
\end{figure*}

\begin{figure*}
\centering
\includegraphics[width=0.7\textwidth]{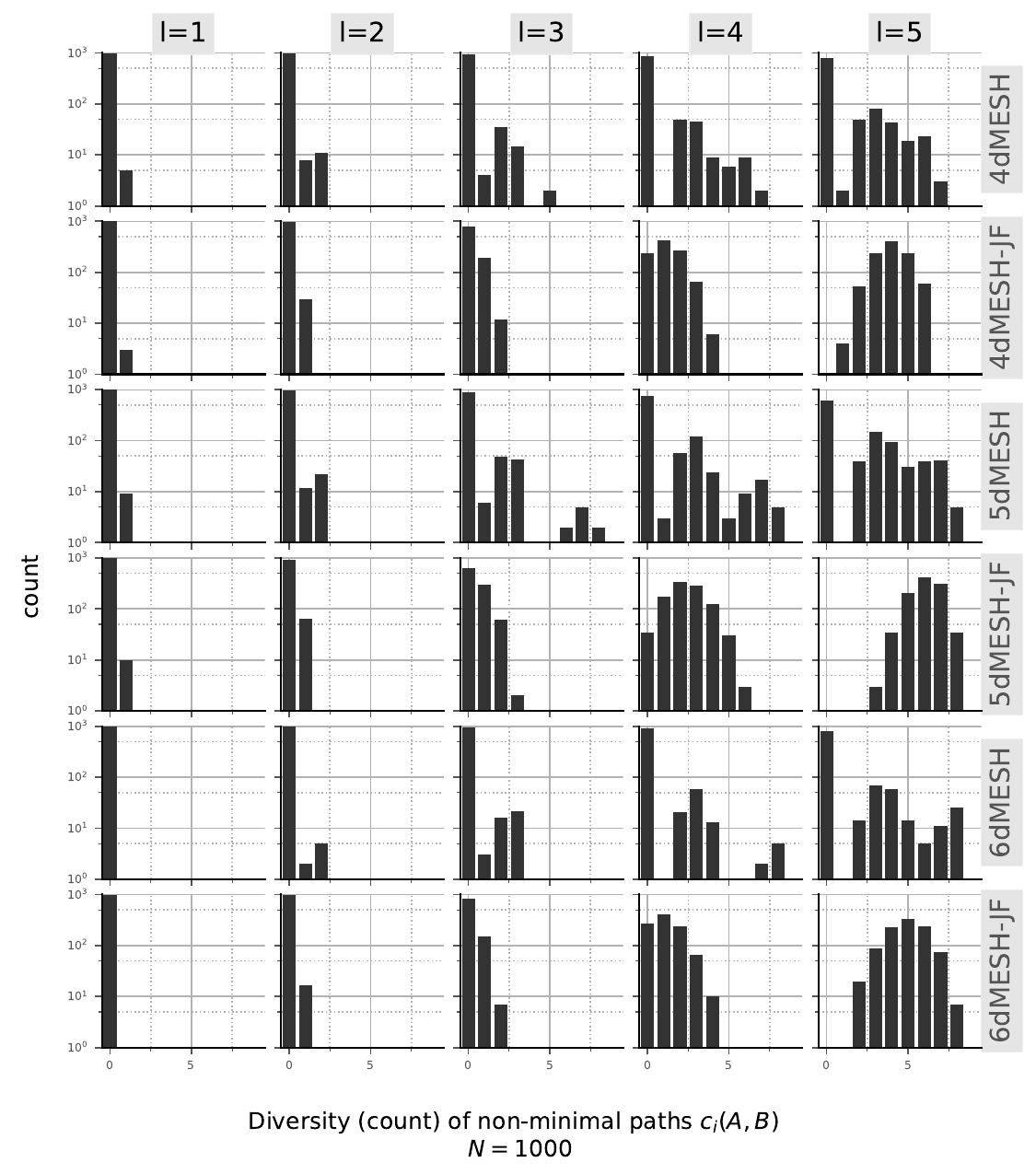}
\caption{Edge disjoint path plots for meshes of different dimensions. For all lengths $l$, the
meshes have mostly zero edge disjoint paths between any two nodes. Furthermore, we can see
that there are almost no node pairs, where there are 2 paths of a given length $l$. In contrast,
none of the Jellyfish equivalent graphs have zero edge disjoint paths for $l$ = 5. The 4D mesh has
a slightly worse path diversity than the 5D and 6D meshes.}
\label{fig:appendix_022}
\end{figure*}

\begin{figure*}
\centering
\includegraphics[width=1.0\textwidth]{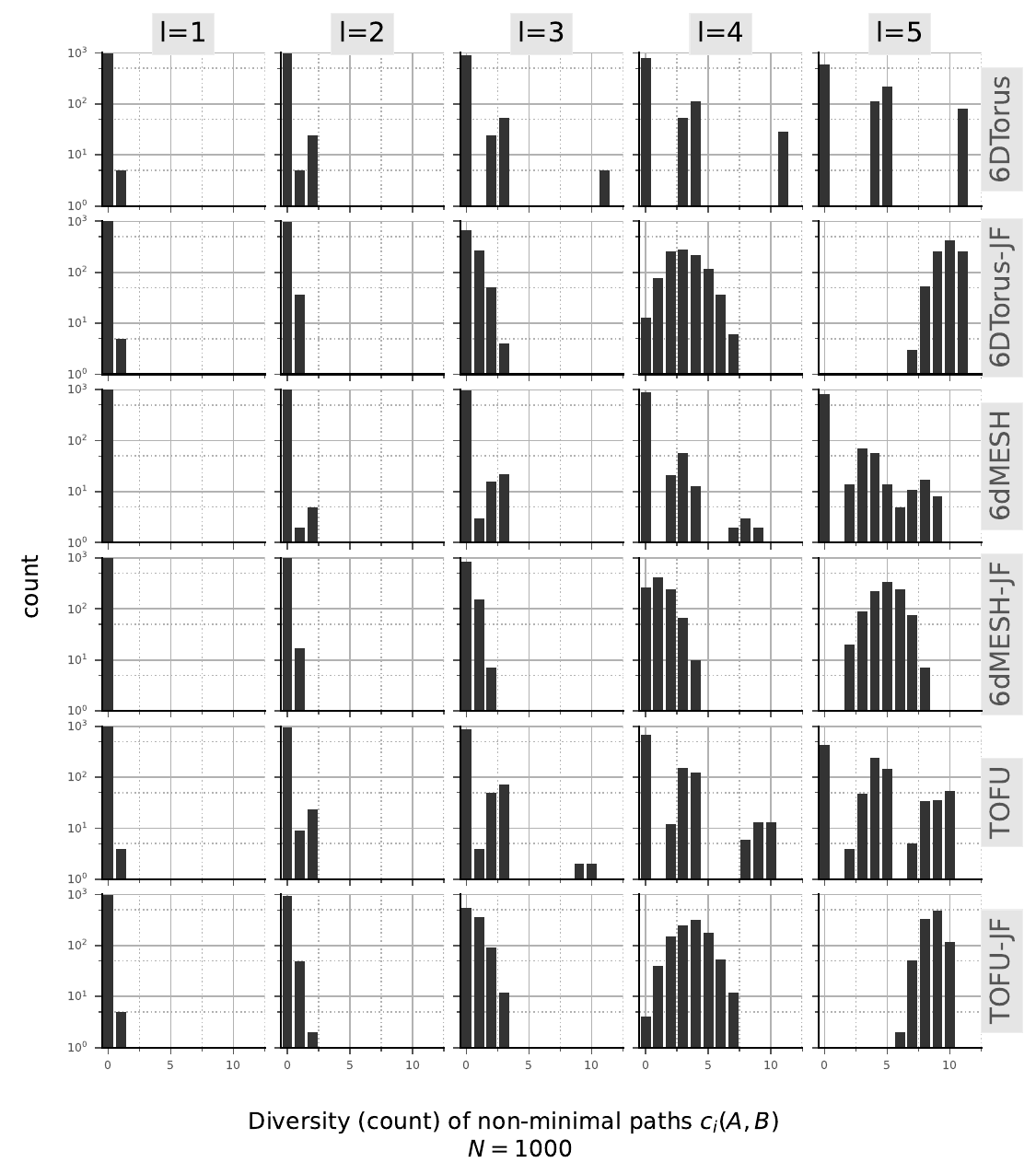}
\caption{Edge disjoint path plots for the Tofu topology, 6D torus and 6D Express Mesh.
Similar to the diversity analysis of the meshes, all the topologies have zero path diversity for
most node pairs of any given length. For both Tofu and the 6D torus network, their path diversity
counts lie in three groups. They have diversity counts of 1, $\approx$ 4 and $\approx$ 10. This is even more
extreme for the 6D torus network and cannot be seen for the 6D mesh graph. Same as before,
the equivalent Jellyfish graphs have a well-distributed diversity count.}
\label{fig:appendix_024}
\end{figure*}

\begin{figure*}
\centering
\includegraphics[width=0.79\textwidth]{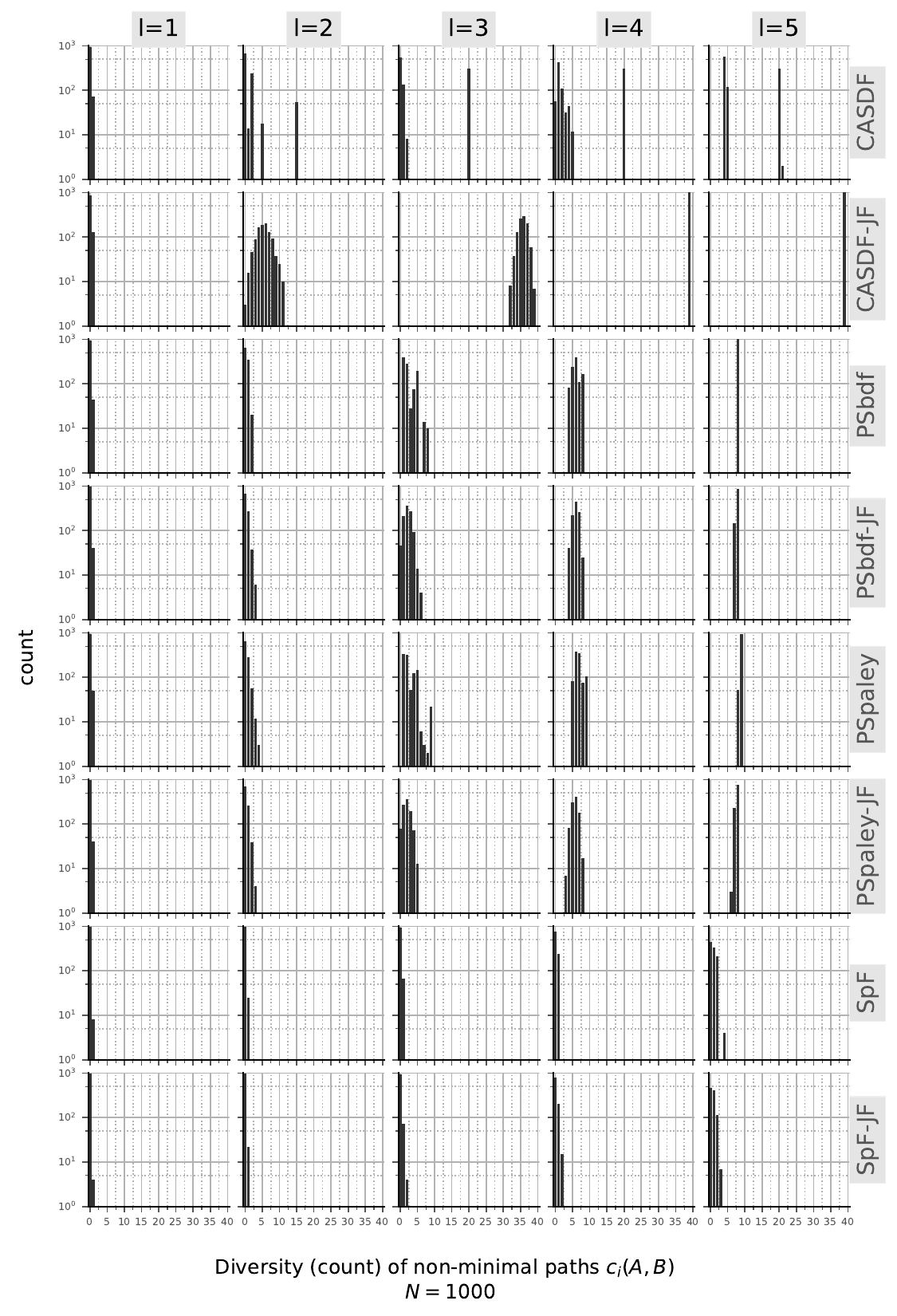}
\caption{Edge disjoint path plots for Cascade Dragonfly, PolarStar and Spectralfly topologies.
Both Spectralfly and Cascade Dragonfly have mostly a low diversity count. Cascade Dragonfly has
an additional peak with node pairs of a higher diversity count. Even for length $l$ = 5, Spectralfly
has only node pairs with 4 edge disjoint paths.}
\label{fig:appendix_025}
\end{figure*}

\begin{figure*}
\centering
\includegraphics[width=0.45\textwidth]{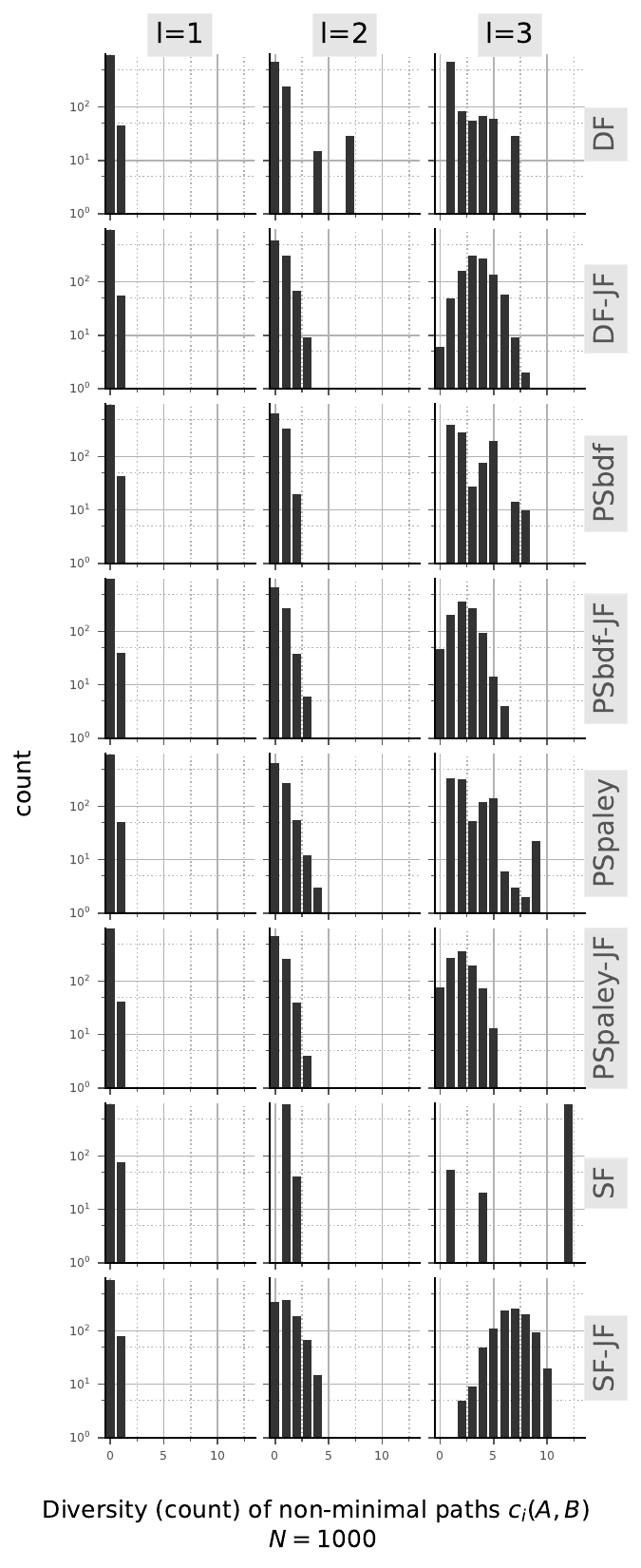}
\caption{Edge disjoint path plots for Dragonfly, PolarStar and Slim Fly topologies. Slim Fly has
the most path diversity by far. It has a diversity count $\geq$ 10 for most nodes, even for paths of
length 3. The diversity analysis of PolarStar is pretty similar to Dragonfly for length 1
to 4. For $l$ = 5 we can see that Dragonfly has a higher diversity count by 2 for all node pairs.
The PolarStar graph with Paley subgraphs has slightly higher edge disjoint paths for $l$ = 5.
Both variants of PolarStar have the same or better diversity compared to their Jellyfish equivalent
for all lengths.}
\label{fig:appendix_026}
\end{figure*}

\begin{figure*}
\centering
\includegraphics[width=0.45\textwidth]{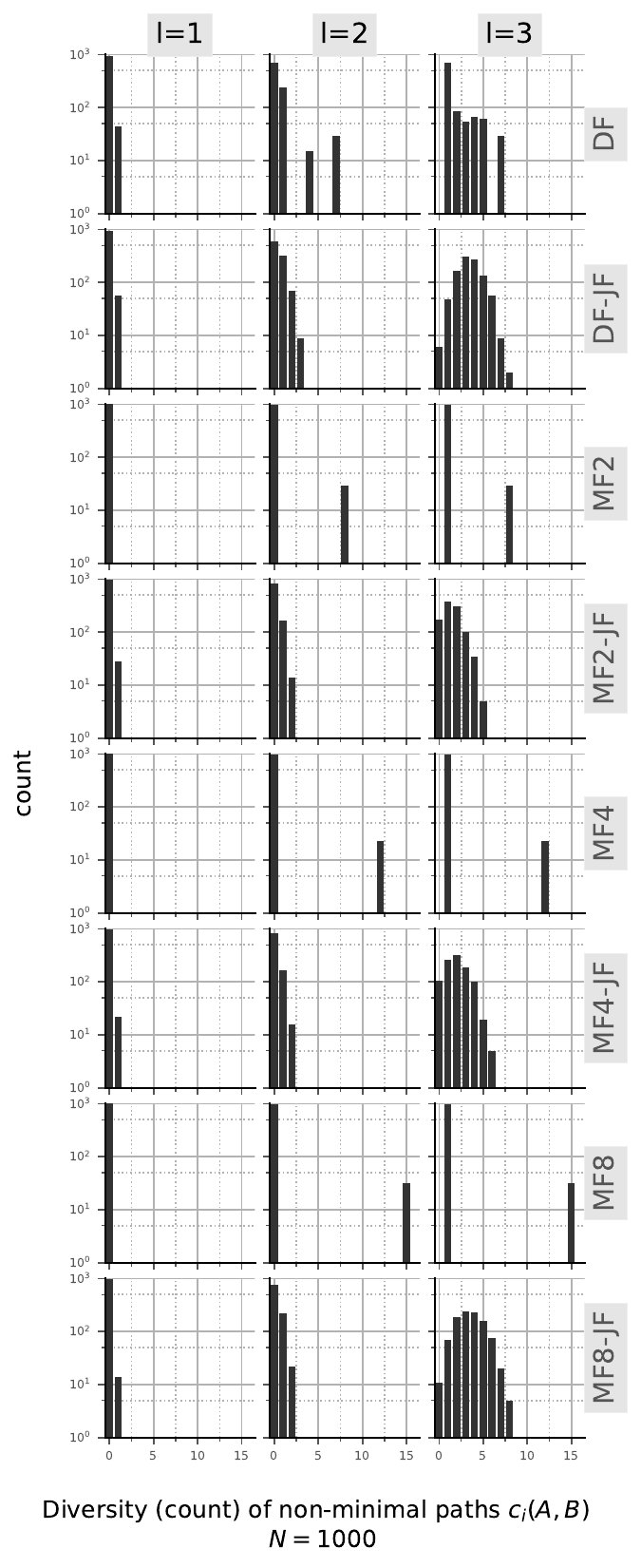}
\caption{Edge disjoint path plots for Dragonfly and Megafly networks with different values
of g. All Megafly networks have a low path diversity of either 0 or 1 for most node pairs and
all lengths. Furthermore, all have a second peak for a very high diversity count. Depending on
the value of g, this peak is between 7 and 15. Although Megafly has node
pairs with a higher diversity count, Dragonfly has a well-distributed diversity count between 1
and 5. It is difficult to compare the values with this plot. The diversity counts of all Jellyfish
equilibriums show a Gauss distribution for $l$ = 3.}
\label{fig:appendix_027}
\end{figure*}

\begin{figure*}
\centering
\includegraphics[width=1.0\textwidth]{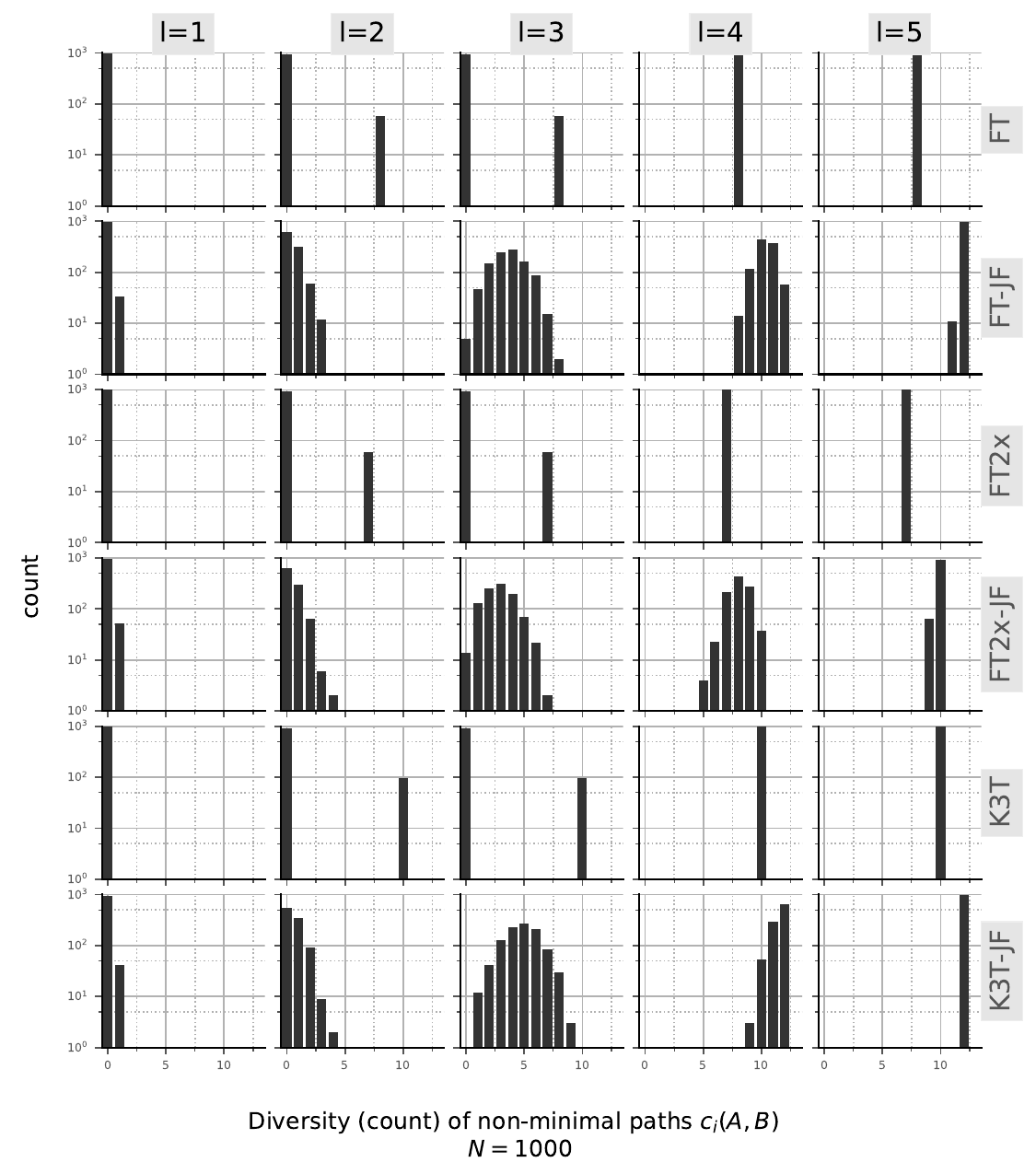}
\caption{Edge disjoint path plots for the Fat Trees and k-ary 3-tree. All the topologies have
only specific values for the path diversity. None of the topologies have any node pairs that have
a diversity count of zero for $l \geq$ 4. The k-ary 3-tree has the best path diversity among all other
topologies.}
\label{fig:appendix_028}
\end{figure*}

\begin{figure*}
\centering
\includegraphics[width=1.0\textwidth]{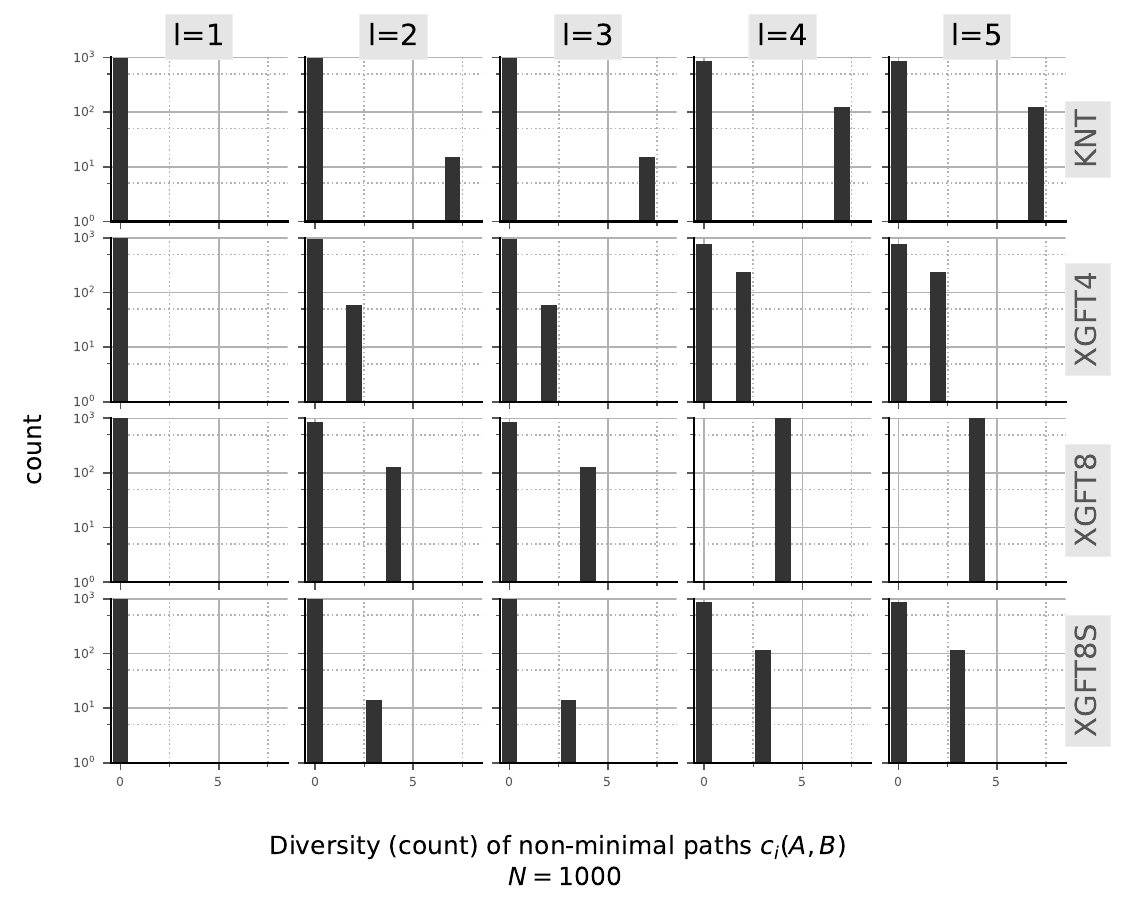}
\caption{Edge disjoint path plots for the xGFTs of different variants.
There exist only edge disjoint paths of specific values. For $l \geq$ 4, xGFT8 does not have
any edge pairs with a diversity count of zero anymore. The k-ary n-tree has edge nodes with a higher
diversity count than any other topology, but it has still node pairs with a diversity count of zero
for $l$ = 5.}
\label{fig:appendix_029}
\end{figure*}

\begin{figure*}
\centering
\includegraphics[width=1.0\textwidth]{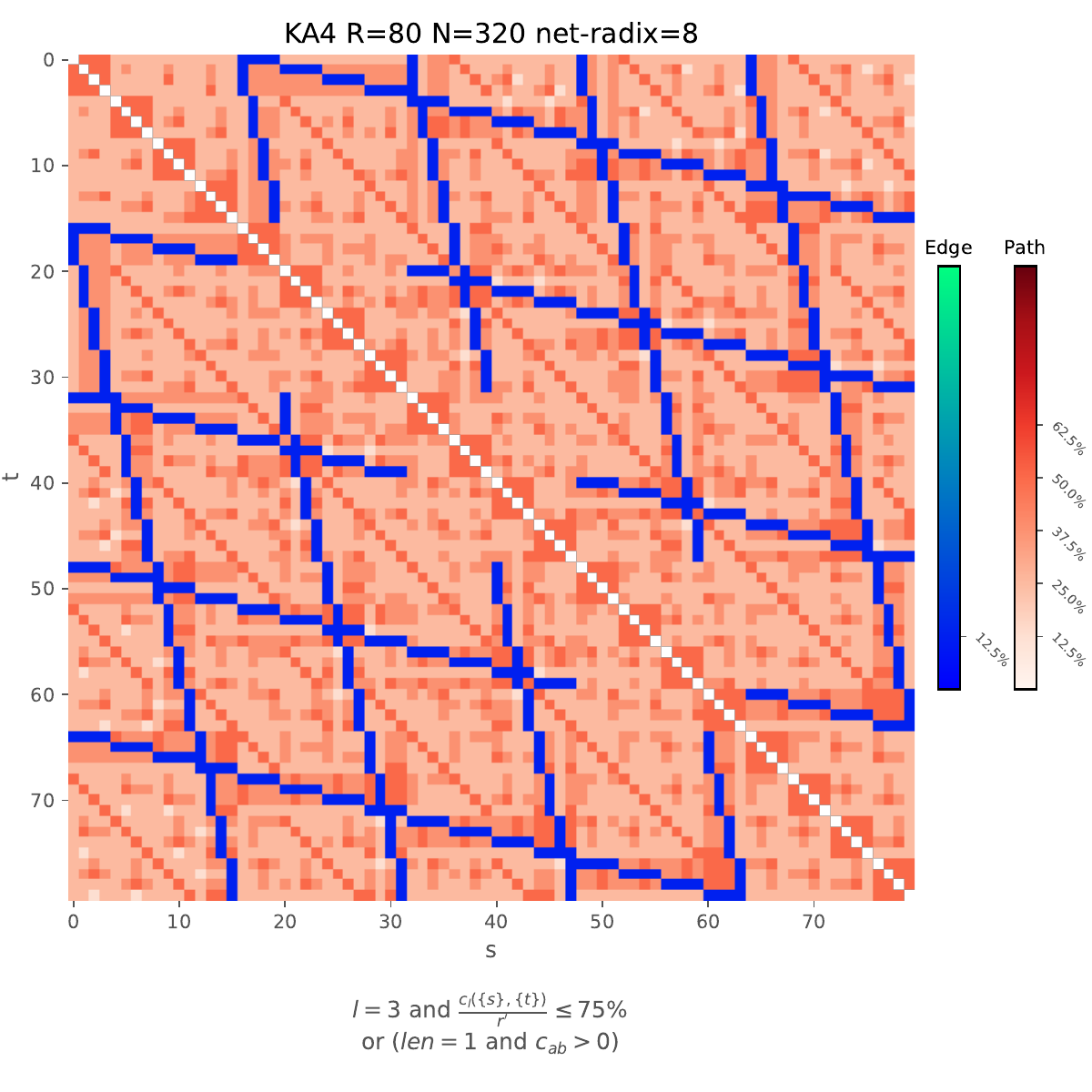}
\caption{Low connectivity plot for $K_{4,3}$ Kautz graph of length $l$ = 3. All router pairs have a
connectivity $\leq$ 75\% for $l$ = 3. We can see the segments (blue) of length 4 connecting a node to
its neighbours. This can be seen horizontally as well as vertically. Due to this mesh of edges, all
nodes are connected and there exist only a couple of node pairs with connectivity $<$ 25\%.}
\label{fig:appendix_030}
\end{figure*}

\begin{figure*}
\centering
\includegraphics[width=1.0\textwidth]{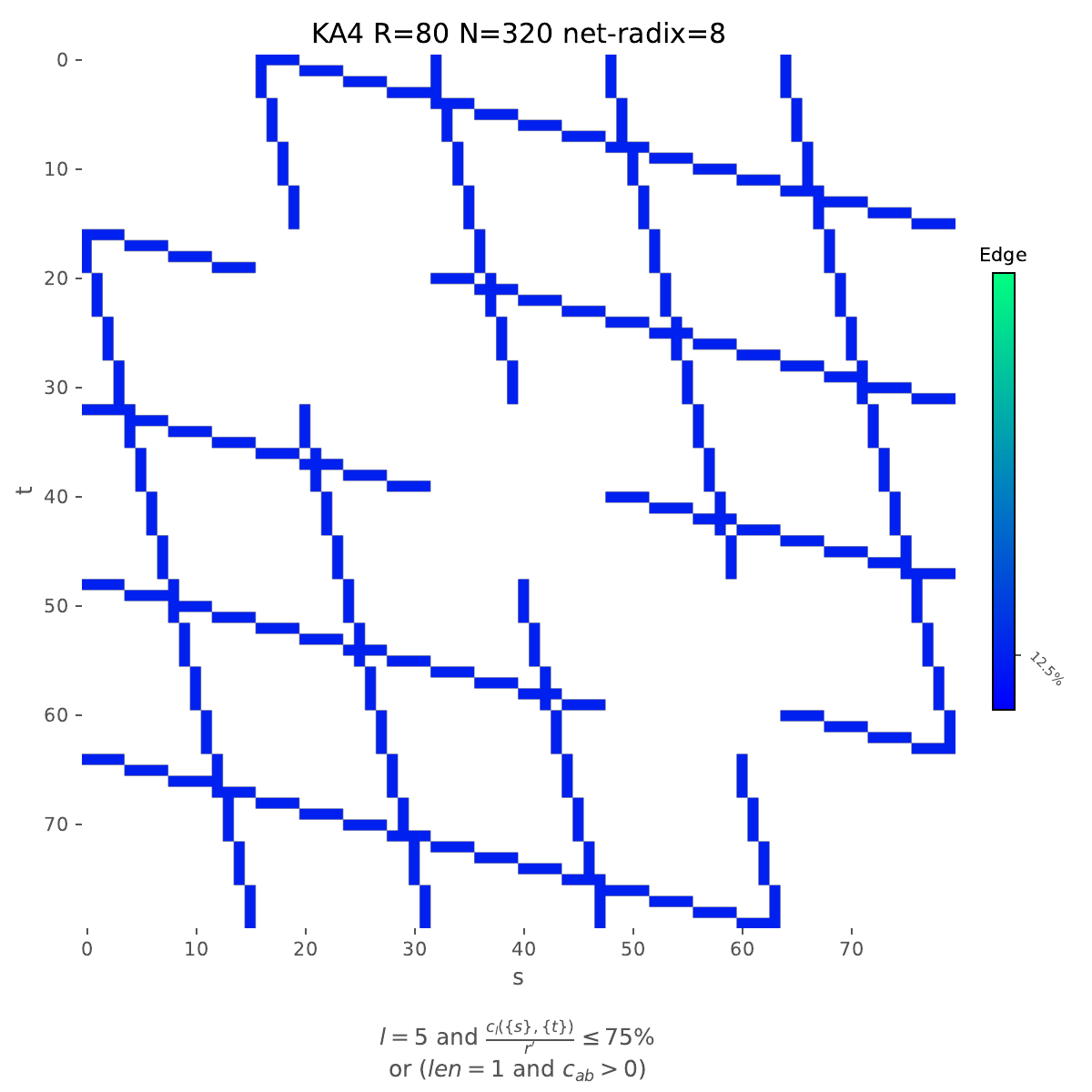}
\caption{Low connectivity plot for $K_{4,3}$ Kautz graph of length $l$ = 5. All router pairs have a
connectivity $\geq$ 75\% for $l$ = 5.}
\label{fig:appendix_031}
\end{figure*}

\begin{figure*}
\centering
\includegraphics[width=1.0\textwidth]{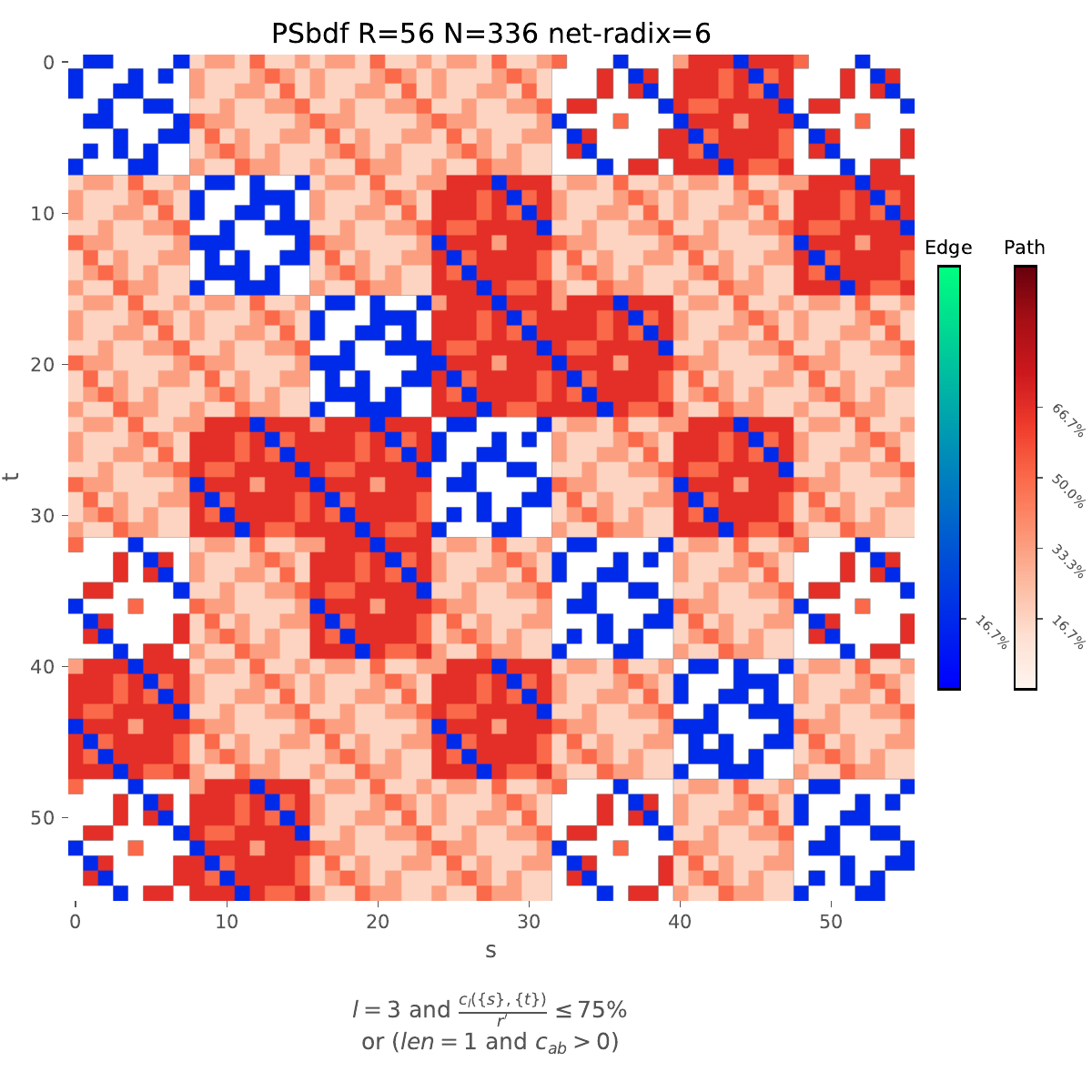}
\caption{Low connectivity plot for the PolarStar graph of length $l$ = 3. The structure
graph is an ER$_2$ graph and the subgraph is a BDF$_3$ graph. We can see the 7 well-connected
BDF$_3$ clusters of each 8 nodes in the diagonal. The other well-connected areas exist due to the
connection of the ER$_2$ graph. Every node in 0 - 7 is connected to a node in 31 - 39, 40 - 47 and
48 - 55.}
\label{fig:appendix_032}
\end{figure*}

\begin{figure*}
\centering
\includegraphics[width=1.0\textwidth]{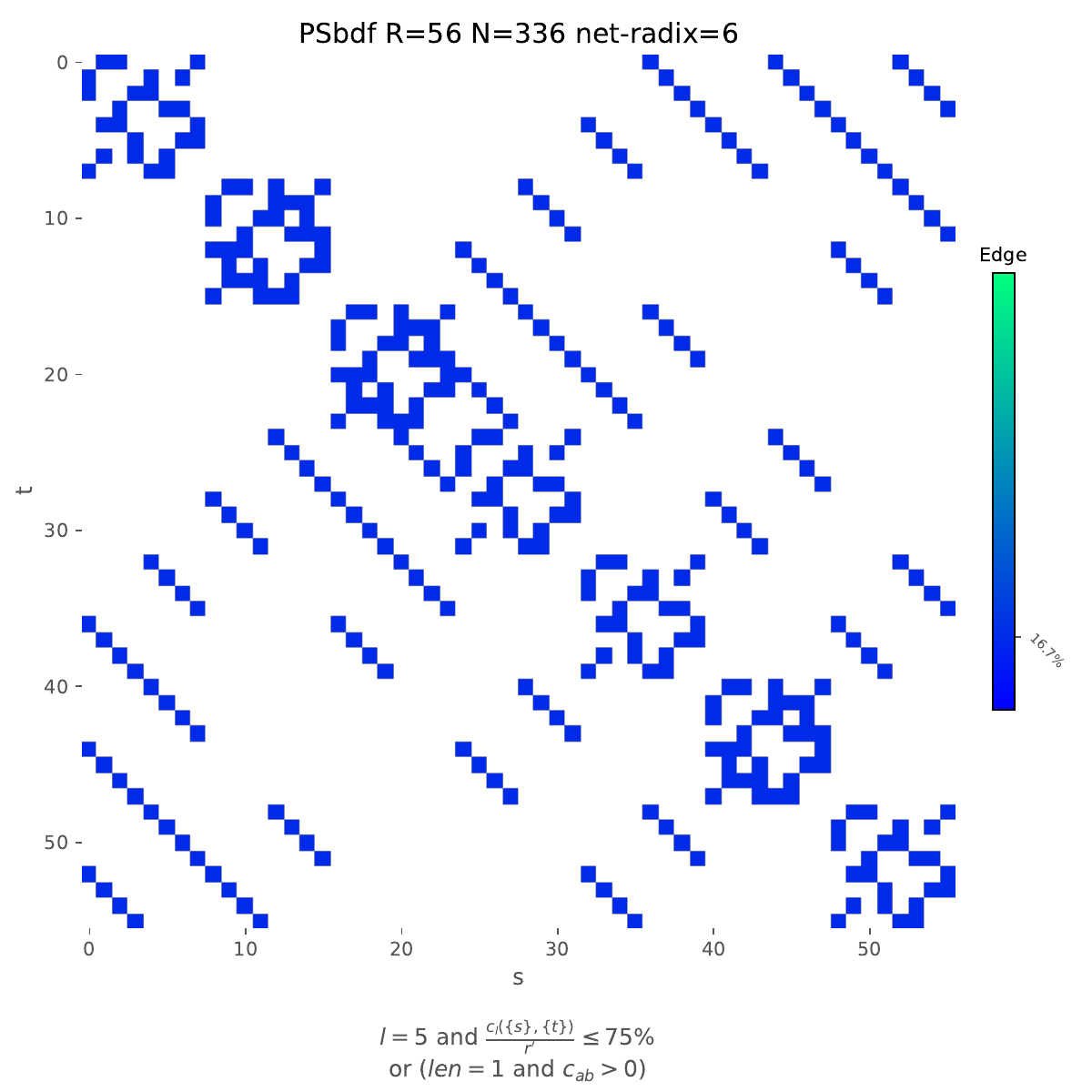}
\caption{Low connectivity plot for the PolarStar graph for the length $l$ = 5. The structure
graph is an ER$_2$ graph and the subgraph is a BDF$_3$ graph. We can clearly see, that for $l$ = 5 this
PolarStar graph has a good connectivity $\geq$ 75\% for all router pairs.}
\label{fig:appendix_033}
\end{figure*}

\begin{figure*}
\centering
\includegraphics[width=1.0\textwidth]{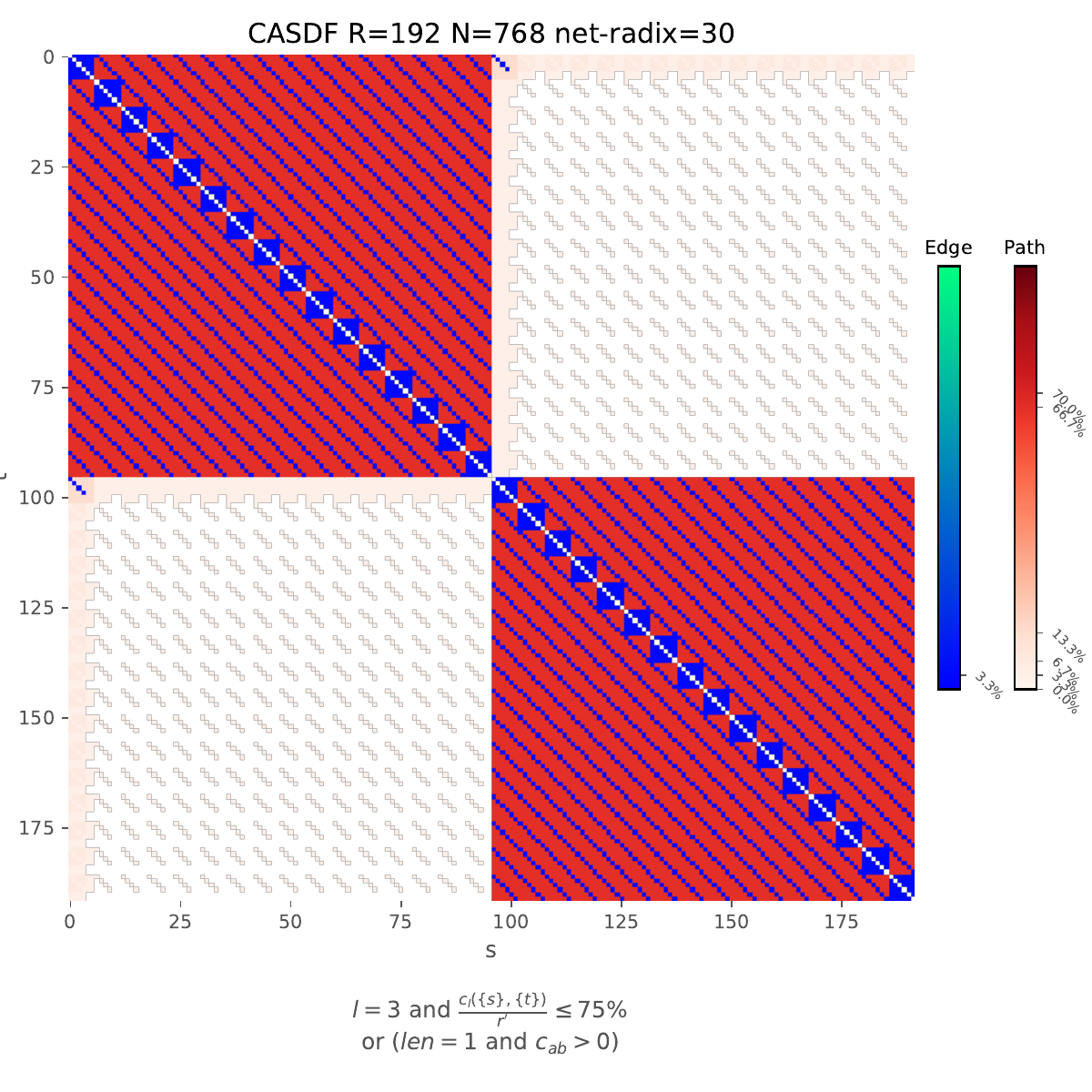}
\caption{Low connectivity plot for the Cascade Dragonfly topology with two clusters for the
length $l$ = 3. Each cluster has a good connectivity within itself. On the diagonal, we can also see
the 16 chassis for the 5 levels. The connectivity between the two clusters is very low and only
about half of the nodes in each cluster have a path to nodes in the other cluster.}
\label{fig:appendix_034}
\end{figure*}

\begin{figure*}
\centering
\includegraphics[width=1.0\textwidth]{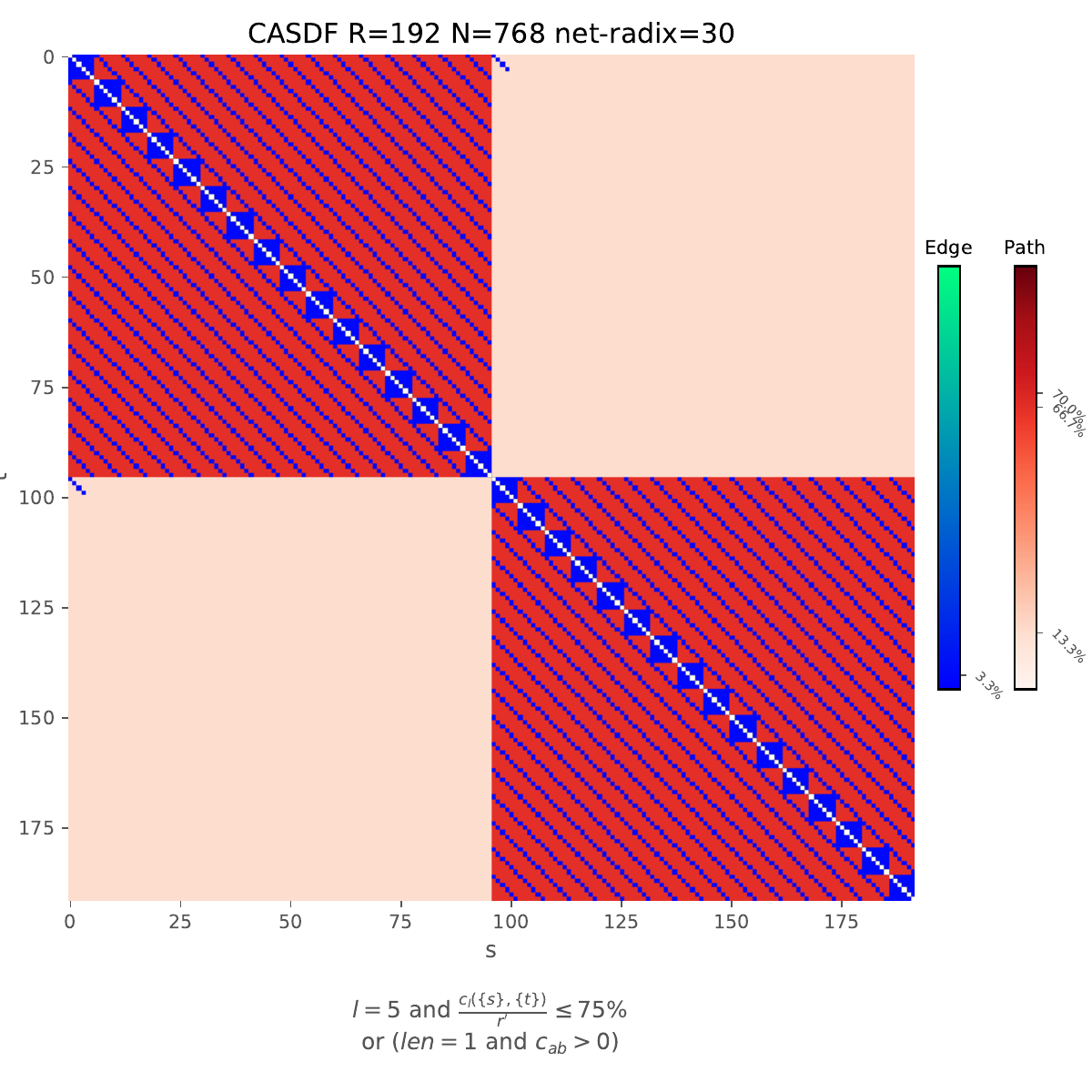}
\caption{Low connectivity plot for the Cascade Dragonfly topology with two clusters for the
length $l$ = 5. Right underneath the first cluster, we can see the four node pairs that connect the
two clusters. Even for paths of $l$ = 5, the connectivity between the two clusters stays low at only
10\%.}
\label{fig:appendix_035}
\end{figure*}

\begin{figure*}
\centering
\includegraphics[width=1.0\textwidth]{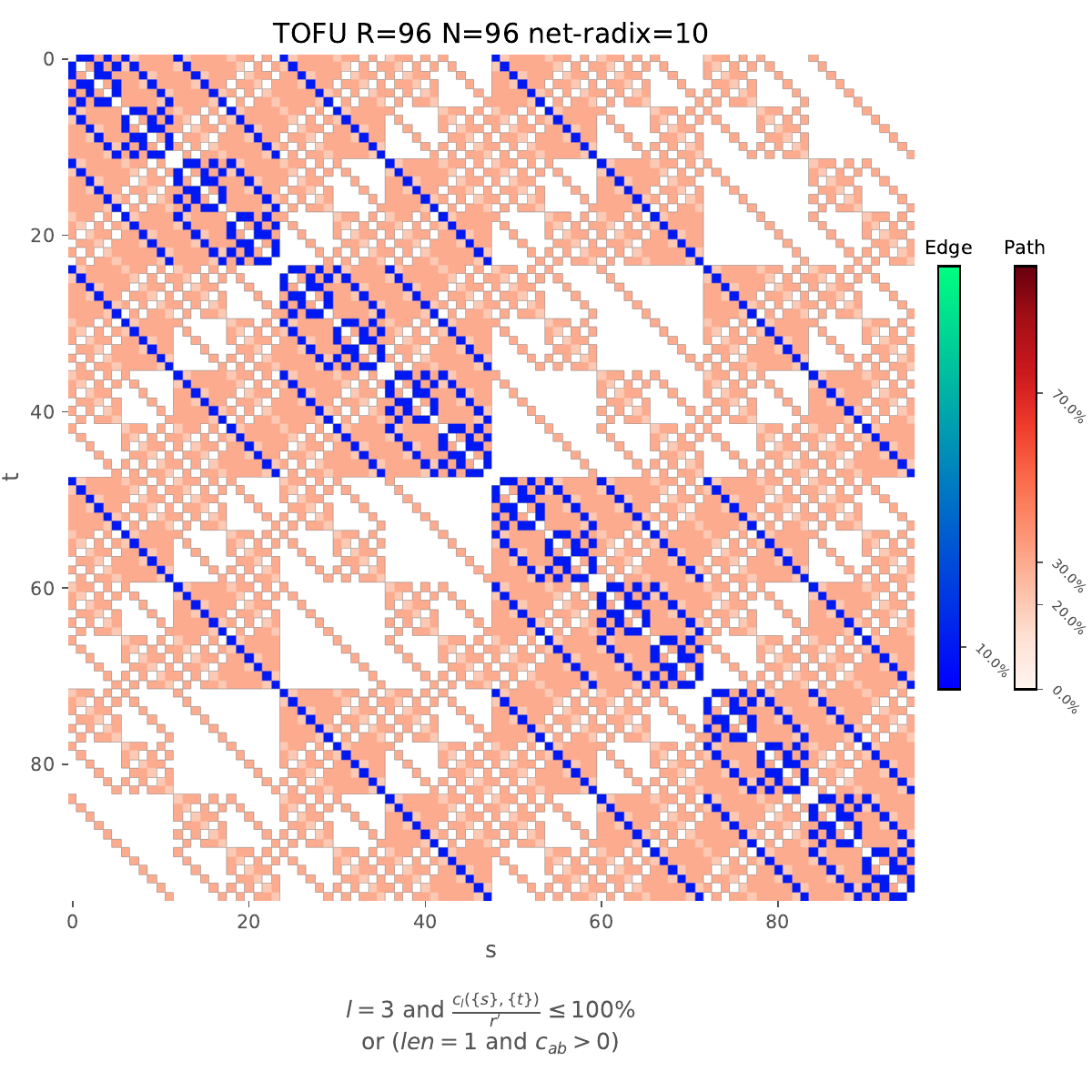}
\caption{Low connectivity plot for the Tofu topology with X = Y = Z = 2 and $l$ = 3. Not all
nodes are connected with each other at length $l$ = 3. The 8 clusters of 12 nodes each are visible
in the diagonal. The small blue diagonal lines show the edges between the nodes in different
clusters. Each row is intersected by 3 of these lines. This is the number of connections each
node has to other clusters.}
\label{fig:appendix_036}
\end{figure*}

\begin{figure*}
\centering
\includegraphics[width=1.0\textwidth]{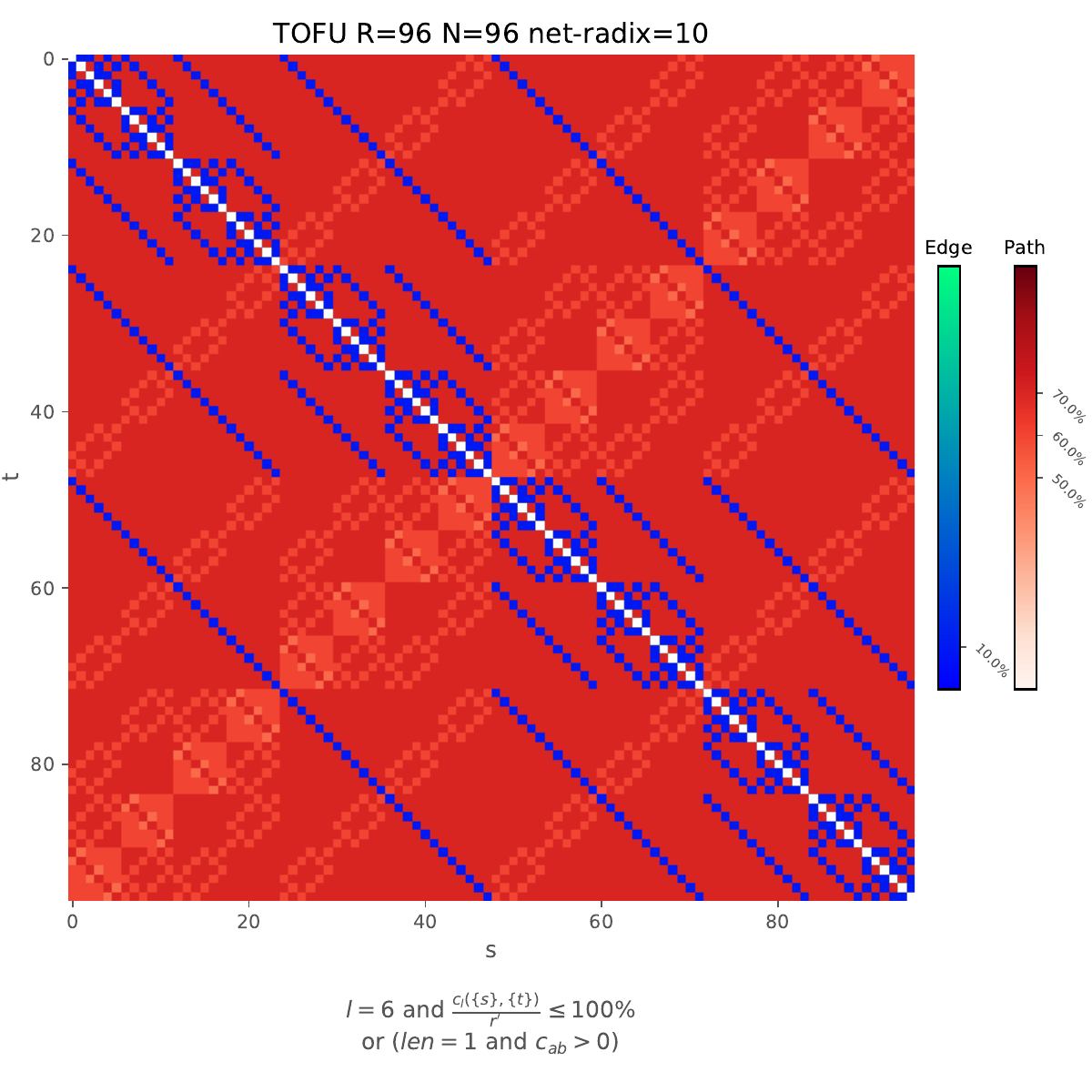}
\caption{Low connectivity plot for the Tofu topology with X = Y = Z = 2 and length $l$ = 6.
There is a good connectivity $\geq$ 70\% for most node pairs.}
\label{fig:appendix_037}
\end{figure*}

\begin{figure*}
\centering
\includegraphics[width=0.63\textwidth]{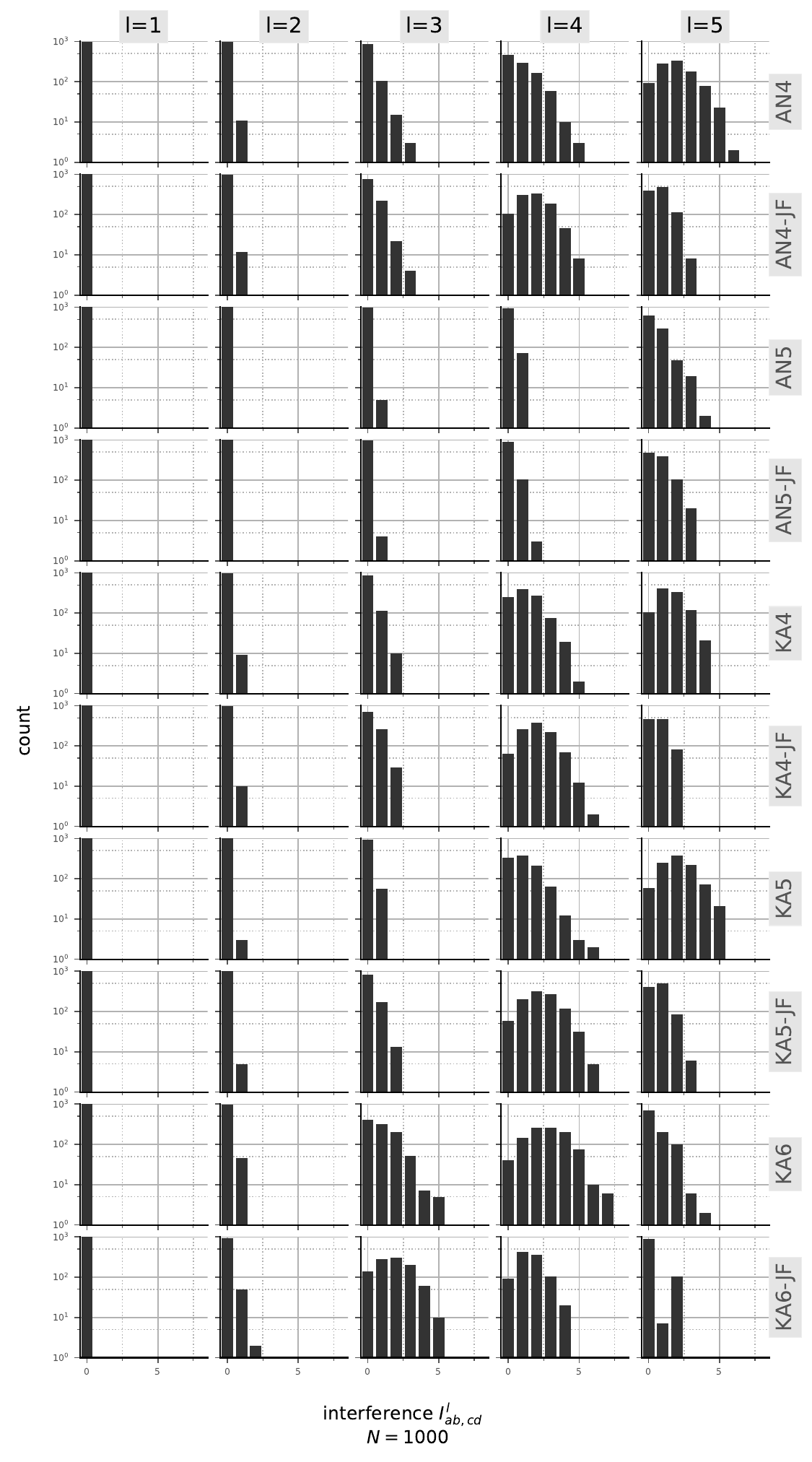}
\caption{Interference plots for Kautz and Arrangement graphs of length $l$ = 5. The
Arrangement networks have the same or less interference than their Jellyfish equivalent counterpart
for $l \leq$ 5. For $l$ = 5, both have slightly more interference. The same can be seen for $K_{4,n}$ and
$K_{5,n}$ Kautz graphs, but not for $K_{6,n}$, which has the most interference for $l$ = 3 and $l$ = 4, but
the least interference for $l$ = 5.}
\label{fig:appendix_038}
\end{figure*}

\begin{figure*}
\centering
\includegraphics[width=0.63\textwidth]{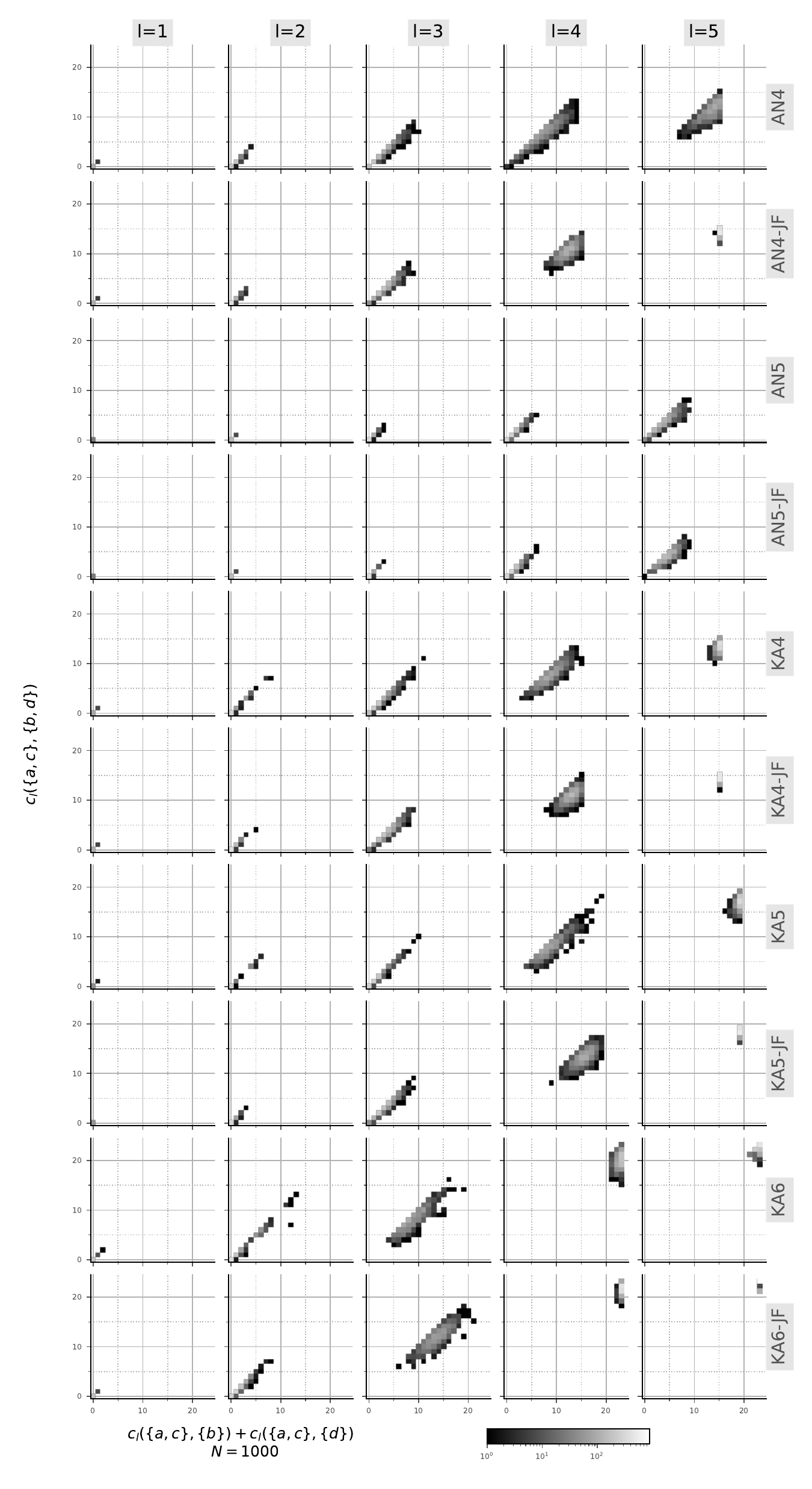}
\caption{Two-dimensional detailed interference plots for Kautz and Arrangement graphs of
length $l$ up to 5. $A_{n,4}$ has more interference than $A_{n,5}$. The different values of k for $K_{b,n}$ Kautz
graphs have less of an impact on the amount of interference, but at which length $l$ they will
occur. Most of these networks have the most interference at $l$ = 4. Their equivalent Jellyfish
have similar interference analysis, but in general a bit less.}
\label{fig:appendix_039}
\end{figure*}

\begin{figure*}
\centering
\includegraphics[width=1.0\textwidth]{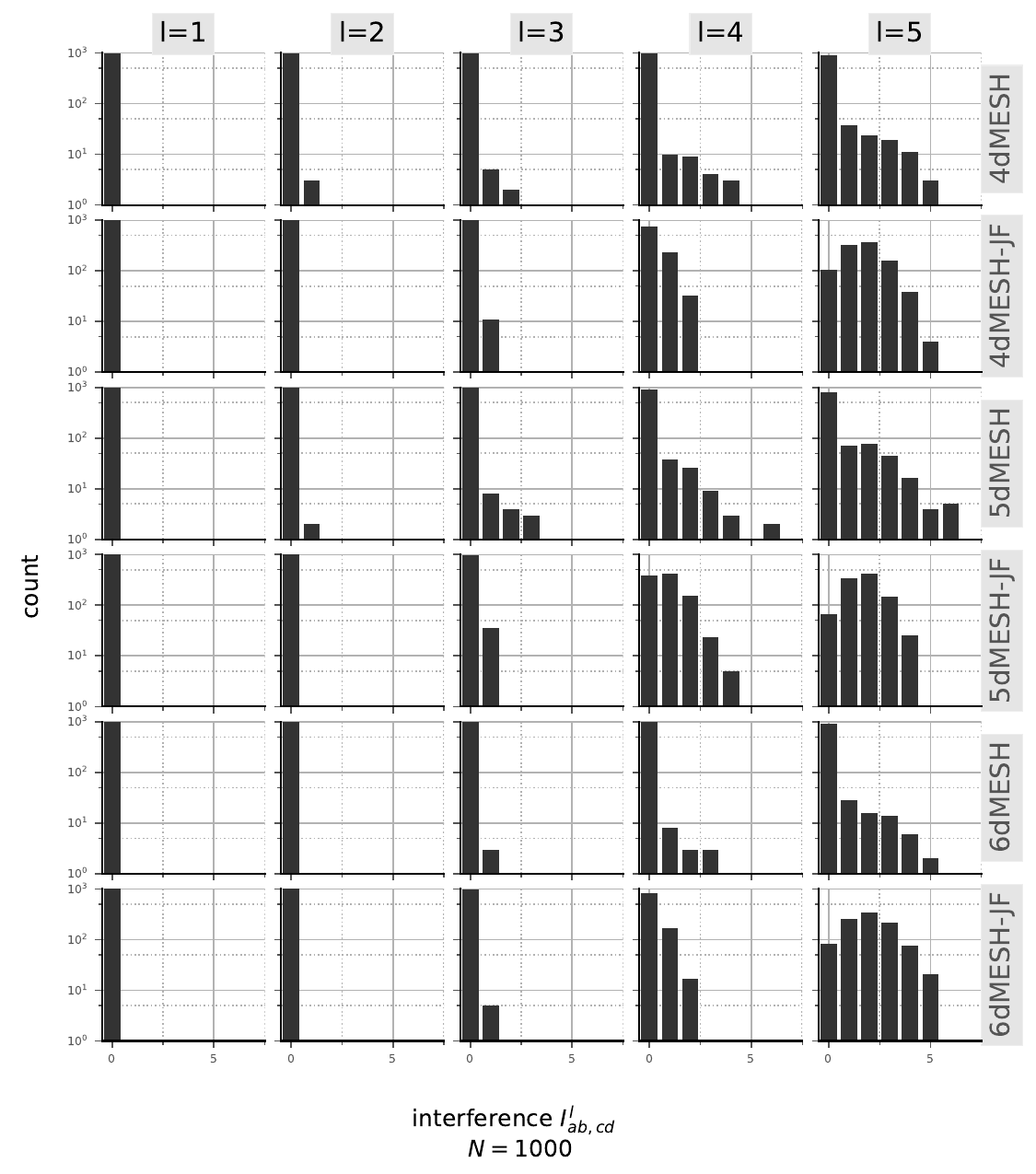}
\caption{Interference plots for meshes of different dimensions of lengths up to $l$ = 5. While
4D and 6D meshes have a pretty similar interference analysis, the 5D mesh has more interference.
Furthermore, for $l \leq$ 4, the base networks of all the meshes have higher or equal interference
than their equivalent Jellyfish. 4D and 6D meshes have better interference for $l$ = 5, which
is not true for the 5D mesh. For $l$ = 5, all meshes have more node pairs with zero interference
compared to their equivalent Jellyfish graphs.}
\label{fig:appendix_040}
\end{figure*}

\begin{figure*}
\centering
\includegraphics[width=0.99\textwidth]{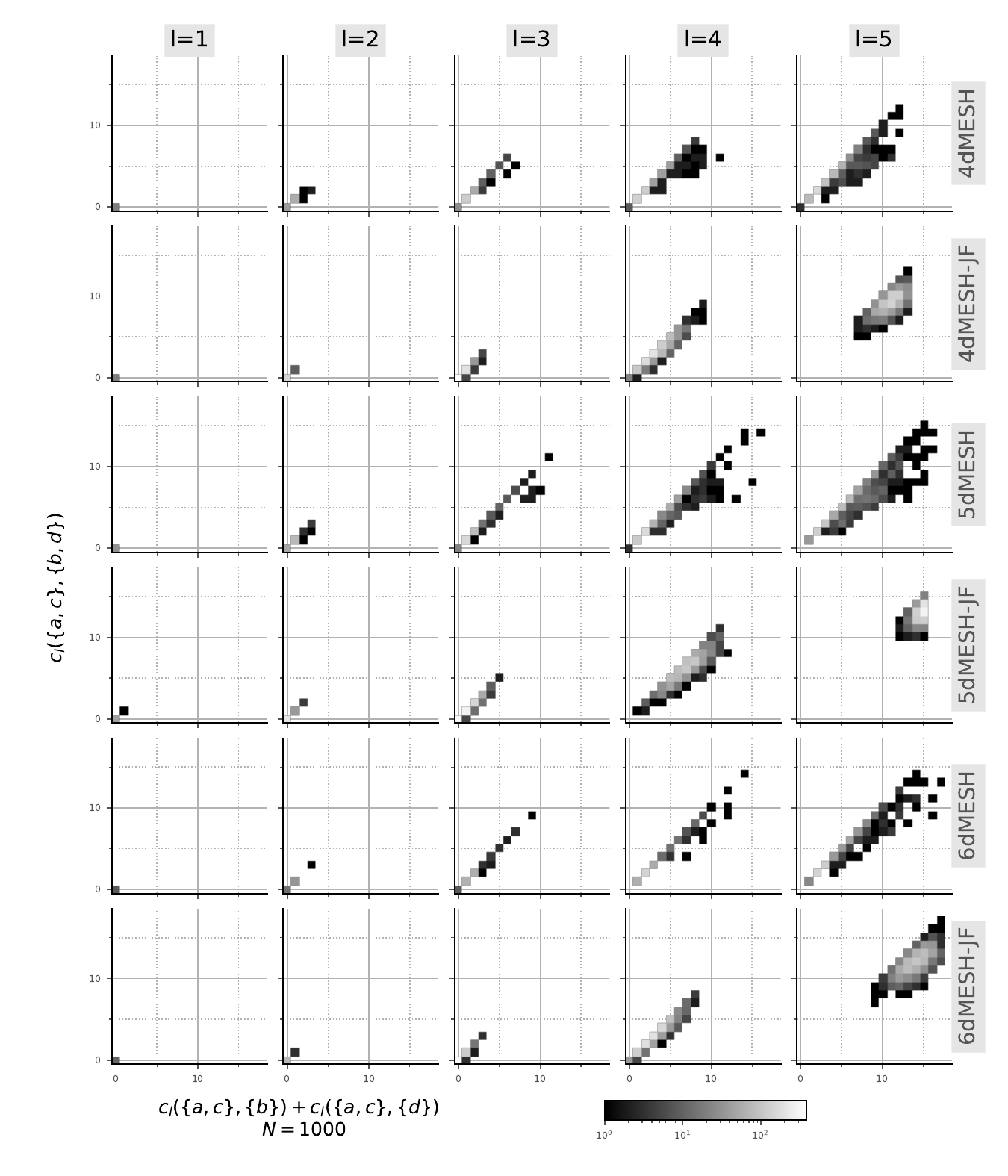}
\caption{Two-dimensional detailed interference plots for meshes of different dimensions of
lengths up to $l$ = 5. We can see, that all meshes have similar interferences analysis. For the base
mesh networks and $l$ = 5, all components are affected by interference, while for the equivalent
Jellyfish, only a subset of them are.}
\label{fig:appendix_041}
\end{figure*}

\begin{figure*}
\centering
\includegraphics[width=1.0\textwidth]{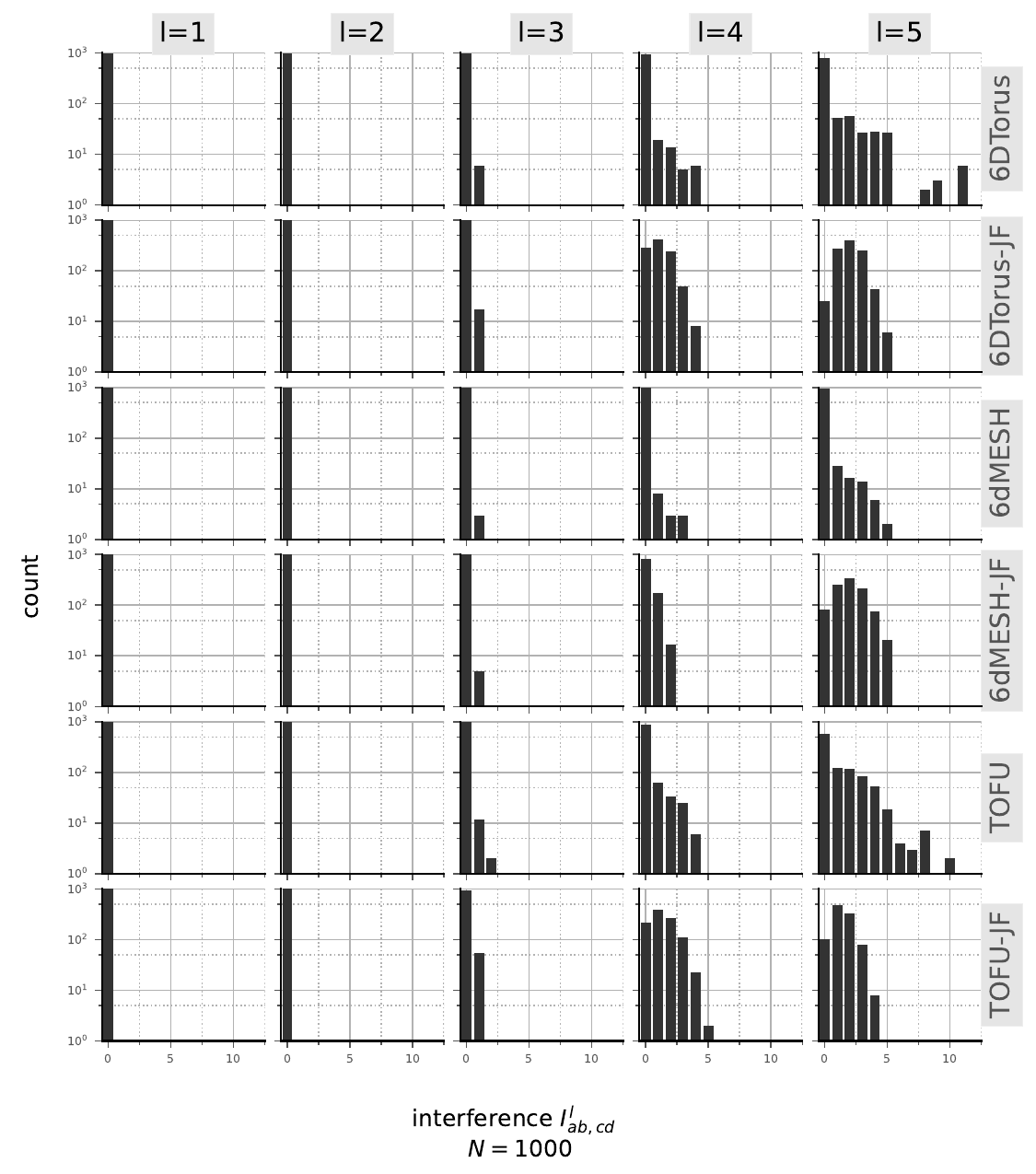}
\caption{Interference plots for Tofu, 6D mesh and 6D torus with lengths up to $l$ = 5. The Tofu
topology has about the same amount of interference as the 6D torus network, but more than the
regular 6D mesh. For $l$ = 5, both torus and the Tofu graph have more interference than the 6D
mesh and more than any of the Jellyfish equivalents.}
\label{fig:appendix_044}
\end{figure*}

\begin{figure*}
\centering
\includegraphics[width=0.45\textwidth]{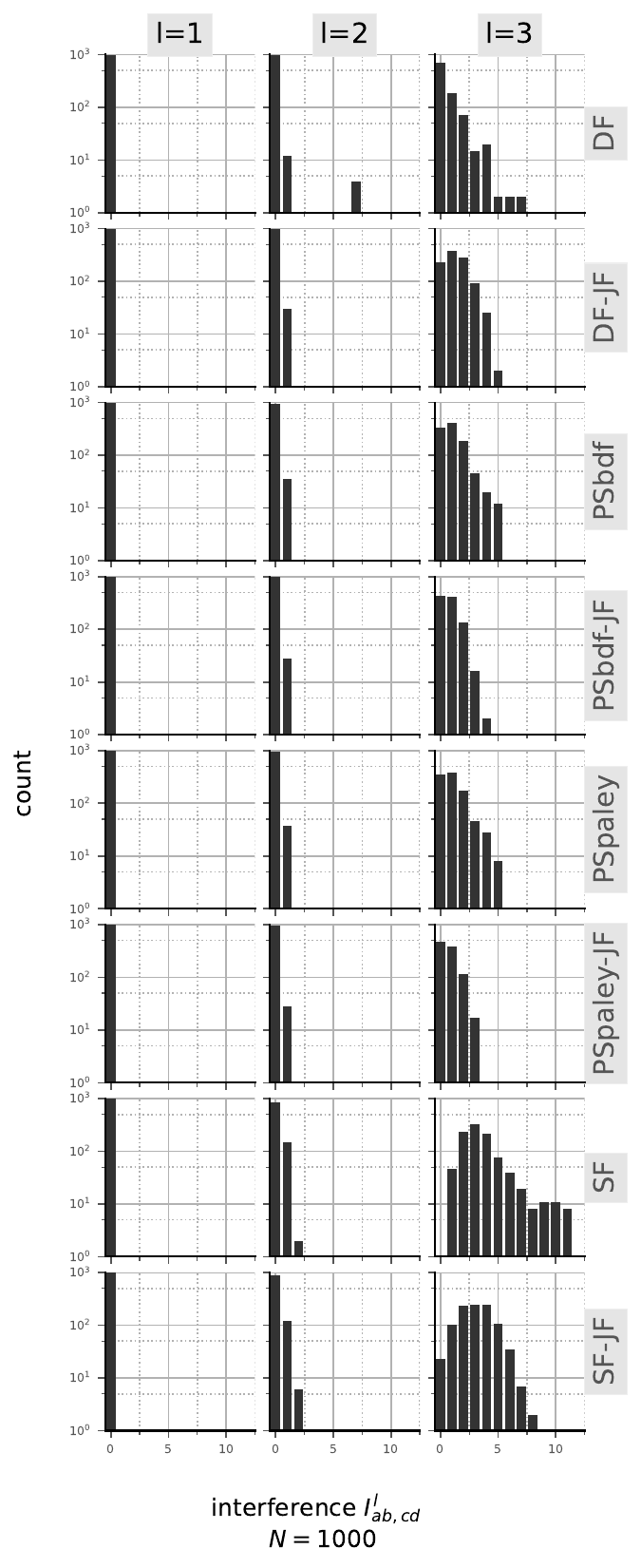}
\caption{Interference plots for PolarStar topology with both subgraphs, Dragonfly and Slim Fly
of length $l$ = 1 up to $l$ = 5. Slim Fly has the most interference at $l$ = 3, but less interference
for all other values of $l$. The other topologies have the most interference for $l$ = 4. The
interference analysis looks similar for the topologies when comparing them to their equivalent
Jellyfish, although all of the base topologies have slightly more interference.}
\label{fig:appendix_045}
\end{figure*}

\begin{figure*}
\centering
\includegraphics[width=0.5\textwidth]{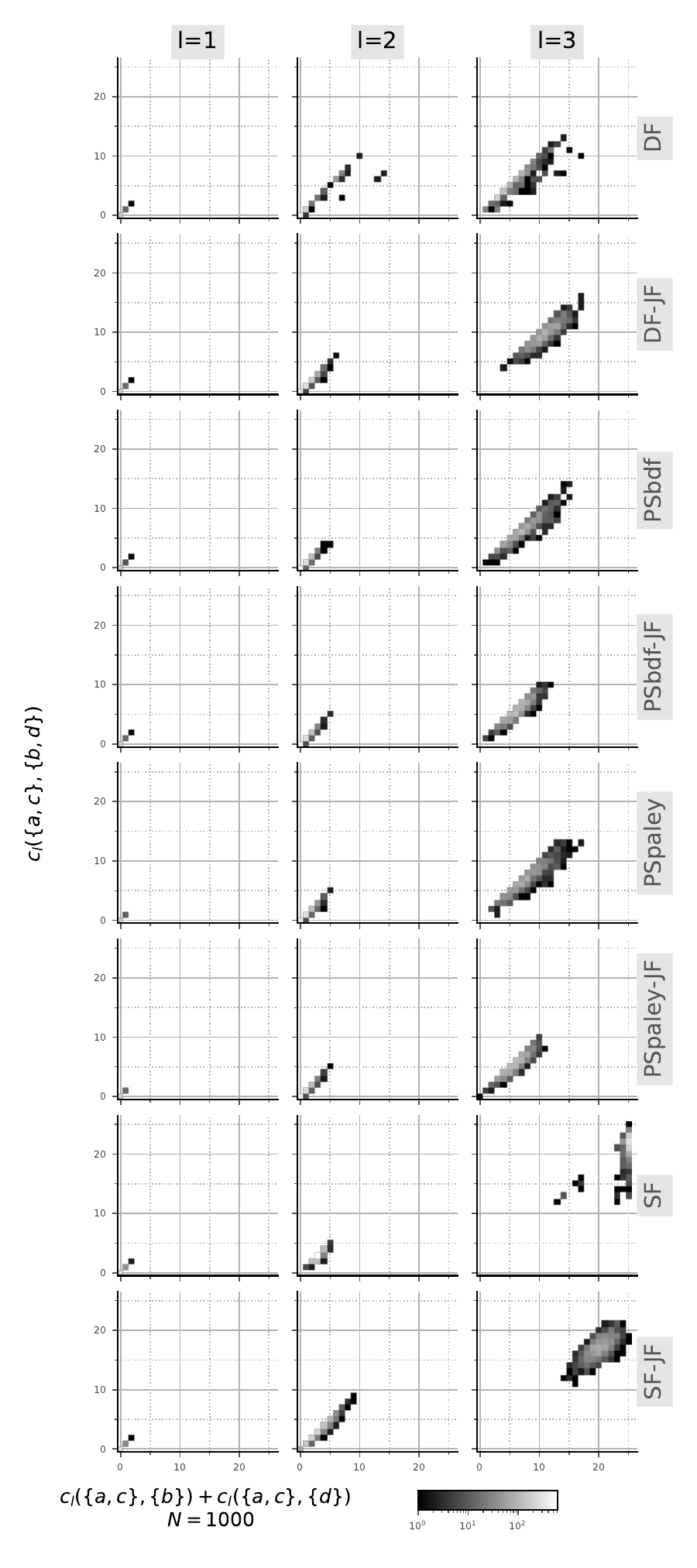}
\caption{Two-dimensional detailed interference plots for PolarStar topology with both
subgraphs, Dragonfly and Slim Fly of length $l \leq$ 5. The detailed analysis of both PolarStar graphs
are fairly similar. While for Dragonfly, the interference for $l$ = 4 is contributed by a lot of
components, only a small amount of components contribute to the high amount of interference
for PolarStar.}
\label{fig:appendix_046}
\end{figure*}

\begin{figure*}
\centering
\includegraphics[width=0.75\textwidth]{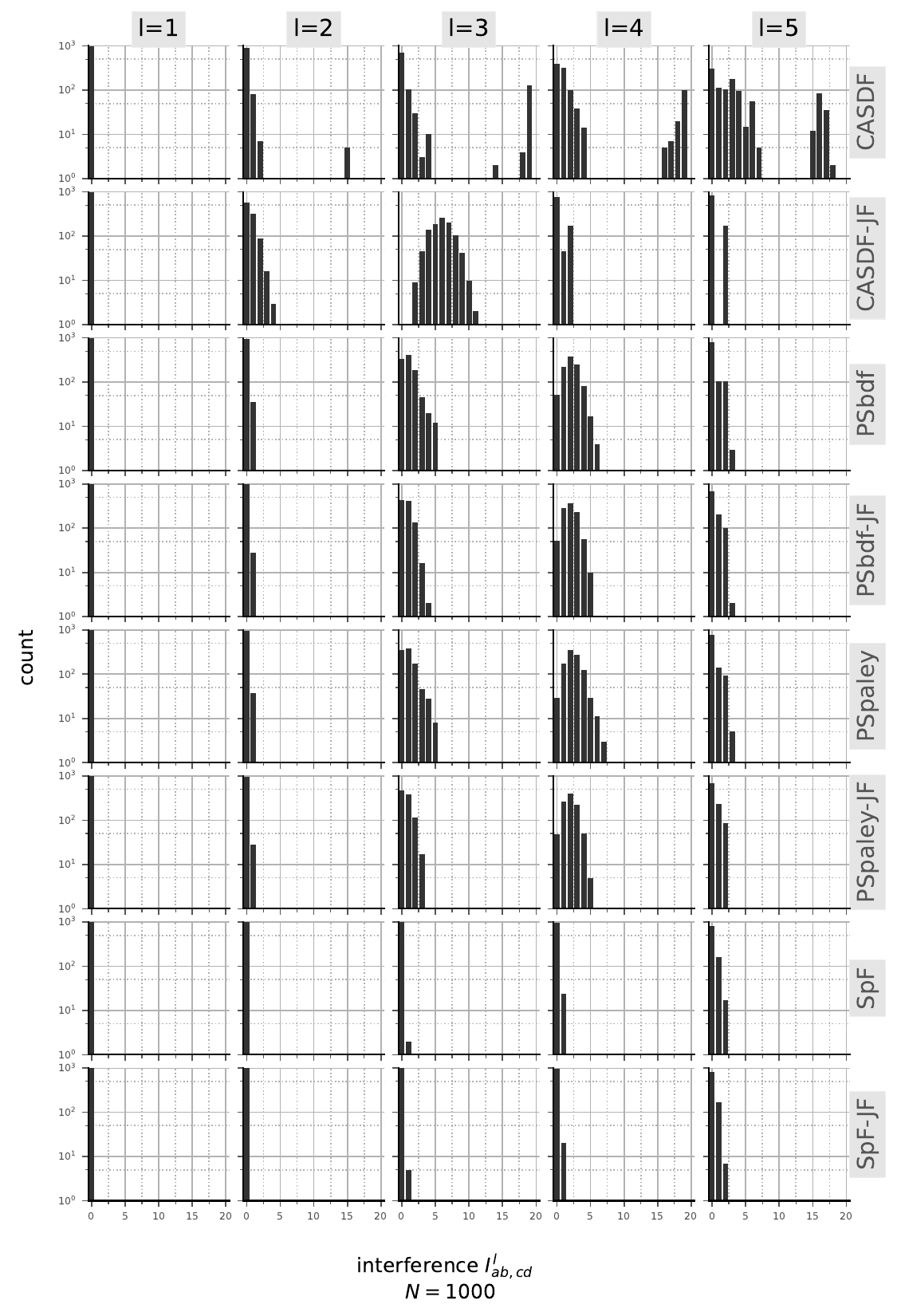}
\caption{Interference plots for Cascade Dragonfly, Spectralfly and PolarStar with both
subgraphs of length $l$ = 5. Spectralfly has little interference compared to the other topologies for
all lengths. It has even about the same amount of interference as its equivalent Jellyfish topology.
In contrast to all the other topologies, Cascade Dragonfly has two groups of node quadruples:
the ones with little interference and the other group with high interference. Due to the second
group of nodes with high interference, Cascade Dragonfly has the most interference of all the
topologies. The most extreme difference when comparing to its equivalent Jellyfish, can be seen
for $l$ = 4 and $l$ = 5. While the Jellyfish graph has at most interference 3 for any node quadruple,
there are groups of nodes with interference values $>$ 15.}
\label{fig:appendix_047}
\end{figure*}

\begin{figure*}
\centering
\includegraphics[width=0.79\textwidth]{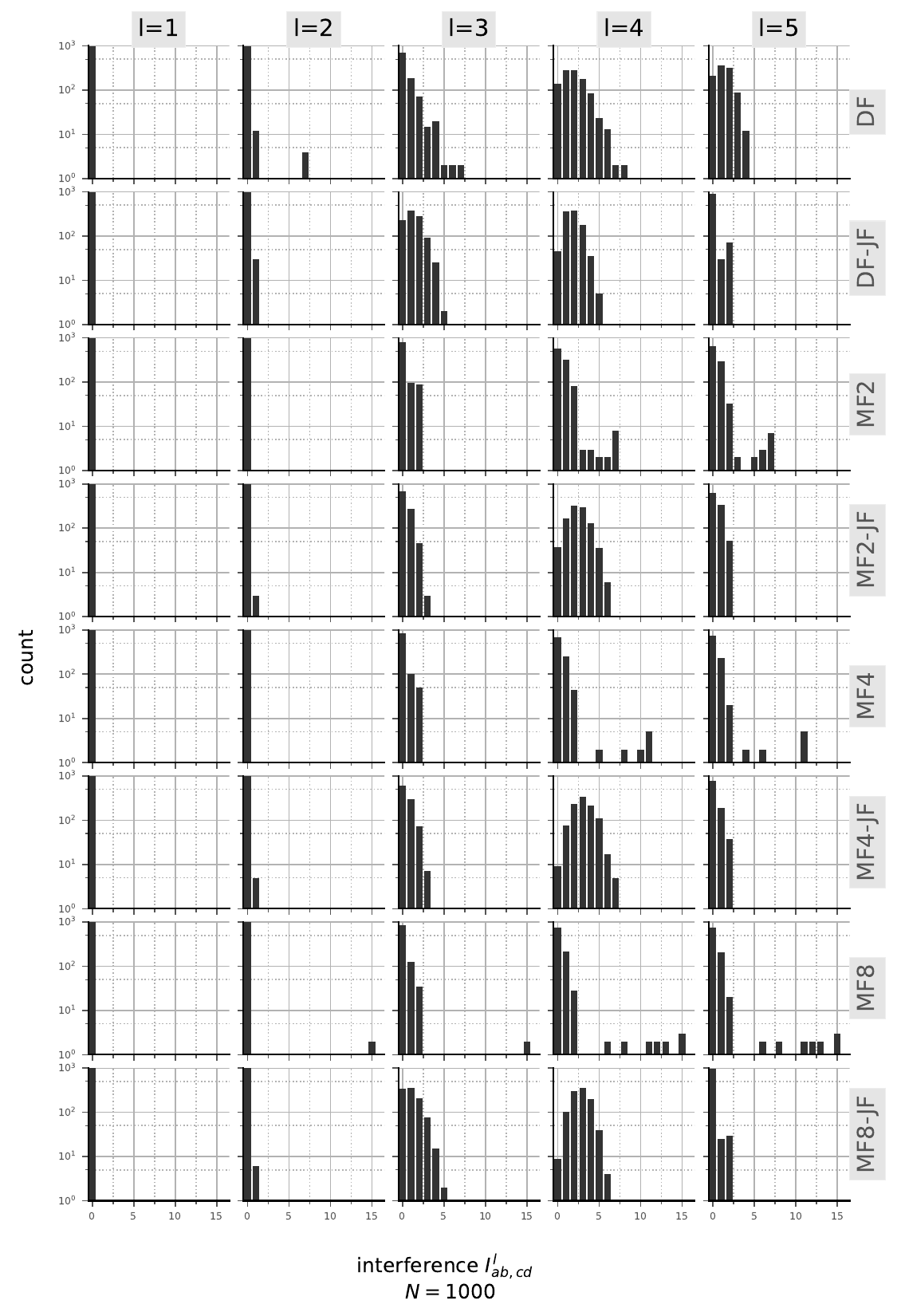}
\caption{Interference plots for Megafly with different gap sizes and Dragonfly of length $\leq$ 5.
In general, Megafly has less interference than Dragonfly for length $l \leq$ 3. Only MF8 has a few
outlying node pairs with a high interference at $l$ = 3. All Megafly variants have most node pairs
with interference between 0 and 3. For $l \geq$ 4, all variants start to have a few $\approx$ 10 outliers with
higher interference. This cannot be seen in the Dragonfly topology.}
\label{fig:appendix_048}
\end{figure*}

\begin{figure*}
\centering
\includegraphics[width=0.7\textwidth]{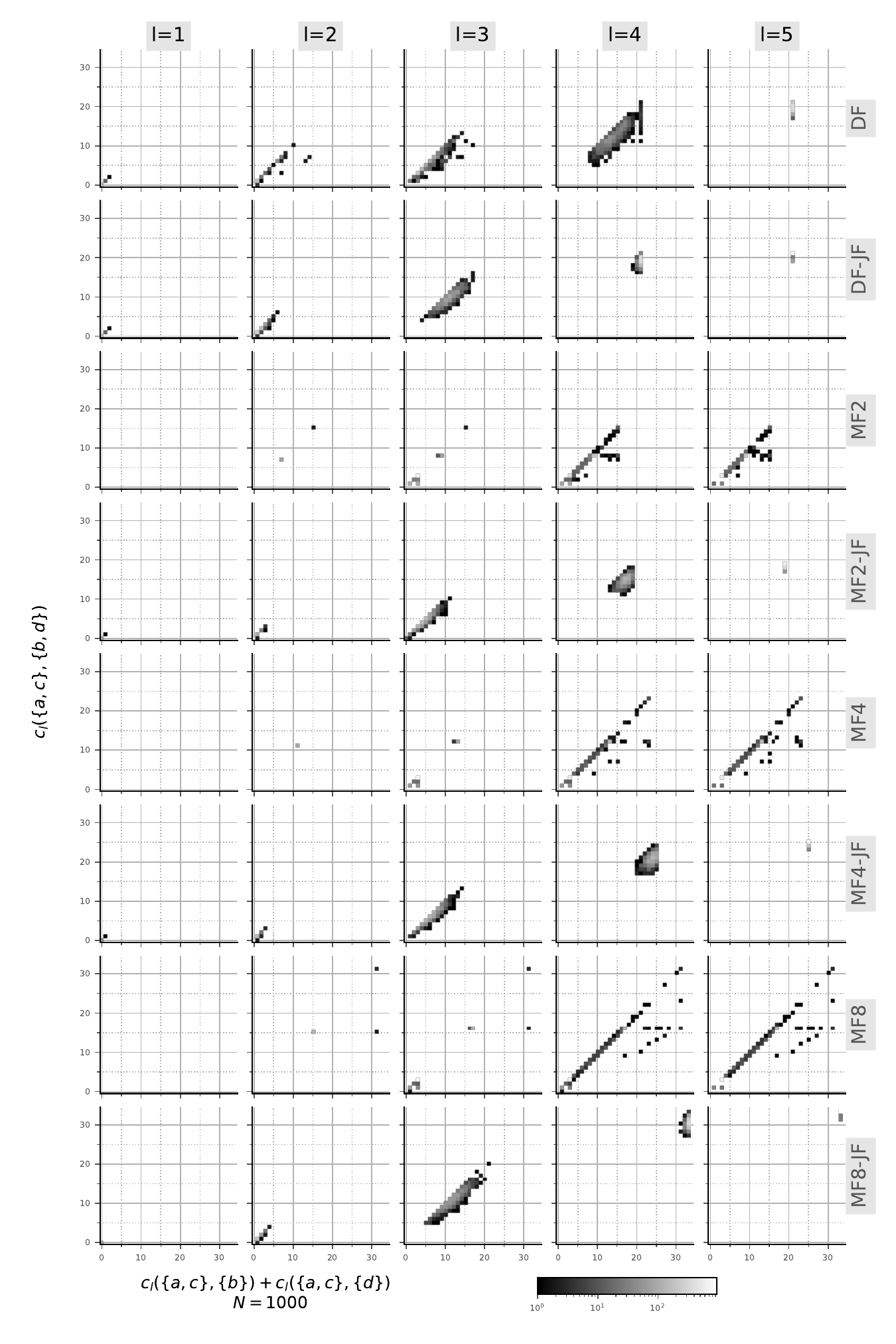}
\caption{Two-dimensional detailed interference plots for Megafly with different gap sizes
and Dragonfly. For lengths of 2 and 3, only a few components contribute to the total interference
in the Megafly networks. This changes for $l >$ 3, where almost all values for $c_l(\{a, c\}, \{b\}) +c_l(\{a, c\}, \{d\})$
are possible. The same cannot be said for the analysis of the Dragonfly graph. For
all lengths only a subset of possible values for the components contribute the total interference.
This is also true for all Jellyfish equivalent graphs.}
\label{fig:appendix_049}
\end{figure*}

\begin{figure*}
\centering
\includegraphics[width=1.0\textwidth]{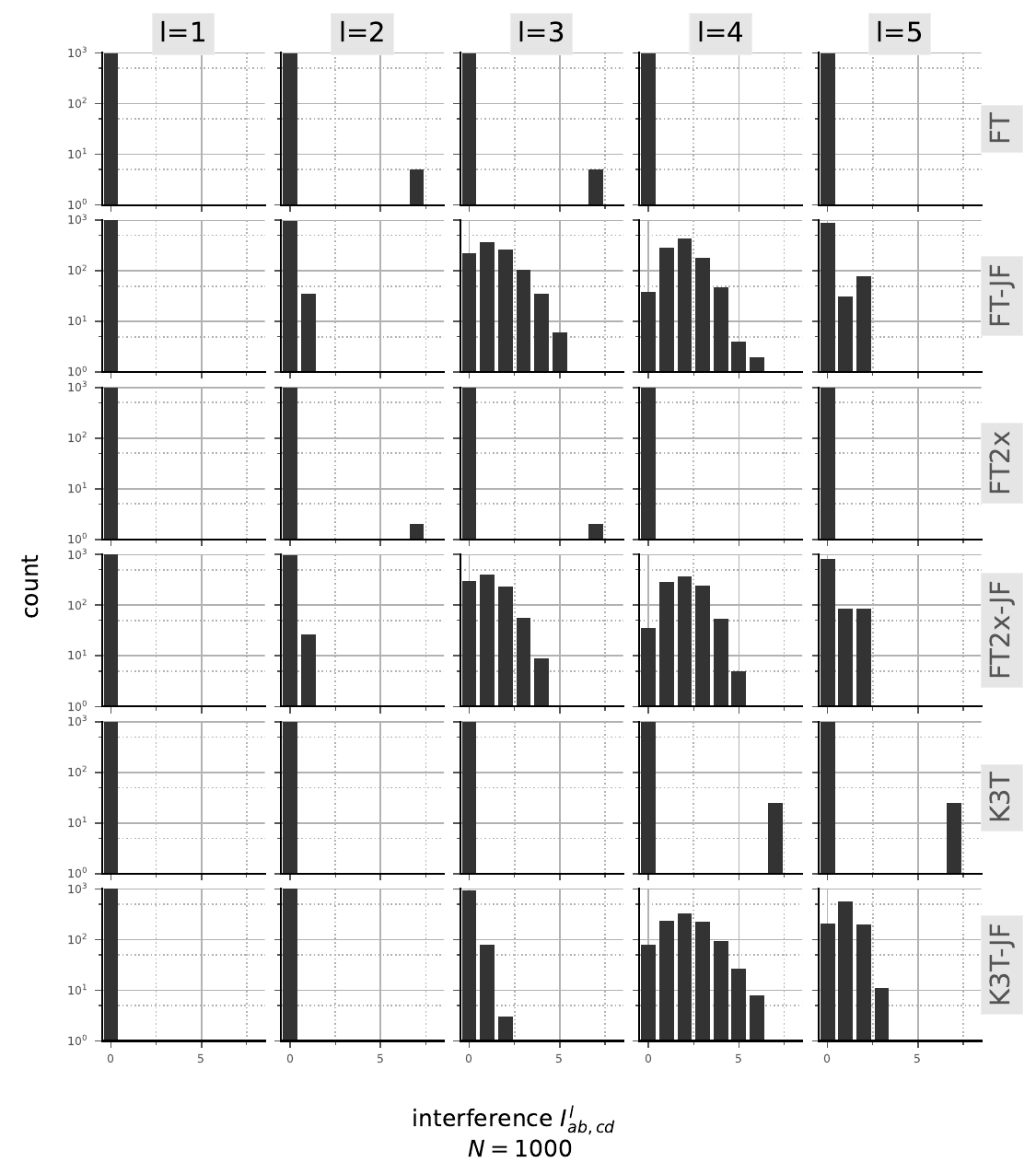}
\caption{Interference plots for k-ary n-trees for different values for k, Fat Trees and Fat Trees2x.
For all base topologies, most node pairs have an interference of zero. Both Fat Tree variants have
a few outliers that have a higher interference for length 2 and 3. The k-ary 3-tree has
these outliers also, although only for $l \geq$ 4. There are also more outlying node pairs than in the
Fat Tree variants.}
\label{fig:appendix_050}
\end{figure*}

\begin{figure*}
\centering
\includegraphics[width=1.0\textwidth]{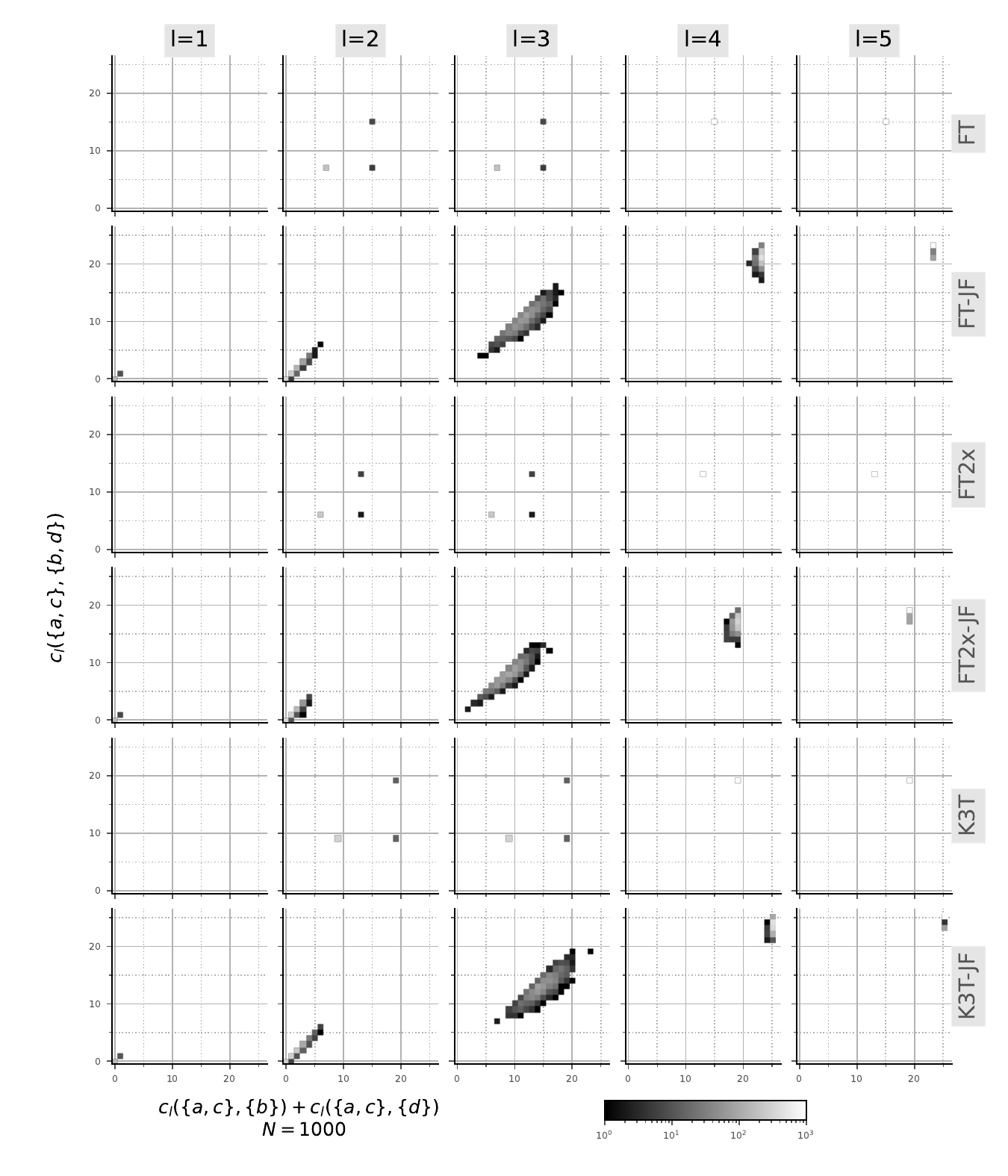}
\caption{Two-dimensional detailed interference plots for the k-ary 3-tree, Fat Tree and Fat
Tree2x. For both components, there exist only specific values they can get. Since values in the
diagonal cancel each other out, we can see that there is only one peak next to interference zero
for all base topologies.}
\label{fig:appendix_051}
\end{figure*}

\begin{figure*}
\centering
\includegraphics[width=1.0\textwidth]{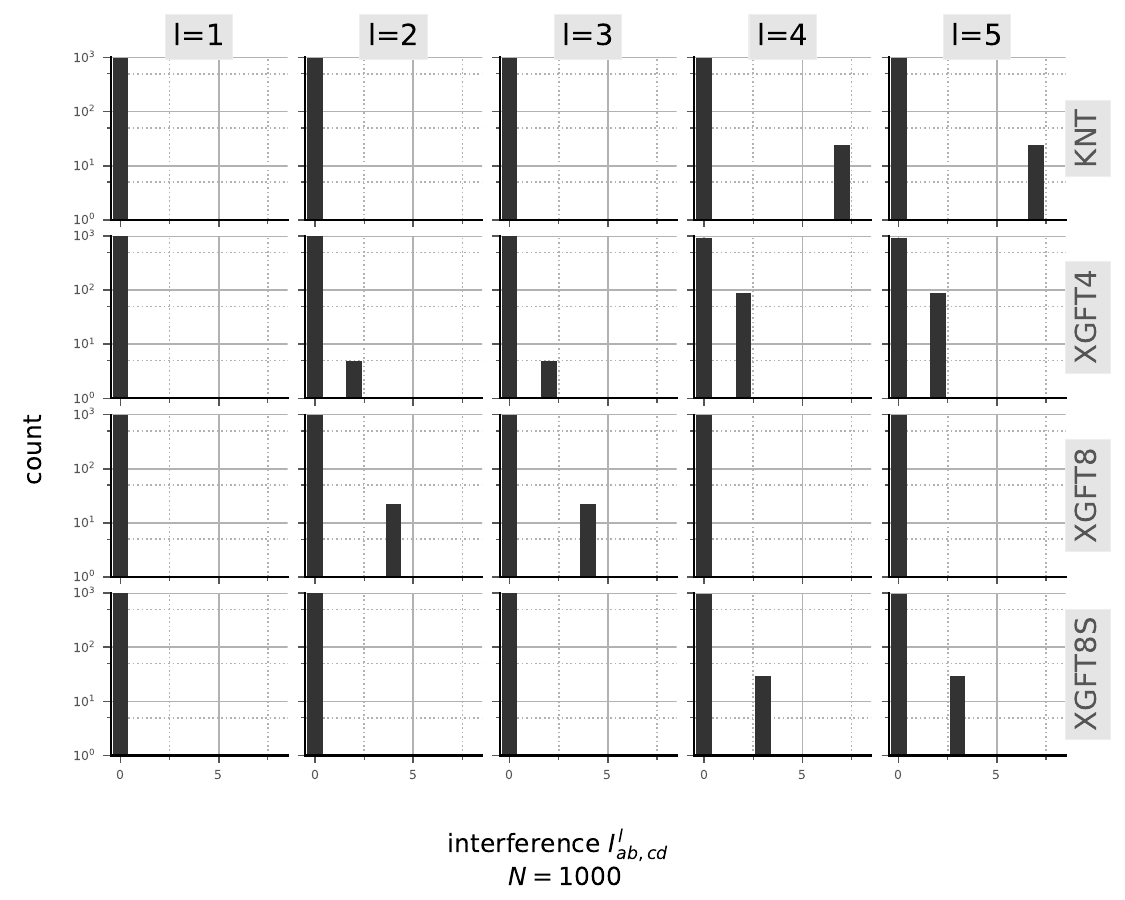}
\caption{Interference plots for xGFTs of different variants and k-ary n-tree. Similar to
before, all topologies have interference of zero for most node pairs. We can also see the peaks
at a specific interference value. xGFT$_8$S has the lowest interference of all of them. It has only
interference higher than zero for $l >$ 3 and the peaks have only interference of 3.
xGFT$_8$ has only an interference $>$ 0 at length 2 and 3.}
\label{fig:appendix_052}
\end{figure*}

\fi

\end{document}